\documentclass[lettersize,journal]{IEEEtran}
\pdfoutput=1
\usepackage{multirow}
\usepackage{makecell}
\usepackage{amsmath,amsfonts}
\usepackage{algorithmic}
\usepackage{algorithm}
\usepackage{array}
\usepackage{textcomp}
\usepackage{stfloats}
\usepackage{url}
\usepackage{verbatim}
\usepackage{graphicx}
\usepackage{cite}

\usepackage{subcaption}
\newsavebox{\bigimage}
\usepackage{float}
\usepackage[shortlabels]{enumitem}
\usepackage{booktabs}
 
\usepackage[hidelinks]{hyperref}
\usepackage{tikz}
\usepackage[T1]{fontenc}
\usepackage[utf8]{inputenc}
\usepackage{pgfplots}
\usepackage{grffile}
\usepackage{colortbl}
\pgfplotsset{compat=newest}
\usetikzlibrary{plotmarks}
\usetikzlibrary{arrows.meta}
\usepgfplotslibrary{patchplots}
\usetikzlibrary{arrows.meta}
\usetikzlibrary{spy}
\newcommand{\midsepremove}{\aboverulesep = 0mm \belowrulesep = 0mm}
\midsepremove
\newcommand{\midsepdefault}{\aboverulesep = 0.605mm \belowrulesep = 0.984mm}
\midsepdefault
\newcommand\mtlarge{\fontsize{24pt}{18pt}\selectfont}

\newcommand\mtlargeTSub{\fontsize{30pt}{20pt}\selectfont}
\newcommand\mtlargeTSubTick{\fontsize{24pt}{18pt}\selectfont}
\newcommand\mtlargeMSub{\fontsize{30pt}{20pt}\selectfont}
\newcommand\mtlargeMSubTick{\fontsize{24pt}{18pt}\selectfont}

\definecolor{pesqcolor}{rgb}{0.00000,0.45000,0.74000}%
\definecolor{ssnrcolor}{rgb}{0.85000,0.30000,0.10000}%
\definecolor{mycolorsr}{rgb}{0.00000,0.44700,0.74100}%

\definecolor{metricgan}{RGB}{242, 133, 0}
\definecolor{demucs}{RGB}{87, 165, 184}
\definecolor{phasen}{rgb}{0.64, 0.76, 0.68}
\definecolor{pfpl}{rgb}{0.6, 0.4, 0.8}
\definecolor{cmgan}{rgb}{0.76, 0.13, 0.28}
\definecolor{arrowred}{rgb}{1,0.141176471,0.18823529411}
\definecolor{arrowblue}{rgb}{0.0156862745,0.51372549019,0.96078431372}
\definecolor{tblcolor}{rgb}{0.75,0.85,0.95}
\colorlet{tblcolortransparent}{tblcolor!85}

\newcommand\figW{4.521in}
\newcommand\figH{1.5in}

\newcommand\figWp{4.0in}
\newcommand\figHp{3.2in}

\newcommand\figWtime{5.421in}
\newcommand\figHtime{3.05in}
\newcommand\SRLineWidthT{1.9pt}

\newcommand\figWbp{3.521in}
\newcommand\figHbp{5.4in}

\newcommand\figWbplt{4.5in}
\newcommand\figHbplt{2.25in}

\newcommand\figWSubMag{5.121in}
\newcommand\figHSubMag{3.5in}

\newcommand\axess{2.5pt}
\newcommand\msize{4.5pt}
\newcommand\bplinewidth{2.5pt}
\newcommand\bpltlinewidth{2.0pt}

\newcommand{\sqdiamond}[1][fill=black]{\tikz [x=1.2ex,y=1.85ex,line 
	width=.1ex,line join=round, yshift=-0.285ex] \draw  [#1]  (0,.5) -- (.5,1) -- 
	(1,.5) -- (.5,0) -- (0,.5) -- cycle;}%
\newcommand{\mydiamond}[1][fill=black]{\mathop{\raisebox{-0.205ex}{$\sqdiamond[#1]$}}}

\newcommand{\metricganrect}
{   \begin{tikzpicture}
		\fill [metricgan](0, 0) rectangle  (0.6, 0.25);
	\end{tikzpicture}
}

\newcommand{\demucsrect}
{   \begin{tikzpicture}
		\fill [demucs](0, 0) rectangle  (0.6, 0.25);
	\end{tikzpicture}
}

\newcommand{\phasenrect}
{   \begin{tikzpicture}
		\fill [phasen](0, 0) rectangle  (0.6, 0.25);
	\end{tikzpicture}
}

\newcommand{\pfplrect}
{   \begin{tikzpicture}
		\fill [pfpl](0, 0) rectangle  (0.6, 0.25);
	\end{tikzpicture}
}

\newcommand{\cmganrect}
{   \begin{tikzpicture}
		\fill [cmgan](0, 0) rectangle  (0.6, 0.25);
	\end{tikzpicture}
}

\newcommand{\myarrowred}[1][0.1pt]
{   \begin{tikzpicture}[overlay]
		\draw [->,>=stealth,line width=0.5mm,arrowred] (0, 0.3) -- (0, -0.1);
	\end{tikzpicture}
}

\newcommand{\myarrowblue}[1][0.1pt]
{   \begin{tikzpicture}[overlay]
		\draw [->,>=stealth,line width=0.5mm,arrowblue] (0, 0.3) -- (0, -0.1);
	\end{tikzpicture}
}

\newcommand{\circledM}{%
	\begin{tikzpicture}[baseline=(char.base)]
		\node[shape=circle,draw,inner sep=0.9pt, minimum size=0.5em] (char) {\fontsize{8}{10}\selectfont M};
	\end{tikzpicture}%
}

\begin{document}
	
	\title{CMGAN: Conformer-Based Metric-GAN for Monaural Speech Enhancement}
	
	\author{Sherif~Abdulatif,~Ruizhe~Cao,~and~Bin~Yang~\IEEEmembership{Senior~Member,~IEEE}\vspace{-6mm}
		\thanks{The authors are with the Institute of Signal Processing and System Theory, University of Stuttgart, 70569~Stuttgart, Germany (e-mail: sherif.abdulatif@
			iss.uni-stuttgart.de; ruizhe.cao96@gmail.com; bin.yang@iss.uni-stuttgart.de).}\vspace{-2mm}
	}
	
	%	\markboth{Journal of \LaTeX\ Class Files,~Vol.~XX, No.~XX, April~2024}%
	%	{Shell \MakeLowercase{\textit{et al.}}: A Sample Article Using IEEEtran.cls for 
	%		IEEE Journals}
	
	
	\maketitle
	
	\begin{abstract}
	In this work, we further develop the conformer-based metric generative adversarial network (CMGAN) model\footnote{A shorter version is available in \textit{\href{https://arxiv.org/abs/2203.15149}
	{https://arxiv.org/abs/2203.15149}} \cite{cao2022cmgan}} for speech enhancement (SE) in the time-frequency (TF) domain. This paper builds on our previous work but takes a more in-depth look by conducting extensive ablation studies on model inputs and architectural design choices. We rigorously tested the generalization ability of the model to unseen noise types and distortions. We have fortified our claims through DNS-MOS measurements and listening tests. Rather than focusing exclusively on the speech denoising task, we extend this work to address the dereverberation and super-resolution tasks. This necessitated exploring various architectural changes, specifically metric discriminator scores and masking techniques. It is essential to highlight that this is among the earliest works that attempted complex TF-domain super-resolution. Our findings show that CMGAN outperforms existing state-of-the-art methods in the three major speech enhancement tasks:  denoising, dereverberation, and super-resolution. For example, in the denoising task using the Voice Bank+DEMAND dataset, CMGAN notably exceeded the performance of prior models, attaining a PESQ score of 3.41 and an SSNR of 11.10 dB. Audio samples and CMGAN implementations are available online\footnote{Open source code is available in \textit{\href{https://github.com/ruizhecao96/CMGAN}{https://github.com/ruizhecao96/CMGAN}}}. 
	\end{abstract}
	
	
	\vspace{-2mm}\section{Introduction}
	\IEEEPARstart{I}{n} real-life speech applications, the perceived speech quality and intelligibility are dependent on the performance of the underlying speech enhancement (SE) systems, e.g., speech denoising, dereverberation and acoustic echo cancellation. As such, SE frameworks are an indispensable component in modern automatic speech recognition (ASR), telecommunication systems and hearing aid devices \cite{weninger2015asr, zheng2021telecommunications, desjardins2014hearingaid}. This is evident by the increasingly large amount of research continuously attempting to push the performance boundaries of current SE systems \cite{wang2018supervised, loizou2007speech}. The majority of these approaches harness the recent advances in deep learning (DL) techniques as well as the increasingly more available speech datasets \cite{valentini2016voicebank, kinoshita2016reverb, barker2015chime, dubey2022dns}.
	
	SE techniques can be roughly categorized into two prominent families of approaches. Chronologically, enhancing the speech time-frequency (TF) representation (spectrogram) constitutes the classical SE paradigm which encompasses the majority of model-based as well as more recent DL approaches \cite{wang2018supervised, fu2019metricgan, yin2020phasen, yu2022dualbranch}. More recently, a new set of approaches were introduced to enhance raw speech time-domain waveform directly without any transformational overheads \cite{pascual2017segan, macartney2018waveunet, wang2021tstnn,defossez2020demucs, kim2021seconformer}. Each paradigm presents unique advantages and drawbacks. Although there are emerging hybrid methods \cite{kim2018fusion,su2021hifi,serra2022universe}, they are not within the scope of this work.
	
	The time-domain paradigm is based on generative models trained to directly estimate fragments of the clean waveform from the distorted counterparts without any TF-domain transformation or reconstruction requirements \cite{macartney2018waveunet, wang2021tstnn}. However, the lack of direct frequency representation hinders these frameworks from capturing speech phonetics in the frequency domain. This limitation is usually reflected as artifacts in the reconstructed speech. Another drawback of this paradigm is the ample input space associated with the raw waveforms, which often necessitates the utilization of deep computationally complex frameworks \cite{pascual2017segan, defossez2020demucs}.
	
	In the TF-domain, most conventional model-based or DL techniques prioritize the magnitude component and often overlook the phase. This omission stems from the complex and unpredictable nature of the phase component, which contrasts the more structured magnitude. Such intricacy in the phase poses challenges to the employed architectures \cite{abdulatif2020aegan,abdulatif2021crossdomain}. To circumvent this challenge, several approaches follow the strategy of enhancing the complex spectrogram (real and imaginary parts), which implicitly enhances both magnitude and phase \cite{williamson2016cirm, tan2019crn}. However, the compensation effect between the magnitude and phase often leads to an inaccurate magnitude estimation \cite{wang2021compensation}. This problem will be discussed in details in Sec.~\ref{sec:denoising}. Recent studies propose enhancing the magnitude followed by complex spectrogram refinement, which can alleviate the compensation problem effectively \cite{yu2022dualbranch, li2022glance}. Furthermore, the commonly used objective function in SE is simply the $\ell^p-$norm distance between the estimated and the target spectrograms. Nevertheless, a lower distance does not always lead to higher speech quality. MetricGAN is proposed to overcome this issue by optimizing the generator with respect to the evaluation metric score that can be learned by a discriminator \cite{fu2019metricgan}. 
	
	In addition, many approaches utilize transformers \cite{vaswani2017attention} to capture the long-term dependency in the waveform or the spectrogram \cite{yu2022dualbranch, wang2021tstnn, dang2021dpt}. Recently, conformers have been introduced as an alternative to transformers in ASR and speech separation tasks due to their capacity of capturing both local context and global context \cite{gulati2020conformer,chen2021ssconformer}. Accordingly, they were also employed for time-domain and TF-domain SE \cite{kim2021seconformer,koizumi2021dfconformer}. 
	
	\begin{figure*}[t!]
		\captionsetup[subfigure]{justification=centering}
		\centering
		\centerline{
			\begin{subfigure}[b]{.23\textwidth}
				\centering
%				{\includegraphics[width=\columnwidth]{Figs/p226_examples/track_clean_p226.pdf}}
				{\resizebox{\columnwidth}{!}{\begin{tikzpicture}
\begin{axis}[%
width=\figWSubMag,
height=\figHSubMag,
at={(0in,0in)},
scale only axis,
axis on top,
xmin=0.0275625,
xmax=4.4690625,
xlabel style={font=\color{white!15!black}, yshift = -4pt},
xlabel={\mtlargeMSub Time [s]},
ylabel={\mtlargeMSub Frequency [kHz]},
ytick={0,2,4,6,8},
ytick style={draw=none},
xtick style={draw=none},
yticklabel style = {font=\mtlargeMSubTick},
xticklabel style = {font=\mtlargeMSubTick, yshift = -5pt},
ymin=-0.00800819625565892,
ymax=8.00018805940326,
ylabel style={font=\color{white!15!black}, yshift = 8pt},
axis background/.style={fill=white},
legend style={legend cell align=left, align=left, draw=white!15!black}
]
\addplot [forget plot] graphics [xmin=0.0275625, xmax=4.4690625, ymin=-0.00800819625565892, ymax=8.00018805940326] {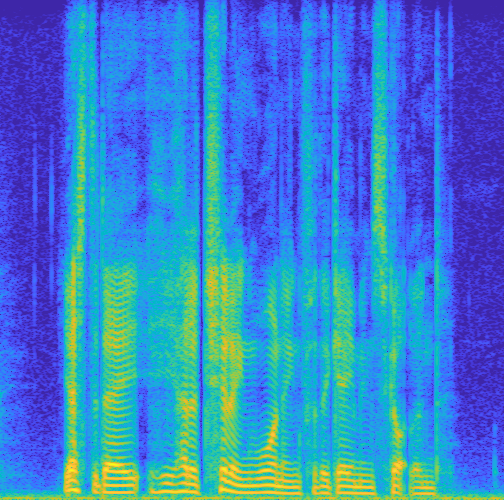};
\end{axis}
\end{tikzpicture}}}
				\captionsetup[subfigure]{justification=centering}\caption{Reference track}
				\label{fig:t_clean}
			\end{subfigure}
			\hspace{3.7mm}
			\begin{subfigure}[b]{.23\textwidth}
				\centering
%				{\includegraphics[width=\columnwidth]{Figs/p226_examples/track_noisy_p226_85th.pdf}}
				{\resizebox{\columnwidth}{!}{\begin{tikzpicture}
\begin{axis}[%
width=\figWSubMag,
height=\figHSubMag,
at={(0in,0in)},
scale only axis,
axis on top,
xmin=0.0275625,
xmax=4.4690625,
xlabel style={font=\color{white!15!black}, yshift = -4pt},
xlabel={\mtlargeMSub Time [s]},
ylabel={\mtlargeMSub Frequency [kHz]},
ytick={0,2,4,6,8},
ytick style={draw=none},
xtick style={draw=none},
yticklabel style = {font=\mtlargeMSubTick},
xticklabel style = {font=\mtlargeMSubTick, yshift = -5pt},
ymin=-0.00800819625565892,
ymax=8.00018805940326,
ylabel style={font=\color{white!15!black}, yshift = 8pt},
axis background/.style={fill=white},
legend style={legend cell align=left, align=left, draw=white!15!black}
]
\addplot [forget plot] graphics [xmin=0.0275625, xmax=4.4690625, ymin=-0.00800819625565892, ymax=8.00018805940326] {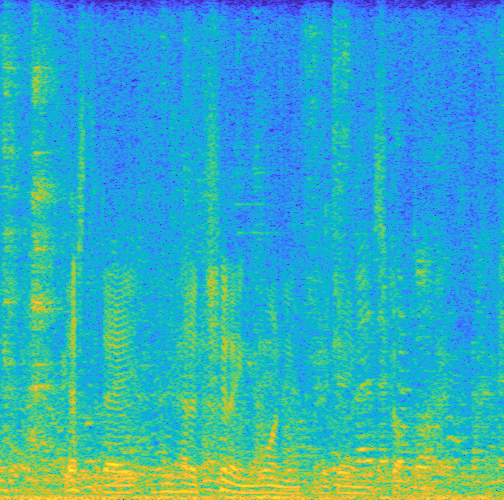};
\end{axis}
\end{tikzpicture}%
}}
				\caption{Noisy track (SNR 0 dB)}
				\label{fig:t_noisy}
			\end{subfigure}
			\hspace{2.7mm}
			\begin{subfigure}[b]{.23\textwidth}
				\centering
%				{\includegraphics[width=\columnwidth]{Figs/p226_examples/track_reverb_p226.pdf}}
				{\resizebox{\columnwidth}{!}{\begin{tikzpicture}
\begin{axis}[%
width=\figWSubMag,
height=\figHSubMag,
at={(0in,0in)},
scale only axis,
axis on top,
xmin=0.0275625,
xmax=4.4690625,
xlabel style={font=\color{white!15!black}, yshift = -4pt},
xlabel={\mtlargeMSub Time [s]},
ylabel={\mtlargeMSub Frequency [kHz]},
ytick={0,2,4,6,8},
ytick style={draw=none},
xtick style={draw=none},
yticklabel style = {font=\mtlargeMSubTick},
xticklabel style = {font=\mtlargeMSubTick, yshift = -5pt},
ymin=-0.00800819625565892,
ymax=8.00018805940326,
ylabel style={font=\color{white!15!black}, yshift = 8pt},
axis background/.style={fill=white},
legend style={legend cell align=left, align=left, draw=white!15!black}
]
\addplot [forget plot] graphics [xmin=0.0275625, xmax=4.4690625, ymin=-0.00800819625565892, ymax=8.00018805940326] {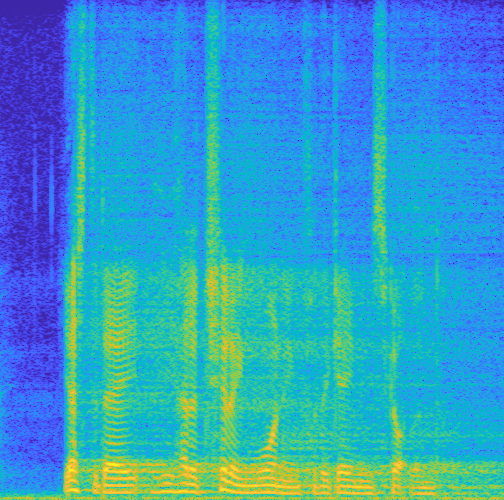};
\end{axis}
\end{tikzpicture}%
}}
				\caption{Reverb. track ($\tau=$ 0.5 s)}
				\label{fig:t_reverb}
			\end{subfigure}
			\hspace{2.7mm}
			\begin{subfigure}[b]{.23\textwidth}
				\centering
%				{\includegraphics[width=\columnwidth]{Figs/p226_examples/track_lr_p226.pdf}}
				{\resizebox{\columnwidth}{!}{\begin{tikzpicture}
\begin{axis}[%
width=\figWSubMag,
height=\figHSubMag,
at={(0in,0in)},
scale only axis,
axis on top,
xmin=0.0275625,
xmax=4.4690625,
xlabel style={font=\color{white!15!black}, yshift = -4pt},
xlabel={\mtlargeMSub Time [s]},
ylabel={\mtlargeMSub Frequency [kHz]},
ytick={0,2,4,6,8},
ytick style={draw=none},
xtick style={draw=none},
yticklabel style = {font=\mtlargeMSubTick},
xticklabel style = {font=\mtlargeMSubTick, yshift = -5pt},
ymin=-0.00800819625565892,
ymax=8.00018805940326,
ylabel style={font=\color{white!15!black}, yshift = 8pt},
axis background/.style={fill=white},
legend style={legend cell align=left, align=left, draw=white!15!black}
]
\addplot [forget plot] graphics [xmin=0.0275625, xmax=4.4690625, ymin=-0.00800819625565892, ymax=8.00018805940326] {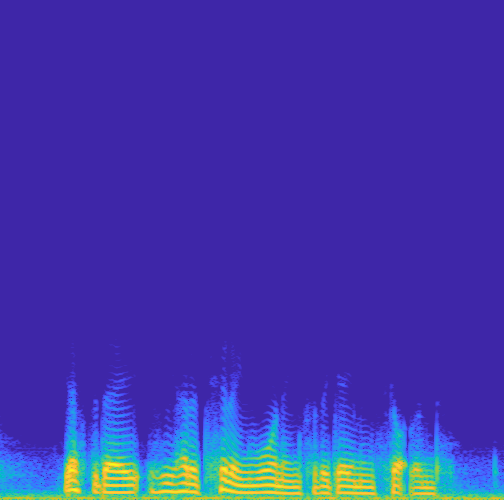};
\end{axis}
\end{tikzpicture}%
}}
				\captionsetup[subfigure]{justification=centering}\caption{Low-res. track ($s=4$)}
				\label{fig:t_low}
			\end{subfigure}
		}
		\caption{The TF-magnitude representation of distorted speech for different SE tasks, i.e., denoising, dereverberation and bandwidth extension (super-resolution). The variable $\tau$ represents the 60 dB reverberation time and $s$ is the bandwidth upscaling ratio. \label{fig:tracks}}
		\vspace{-5mm}
	\end{figure*}
	
	Inspired by the stated problems and previous works, we propose the first conformer-based MetricGAN (CMGAN) for various monaural speech enhancement tasks. At its core, the CMGAN combines a generator and a metric discriminator. The latter excels in estimating and optimizing a black-box non-differentiable metric without adversely affecting other metrics. Drifting from traditional paradigms where dual-branch models employed separate networks for magnitude masking and complex refinement, often necessitating resource-intensive interaction modules. Instead, our generator utilizes a shared encoder that ingests the concatenated magnitude and complex (both real and imaginary) parts. To mitigate computational intensity of the conformer, we integrate dual-path transformers~\cite{wang2021tstnn, dang2021dpt,chen2020dualpath} in a two-stage block, sequentially capturing temporal and frequency dependencies from shared encoded representations. Subsequently, the architecture splits into a dedicated mask decoder for magnitude interpretation and another branch to refine both real and imaginary facets. Our contributions can be summarized as follows:
	\begin{itemize}[label=$\bullet$,wide = 0pt]
		\item We present a new generator leveraging a shared encoder that takes concatenated magnitude and complex components. Additionally, It employs a dedicated mask decoder and a shared decoder for both real and imaginary parts, optimizing enriched shared representation learning with fewer parameters. 
		\item  We employ two-stage conformer blocks, capitalizing on their capacity to discern both local and global dependencies across time and frequency domains. 
		\item Instead of solely optimizing for point-wise loss functions, we integrate a metric discriminator to also include perceptual elements, thereby enhancing the resultant speech quality.
	\end{itemize}
	
	In this follow-up research, we have broadened the scope beyond the focus on speech denoising in our conference paper to include dereverberation and super-resolution. 
	This expansion facilitates a detailed exploration of our original model, leading to several key contributions:
	\begin{itemize}[label=$\bullet$,wide = 0pt]
		\item \textbf{Denoising:} We study the generalization capability of the proposed CMGAN relative to state-of-the-art methods under unseen distortions. Specifically, CMGAN is trained on the Voice Bank+DEMAND dataset and tested on three different datasets spanning both real and simulated distortions. Additionally, we demonstrated the superior performance of CMGAN across various opinion score evaluations.
		\item \textbf{Dereverberation:} Beyond merely testing the model on new data, our exploration was profound. We conducted an in-depth comparative analysis focusing on the metric discriminator. This involved evaluating multiple objective scoring metrics and an examination of the trade-offs associated with each.
		\item \textbf{Super-resolution:} Our research delves into an area not extensively covered in recent works: super-resolution within a complex TF representation. While traditional mapping-based methods are used for super-resolution, we innovatively incorporate masking by adding a reconstructed mask. This enabled our trained network to focus mainly on estimating the missing high-frequency bands. Our ablation studies in super-resolution manifested the strength of complex TF-domain approaches.
	\end{itemize}
	\vspace{-2mm}
	\section{Problem Statement \& Relevant Literature}
	In this paper, the proposed CMGAN will be evaluated on different SE tasks, namely speech denoising, dereverberation and super-resolution. Accordingly, for any acoustic environment the aforementioned SE tasks can be modeled as follows:
	\begin{equation}\label{eq:model}
		y(t) = x(t)*h(t) + n(t),
	\end{equation}
	where $y(t)$ is the distorted speech, $x(t)$ is the required clean speech, $n(t)$ is a background noise and `*' is a convolution operation with a filter $h(t)$. However, due to space constraints, this study will focus on evaluating each task alone and not the superimposed effects, as shown in Fig.~\ref{fig:tracks}. Hence, for denoising the additive background noise $n(t)$ will only be considered (Fig.~\ref{fig:t_noisy}). For dereverberation (Fig.~\ref{fig:t_reverb}), the filter $h(t)$ will represent a room impulse response (RIR) filter. Finally, $h(t)$ will function as a low pass filter (LPF) in the super-resolution task to simulate the impact of low sampling frequency (Fig.~\ref{fig:t_low}). The pertinent literature for each task will be presented in the following subsections.
	
	\vspace{-3mm}
	\subsection{Denoising}\label{sec:denoising}
	Speech denoising is considered as a source separation problem, where the objective is to suppress the background noise $n(t)$ and predict the desired speech $\hat{x}(t)$ with maximum possible quality and intelligibility. Accordingly, the difficulty of this problem would highly depend on the nature of both the desired speech and the background noise. For instance, speech signals are highly non-stationary. As for the noise component, it can be divided into stationary scenarios (e.g., computer fan noise and air conditioners) and non-stationary scenarios (e.g., babble and street noise). Usually, the latter scenario is more challenging, as in these cases, the noise would occupy similar frequency bands as the desired speech \cite{abdulatif2020aegan}.
	
	In the speech denoising literature, due to the non-stationary nature of the problem, exploring the TF representations of the superimposed signal to reflect the time-varying frequency properties of the waveform is the typical approach \cite{wang2018supervised,purwins2019overview,michelsanti2021overview}. The only limitation arising from the TF-domain denoising is the unstructured phase representation. However, for a long time phase was considered insensitive to noise \cite{wang1982unimportancephase}. As a result, research mostly focused on magnitude denoising while maintaining the noisy phase \cite{loizou2007speech}. Recently, many studies pointed out the importance of the phase on the denoised speech quality \cite{williamson2016cirm,paliwal2011importancephase}. To this end, TF speech denoising can be categorized into mapping-based and masking-based methods.
	
	For mapping-based methods, a non-linear function is utilized to map the noisy speech to a corresponding denoised speech. These methods were first visited in time-domain speech denoising \cite{macartney2018waveunet,rethage2018wavenet,fu2018waveform,pandey2019tcnn}. For instance, SEGAN \cite{pascual2017segan} is introduced as an adversarial framework to map the noisy waveform to a corresponding denoised speech. Variants of SEGANs are also proposed to increase the capacity of the generator \cite{phan2020segan}, or using an additional TF-domain loss to benefit from both domains \cite{pascual2019segan}. Building upon these trials, different mapping-based adversarial frameworks are also investigated on TF-domain speech denoising and they achieved more promising results \cite{donahue2018fsegan,michel2017cgan,meng2018ganmapping,abdulatif2020aegan}. 
	
	On the other hand, masking-based methods are mostly utilized in TF-domain with few trials on time-domain speech denoising \cite{luo2019convtasnet}. TF-domain masking-based methods operate under the assumption that two signals are considered to be W-disjoint orthogonal if their short-time Fourier transformations (STFT) do not overlap \cite{yilmaz2002wdisjoint}. Accordingly, it is possible to demix the signals by determining the active source in each TF unit. Inspired by the auditory masking phenomenon and the exclusive allocation principle in auditory scene analysis \cite{bregman1994asc}, ideal binary masking (IBM) is the first masking-based method utilized in supervised speech denoising \cite{yilmaz2004bss}. In IBM, a mask is generated by assigning a value of 1 for a TF unit if the signal-to-noise ratio (SNR) in this unit exceeds a predefined threshold (required speech) and 0 otherwise (noise to suppress). In other words, IBM can be treated as a binary classification problem \cite{wang2013ibm,healy2013ibm}. Although IBM has been shown to considerably improve speech intelligibility, it can degrade the speech quality by introducing musical noise distortions \cite{hummer2014irm}. Ideal ratio masking (IRM) is introduced as a remedy and it can be viewed as a soft version of the IBM, where each TF unit can take a value between 0 and 1 depending on the corresponding signal and noise powers \cite{Sirni2006irm,narayanan2013irm}. Spectral magnitude mask (SMM) is considered as an unbounded variant of IRM \cite{wang2014smm}. 
	
	The aforementioned masking-based methods would solely enhance the magnitude and keep the noisy phase unaltered. Subsequently, tackling the phase is divided into phase reconstruction and phase denoising approaches. For phase reconstruction, deep neural networks (DNNs) are trained to estimate the magnitude, which is then used for iterative phase reconstruction (IPR) \cite{han2015mapping,wang2018phaserec,wang2019phaserec,zhao2019phaserec}. As for phase denoising, authors in \cite{erdogan2015psm} are the first to introduce a phase-sensitive mask (PSM) as a variant of SMM and they claimed a considerable improvement in speech quality. Using IRM as a foundation, a complex ideal ratio masking (cIRM) approach is proposed that can operate on the real and imaginary parts, implicitly addressing both magnitude and phase denoising \cite{williamson2016cirm}. Nevertheless, since the real and imaginary parts are not necessarily positive, the authors would compress the cIRM with a tanh activation to obtain values between $-\textrm{1}$ and 1. The idea of cIRM is further extended by incorporating a deep complex-valued recurrent neural network (DCCRN) and new loss functions to estimate the relevant masks \cite{hu2020dccrn}.
	
	The main drawback behind these approaches is the magnitude and phase compensation effect discussed in \cite{wang2021compensation}. In this case, denoising the complex representations using only a complex loss (penalizing real and imaginary parts) would implicitly provide the trained model with a certain degree of freedom in estimating the magnitude and phase. Since the phase is unstructured and always challenging to estimate, this might result in an inaccurate magnitude estimation to compensate for the challenging phase. This problem can be mitigated by including both complex and magnitude losses or by complex refinement approaches, which basically decouple the problem into estimating a bounded mask for the magnitude followed by a complex refinement branch to further improve the magnitude and estimate the phase from the denoised complex representations \cite{yu2022dbt,yu2022dualbranch,li2021ts,li2021sim,li2022glance}. However, since recent studies recommended mapping-based methods over the preceding masking-based approaches for complex spectrogram estimation \cite{tan2019crn,tan2020crn}, the complex refinement branch would follow a mapping-based approach. In this sense, the model can combine the fragmented benefits of both masking-based and mapping-based methods.
	
	\vspace{-3mm}
	\subsection{Dereverberation} \label{sec:reverb}
	In an enclosed acoustic environment, the sound is perceived as a superposition of three distinct components: direct path, early reflections and late reverberations, which can be modeled by the convolutive RIR filter $h(t)$ in Eq.~\ref{eq:model} \cite{kuttruff2016acoustics}. Thus, speech dereverberation would mainly focus on suppressing the unwanted reflections and maintaining the direct path representing the estimated desired speech $\hat{x}(t)$. Early reflections usually arrive shortly (50 ms) at the microphone as they come from a specific direction, thus they can be addressed as an attenuated copy of the direct path. In contrast, late reverberations arrive later as they represent delayed and attenuated superimposed signals from different directions. The difficulty of the dereverberation problem is accounted to different factors. For instance, room size and surface properties mainly contribute to the amount of reflections and degree of attenuation \cite{schultz1971room}. Additionally, the distance between the microphone and the speaker would affect the reflection strength, i.e., the longer the distance, the stronger the reflections \cite{gelbart2002far}.
	
	To the best of our knowledge, the dereverberation problem is usually addressed in TF-domain with limited trials on time-domain \cite{defossez2020demucs,luo2018reverb}. This is due to the fact that time-domain models are prone to temporal distortions, which are severe in reverberant conditions. Similar to denoising, TF-domain masking-based methods are also extended to dereverberation. For instance, in \cite{roman2011reverb}, direct path and early reflections are considered as the desired speech and an IBM is utilized to suppress late reverberations. Unlike denoising, the SNR criteria for assigning 0 and 1 in each TF unit is modified in \cite{may2014reverb} to address the speech presence probability. However, IBM is originally defined for additive noise under anechoic conditions. In reverberation, temporal smearing of speech is observed in the resultant TF representation, as shown in Fig.~\ref{fig:t_reverb}. Hence, IBM with hard boundaries can cause a degradation in the resultant speech quality \cite{jin2007reverb} and soft IRM is usually the preferred method in this case \cite{wang2014smm,zhao2016reverb,zhang2016reverb,li2017reverb}. Following the denoising path, IRM is extended with cIRM to include phase in the dereverberation process \cite{williamson2017reverb,williamson2017reverbj,kothapally2022reverb}. 
	
	Furthermore, mapping-based methods are also investigated in speech dereverberation. For instance, Han \emph{et~al.} \cite{han2015mapping} is one of the first to investigate spectral mapping on dereverberation using a simple fully connected  network. Later, authors in \cite{ernst2018reverb} applied a fully convolutional U-Net (encoder-decoder) architecture with intermediate skip connections for this task. The SkipConvNet changed the U-Net architecture by replacing each skip connection with multiple convolutional modules to provide the decoder with intuitive feature maps \cite{kothapally2020reverb}. Additionally, a wide residual network is introduced in \cite{ribas2019reverb} to process different speech representations in the TF-domain, namely the magnitude of the STFT, Mel filterbank and cepstrum. Some approaches are able to provide significant performance gain by combining DNNs with conventional methods such as delay-and-sum beamforming and late reverberation reduction by spectral subtraction \cite{xiao2014reverb}.

	\vspace{-3.5mm}
	\subsection{Super-resolution} \label{sec:sr}
	The super-resolution problem is slightly different from prior SE use cases. In denoising and dereverberation, the desired speech is available with superimposed unwanted noise or reflections and the task is to suppress these effects while preserving the speech. In contrast, super-resolution would reconstruct the missing samples from a low sampling frequency input signal. Accordingly, this problem can be formulated from two different perspectives based on the input domain. In the time-domain, the problem is closely related to super-resolution in natural images \cite{dong2016sr}, where the task is to upsample an input signal of $K\!\times\!1$ samples to an output signal of $M\!\times\!1$ samples ($K\!<\!M$). In this case, a DNN can be trained for an interpolation task. On the other hand, for TF-domain, the task would rather resemble natural image inpainting \cite{yu2018inpaint}, where a part of the image or spectrogram is missing and the DNN is trained to complete the image or reconstruct the missing high-frequency bands, as shown in Fig.~\ref{fig:t_clean} and \ref{fig:t_low}. Based on the previous description, it can be deduced that mapping-based is the only relevant approach in super-resolution.
	
	In conventional audio processing, super-resolution has been investigated under the name of bandwidth extension \cite{ekstrand2002bw}. Recently, DL-based audio super-resolution studies demonstrated superior performance compared to traditional methods. In 2017, Kuleshov \textit{et al.} \cite{kuleshov2017sr} proposed to use U-Net with skip connection architecture to reconstruct the waveform. TFiLM \cite{birnbaum2019sr} and AFiLM \cite{rakotonirina2021sr} utilized recurrent models and attention blocks to capture the long-range time dependencies, respectively. However, the lack of frequency components limits further improvements in the performance. TFNet \cite{lim2018sr} utilized both time and frequency domain by employing two branches, one branch models the reconstruction of spectral magnitude and the other branch models the waveform. However, the phase information is ignored in the frequency branch. Wang \textit{et al.} \cite{wang2021sr} proposed a time-domain modified autoencoder (AE) and a cross-domain loss function to optimize the hybrid framework. Recently, authors in \cite{liu2022nvsr} proposed a neural vocoder based framework (NVSR) for the super-resolution task. While the above studies show promising results, many of them focus on the time-domain or hybrid time-domain and TF-domain magnitude representations. Nevertheless, the research on complex TF-domain super-resolution is not yet addressed.
		\begin{figure*}[t!]
		\includegraphics[width=\textwidth]{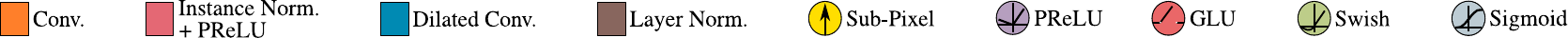}
		\vspace{-1mm}
		\centering
		\sbox{\bigimage}{
			\begin{subfigure}[b]{.485\textwidth}
				\centering
				\includegraphics[width=\textwidth]{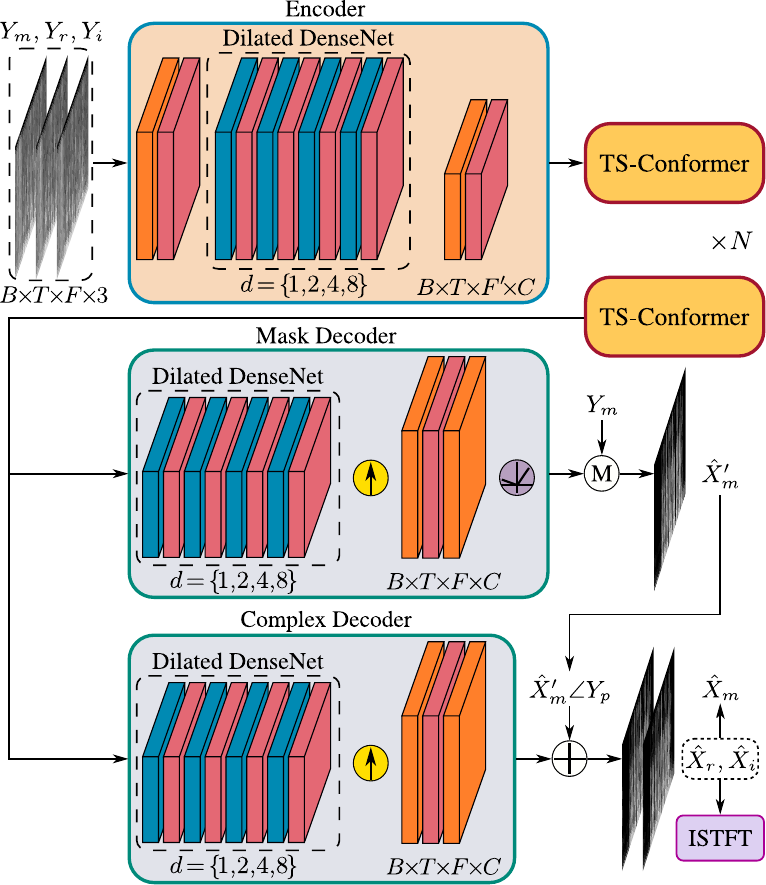}
				\caption{Encoder-decoders generator architecture}
				\label{fig:generator}
				\vspace{0pt}% reference point at the very bottom
			\end{subfigure}
		}
		
		\usebox{\bigimage}\hfill
		\begin{minipage}[b][\ht\bigimage][s]{.485\textwidth}
			\begin{subfigure}{\textwidth}
				\centering
				\includegraphics[height=6.476cm,width=\textwidth]{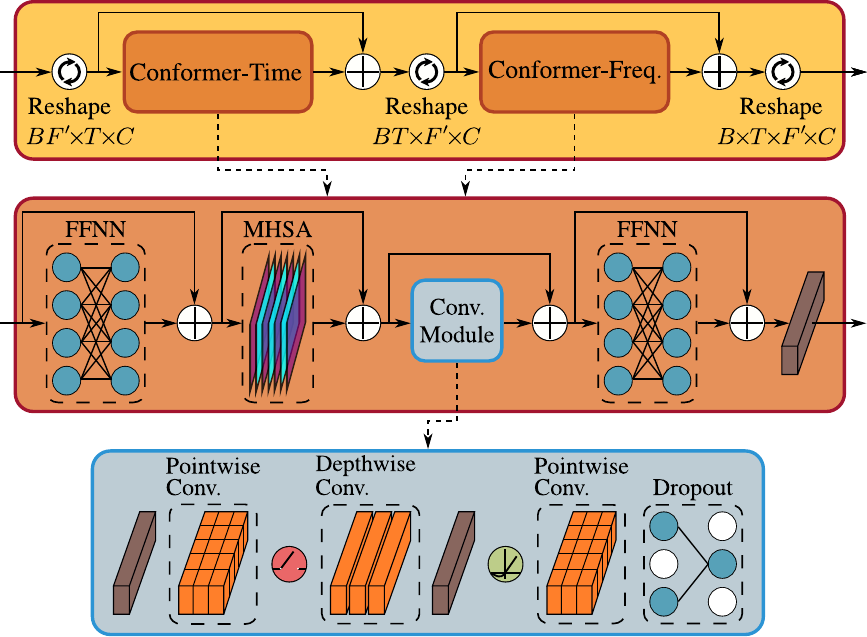}
				\caption{Two-stage conformer (TS-Conformer)}
				\label{fig:TS-Conformer}
			\end{subfigure}
			\vfill
			\begin{subfigure}{\textwidth}
				\centering
				\includegraphics[height=2.418cm,width=\textwidth]{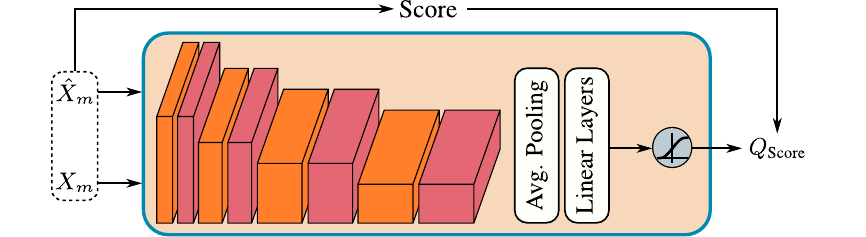}
				\caption{Metric discriminator}
				\label{fig:discriminator}
			\end{subfigure}
			
			\vspace{0pt}% reference point at the very bottom
		\end{minipage}
		\caption{An overview of the proposed CMGAN architecture. The masking operator \protect\circledM{} in (a) denotes task-specific operations: element-wise multiplication for speech denoising/dereverberation and addition for super-resolution (details in Sec.~\ref{sec:sr_results}).}
		\label{fig:overview}
		\vspace{-5mm}
	\end{figure*}
	\vspace{-2mm}
	\section{Methodology}
	\subsection{Generator architecture}\label{sec:generator}
	An overview of the generator architecture of CMGAN is shown in Fig.~\ref{fig:generator}. For a distorted speech waveform $y\in \mathbb{R}^{L\times 1}$, an STFT operation first converts the waveform into a complex spectrogram $Y_{o}\in \mathbb{R}^{T \times F\times 2}$, where $T$ and $F$ denote the time and frequency dimensions, respectively. Then the compressed spectrogram $Y$ is obtained by the power-law compression:
	\begin{equation}
		Y=|Y_o|^ce^{jY_p} = Y_me^{jY_p}=Y_r+jY_i,
	\end{equation}
	where $Y_m$, $Y_p$, $Y_r$ and $Y_i$ denote the magnitude, phase, real and imaginary components of the compressed spectrogram, respectively. $c$ is the compression exponent which ranges from 0 to 1, here we follow Braun \textit{et al.} \cite{braun2020loss} to set $c=0.3$. The power-law compression of the magnitude equalizes the importance of quieter sounds relative to loud ones, which is closer to human perception of sound \cite{lee2018phase, wilson2018exploring}. The real and imaginary parts $Y_r$ and $Y_i$ are then concatenated with the magnitude $Y_m$ as an input to the generator.
	
	\subsubsection{Encoder}
	Given the input feature $Y_{in} \in \mathbb{R}^{B\times T\times F\times 3}$, where $B$ denotes the batch size, the encoder consists of two convolution blocks with a dilated DenseNet \cite{pandey2020densely} in between. Each convolution block comprises a convolution layer, an instance normalization \cite{ulyanov2016instancenorm} and a PReLU activation \cite{he2015prelu}. The first convolution block is used to extend the three input features to an intermediate feature map with $C$ channels. The dilated DenseNet contains four convolution blocks with dense residual connections, the dilation factors of each block are set to \{1, 2, 4, 8\}. The dense connections can aggregate all previous feature maps to extract different feature levels. As for the dilated convolutions, they serve to increase the receptive field effectively while preserving the kernels and layers count. The last convolution block is responsible for halving the frequency dimension to $F'=F/2$ to reduce the complexity.
	
	\subsubsection{Two-stage conformer block}
	Conformers \cite{gulati2020conformer,chen2021ssconformer} achieved great success in speech recognition and separation as they combine the advantages of both transformers and convolutional neural networks (CNNs). Transformers can capture long-distance dependencies, while CNNs exploit local features effectively. Here we employ two conformer blocks sequentially to capture the time dependency in the first stage and the frequency dependency in the second stage. As shown in the Fig.~\ref{fig:TS-Conformer}, given a feature map $D\in \mathbb{R}^ {B \times T \times F' \times C}$, the input feature map $D$ is first reshaped to $D^{T} \in \mathbb{R}^{BF'\times T \times C}$ to capture the time dependency in the first conformer block. Then the output $D^{T}_o$ is element-wise added with the input $D^T$ (residual connection) and reshaped to a new feature map $D^{F}\in \mathbb{R}^{BT\times F' \times C}$. The second conformer thus captures the frequency dependency. After the residual connection, the final output $D_o$ is reshaped back to the input size.
	
	Similar to \cite{gulati2020conformer}, each conformer block utilizes two half-step feed-forward neural networks (FFNNs). Between the two FFNNs, a multi-head self-attention (MHSA) with four heads is employed, followed by a convolution module. The convolution module depicted in Fig.~\ref{fig:TS-Conformer} starts with a layer normalization, a point-wise convolution layer and a gated linear unit (GLU) activation to diminish the vanishing gradient problem. The output of the GLU is then passed to a 1D-depthwise convolution layer with a swish activation function, then another point-wise convolution layer. Finally, a dropout layer is used to regularize the network. Also, a residual path connects the input to the output.
	
	\subsubsection{Decoder} 
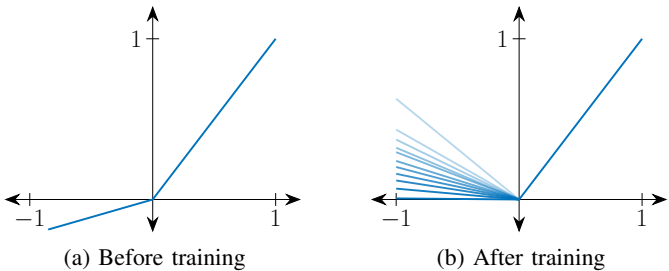
\begin{figure}[b!]
		\vspace{-2mm}
		\centering
		\begin{subfigure}[b]{0.45\columnwidth}
			\centering
			\resizebox{\columnwidth}{!}{\begin{tikzpicture}
	\pgfmathsetlengthmacro\MajorTickLength{
		\pgfkeysvalueof{/pgfplots/major tick length} * 2.5
	}
	\begin{axis}[
		width=\figWp,
		height=\figHp,
		axis lines=middle,
		xmax=1.2,
		xmin=-1.2,
		ymin=-0.2,
		ymax=1.2,
		xtick={-1,1},
		ytick={1},
		major tick length=\MajorTickLength,
		every tick/.style={
			black,
			semithick,
		},
		every x tick label/.append style={font=\huge},
		every y tick label/.append style={font=\huge},
		axis line style={{Stealth[scale=\axess]}-{Stealth[scale=\axess]}}
		]
		\addplot [domain=-0.85:1, samples=500,
		ultra thick, pesqcolor] {max(0.22 * x, x)};
	\end{axis}
\end{tikzpicture}}
			\caption{Before training}
		\end{subfigure}
		\hfill
		\begin{subfigure}[b]{0.45\columnwidth}
			\centering
			\resizebox{\columnwidth}{!}{\begin{tikzpicture}
	\pgfmathsetlengthmacro\MajorTickLength{
		\pgfkeysvalueof{/pgfplots/major tick length} * 2.5
	}
	\begin{axis}[
		width=\figWp,
		height=\figHp,
		axis lines=middle,
		xmax=1.2,
		xmin=-1.2,
		ymin=-0.2,
		ymax=1.2,
		xtick={-1,1},
		ytick={1},
		major tick length=\MajorTickLength,
		every tick/.style={
			black,
			semithick,
		},
		every x tick label/.append style={font=\huge},
		every y tick label/.append style={font=\huge},
		axis line style={{Stealth[scale=\axess]}-{Stealth[scale=\axess]}}
		]
		\addplot [domain=-1:1, samples=5, ultra thick, pesqcolor, opacity=1] {max(-0.0067 * x, x)};
		\addplot [domain=-1:0, samples=5, ultra thick, pesqcolor, opacity=0.92] {max(-0.0676 * x, x)};
%		\addplot [domain=-1:0, samples=5, ultra thick, blue, opacity=0.92] {max(-0.0945 * x, x)};
%		\addplot [domain=-1:0, samples=5, ultra thick, blue, opacity=0.92] {max(-0.1118 * x, x)};
%		\addplot [domain=-1:0, samples=5, ultra thick, blue, opacity=0.92] {max(-0.1118 * x, x)};
		\addplot [domain=-1:0, samples=5, ultra thick, pesqcolor, opacity=0.84] {max(-0.1183 * x, x)};
%		\addplot [domain=-1:0, samples=5, ultra thick, blue, opacity=0.92] {max(-0.1341 * x, x)};
%		\addplot [domain=-1:0, samples=5, ultra thick, blue, opacity=0.92] {max(-0.1458 * x, x)};
		\addplot [domain=-1:0, samples=5, ultra thick, pesqcolor, opacity=0.76] {max(-0.16 * x, x)};
%		\addplot [domain=-1:0, samples=5, ultra thick, blue, opacity=0.92] {max(-0.1699 * x, x)};
%		\addplot [domain=-1:0, samples=5, ultra thick, blue, opacity=0.92] {max(-0.1852 * x, x)};
		\addplot [domain=-1:0, samples=5, ultra thick, pesqcolor, opacity=0.68] {max(-0.2023 * x, x)};
%		\addplot [domain=-1:0, samples=5, ultra thick, blue, opacity=0.92] {max(-0.2161 * x, x)};
		\addplot [domain=-1:0, samples=5, ultra thick, pesqcolor, opacity=0.6] {max(-0.2399 * x, x)};
%		\addplot [domain=-1:0, samples=5, ultra thick, blue, opacity=0.92] {max(-0.2700 * x, x)};
		\addplot [domain=-1:0, samples=5, ultra thick, pesqcolor, opacity=0.52] {max(-0.2944 * x, x)};
		\addplot [domain=-1:0, samples=5, ultra thick, pesqcolor, opacity=0.44] {max(-0.3220 * x, x)};
		\addplot [domain=-1:0, samples=5, ultra thick, pesqcolor, opacity=0.36] {max(-0.3727 * x, x)};
		\addplot [domain=-1:0, samples=5, ultra thick, pesqcolor, opacity=0.3] {max(-0.4348 * x, x)}; %0.28
		\addplot [domain=-1:0, samples=5, ultra thick, pesqcolor, opacity=0.28] {max(-0.6268 * x, x)};%0.2
		
	\end{axis}
\end{tikzpicture}}
			\caption{After training}
		\end{subfigure}
		\caption{PReLU slopes of the resultant magnitude mask.}	\vspace{-0.5mm}
		\label{fig:prelu_slope}
	\end{figure}

	The decoder extracts the output from $N$ two-stage conformer blocks in a decoupled way, which includes two paths: the mask decoder and the complex decoder. The mask decoder aims to predict a mask that will be element-wise multiplied by the input magnitude $Y_m$ to predict $\hat{X}'_m$. On the other hand, the complex decoder directly predicts the real and imaginary parts. Both mask and complex decoders consist of a dilated DenseNet, similar to the one in the encoder. The sub-pixel convolution layer is utilized in both paths to upsample the frequency dimension back to $F$ \cite{shi2016subpixel}. For the mask decoder, a convolution block is used to squeeze the channel number to 1, followed by another convolution layer with PReLU activation to predict the final mask. Note that the PReLU activation learns different slopes for each frequency band and initially the slopes are defined as a fixed positive value (0.2). Post-training evaluation indicates that all the slopes reflect different negative values, i.e., the output mask is always projected in the positive 1\textsuperscript{st} and 2\textsuperscript{nd} quadrants, as depicted in Fig.~\ref{fig:prelu_slope}. For the complex decoder, the architecture is identical to the mask decoder, except no activation function is applied for the complex output.
	
\begin{figure*}[t!]
		\centering
		
		\sbox{\bigimage}{%
			\begin{subfigure}[b]{.38627\textwidth}
				\centering
				\includegraphics[width=\textwidth]{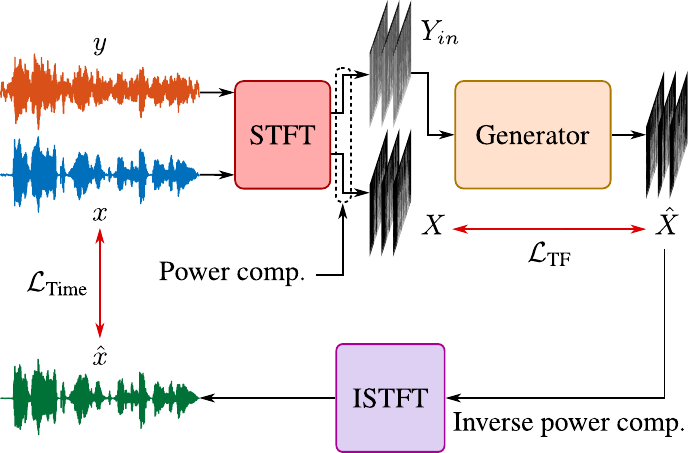}
				\caption{Non-adversarial generator losses}
				\label{fig:gen_loss}
				\vspace{0pt}% reference point at the very bottom
			\end{subfigure}%
		}
		
		\usebox{\bigimage}\hfill 
		\begin{minipage}[b][\ht\bigimage][s]{.5345\textwidth}
			\begin{subfigure}{\textwidth}
				\centering
				\includegraphics[height=2.59cm,width=\textwidth]{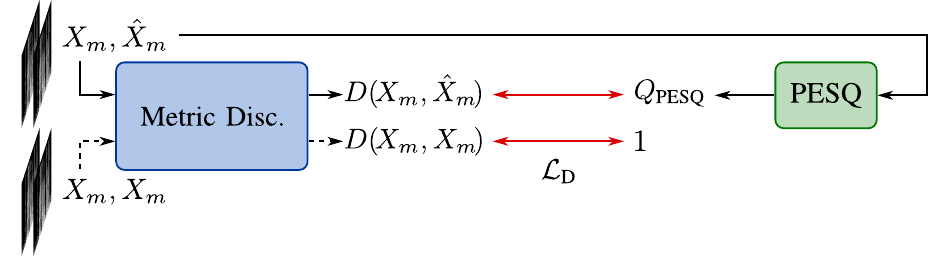}
				\vspace{-8mm}
				\caption{Discriminator loss}
				\label{fig:disc_loss}
			\end{subfigure}%
			\vspace{3mm}
			\begin{subfigure}{\textwidth}
				\centering
				\includegraphics[height=1.578cm,width=\textwidth]{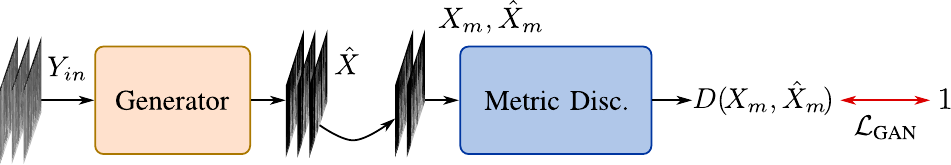}
				\caption{Adversarial generator loss}
				\label{fig:adv_loss}
			\end{subfigure}
			
			\vspace{0pt}% reference point at the very bottom
		\end{minipage}	
		\caption{An illustration of the propagated loss functions in the CMGAN architecture. For simplicity, $X$ and $\hat{X}$ denote the three-channel magnitude and complex representations of the clean target and the estimated output spectrograms, respectively.  }
		\label{fig:loss_overview}
		\vspace{-5mm}
	\end{figure*}

	Same as in \cite{yu2022dualbranch, li2022glance}, the masked magnitude $\hat{X}'_m$ is first combined with the noisy phase $Y_p$ to obtain the magnitude-enhanced complex spectrogram. Then it is element-wise summed with the output of the complex decoder $(\hat{X}'_r, \hat{X}'_i)$ to obtain the final complex spectrogram:
	\begin{equation} \label{eq:comb}
		\hat{X}_r=\hat{X}'_m\cos(Y_p)+\hat{X}'_r \hspace{8mm}
		\hat{X}_i=\hat{X}'_m\sin(Y_p)+\hat{X}'_i
	\end{equation}
	The power-law compression is then inverted on the complex spectrogram $(\hat{X}_r, \hat{X}_i)$ and an inverse short-time Fourier transform (ISTFT) is applied to get the time-domain signal $\hat{x}$, as shown in Fig.~\ref{fig:gen_loss}. To further improve the magnitude component and propagate magnitude loss on both decoder branches, we compute the magnitude loss on $\hat{X}_m$ expressed by:
	\begin{equation} \label{eq:mag_comb}
		\hat{X}_m=\sqrt{\hat{X}_r^2 + \hat{X}_i^2}
	\end{equation}
	
	\vspace{-4mm}
	\subsection{Metric discriminator}
	In SE, the objective functions are often not directly correlated to the 
	evaluation metrics. Consequently, even if the objective loss is optimized, the 
	evaluation score is still not satisfied. Furthermore, some evaluation metrics 
	like perceptual evaluation of speech quality (PESQ) \cite{rix2001pesq} and 
	short-time objective intelligibility (STOI) \cite{taal2010stoi} cannot be used 
	as loss functions because they are non-differentiable. Hence, the discriminator 
	in CMGAN aims to mimic the metric score and use it as a part of the loss 
	function. Here we follow the MetricGAN to use the PESQ score as a label 
	\cite{fu2019metricgan}. As shown in Fig.~\ref{fig:discriminator}, the 
	discriminator consists of four convolution blocks. Each block starts with a 
	convolution layer, followed by instance normalization and a PReLU activation. 
	After the convolution blocks, a global average pooling is followed by two 
	feed-forward layers and a sigmoid activation. The discriminator is then trained 
	to estimate the maximum normalized PESQ score ($=1$), by using the clean magnitude as both reference and degraded inputs. Additionally, the discriminator is trained to estimate the 
	enhanced PESQ score by taking both clean and enhanced spectrum as an input 
	together with their corresponding PESQ label, as shown in 
	Fig.~\ref{fig:disc_loss}. On the other hand, the generator is trained to render 
	an enhanced speech resembling the clean speech, thus approaching a PESQ label 
	of 1, as shown in Fig.~\ref{fig:adv_loss}.
	\vspace{-3mm}
	\subsection{Loss functions}
	Inspired by Braun \textit{et al.} \cite{braun2020loss}, we use a linear combination of magnitude loss $\mathcal{L}_{\small\textrm{Mag.}}$ and complex loss $\mathcal{L}_{\small\textrm{RI}}$ in the TF-domain:
	\begin{equation}
	\vspace{-0.5mm}
		\begin{aligned}
			&\mathcal{L}_{\small\textrm{TF}} =\alpha\, \mathcal{L}_{\small\textrm{Mag.}}+ (1-\alpha)\,\mathcal{L}_{\small\textrm{RI}} \\
			&\mathcal{L}_{\small\textrm{Mag.}} =\mathbb{E}_{X_m,\hat{X}_m} \big[\| X_m - \hat{X}_m  \|_2 \big]  \\
			&\mathcal{L}_{\small\textrm{RI}} =\mathbb{E}_{X_r,\hat{X}_r} \big[\| X_r - \hat{X}_r \|_2\big] + \mathbb{E}_{X_i,\hat{X}_i} \big[\| X_i - \hat{X}_i \|_2\big],
		\end{aligned}
		\label{TF loss}
			\vspace{-0.5mm}
	\end{equation}
	where $\alpha$ is a chosen weight. Based on grid search, $\alpha=0.7$ leads to the best performance.
	Similar to least-square GANs \cite{mao2017lsgans}, the adversarial training is following a min-min optimization task over the discriminator loss $\mathcal{L}_{\small\textrm{D}}$ and the corresponding generator loss $\mathcal{L}_{\small\textrm{GAN}}$ expressed as follows:
	\begin{equation}
		\begin{aligned}
			&\mathcal{L}_{\small\textrm{GAN}} =\mathbb{E}_{X_m,\hat{X}_m} \big[\| D(X_m,\hat{X}_m) - 1 \|_2\big]\\
			&
			\begin{split}
				\mathcal{L}_{\small\textrm{D}} &= \mathbb{E}_{X_m} \big[\| D(X_m, X_m) - 1\|_2\big] \\&+ \mathbb{E}_{X_m,\hat{X}_m}\big[\,\| D(X_m,\hat{X}_m)- Q_{\small\textrm{PESQ}} \|_2\big],
			\end{split}
		\end{aligned}
		\label{eq:gan_loss}
	\end{equation}
	where $D$ refers to the discriminator and $Q_{\small\textrm{PESQ}}$ refers to the normalized PESQ score. Here we normalize the PESQ score to the range [0,1]. Moreover, an additional penalization in the resultant waveform $\mathcal{L}_{\small\textrm{Time}}$ is proven to improve the restored speech quality \cite{abdulatif2021crossdomain}:
	\begin{equation}
		\mathcal{L}_{\small\textrm{Time}}= \mathbb{E}_{x,\hat{x}} \big[\| x-\hat{x} \|_1\big],
		\label{eq:time_loss}
	\end{equation}
	where $\hat{x}$ is the enhanced waveform, and $x$ is the clean target waveform. In this context, $\|\cdot\|_1$ denotes the $\ell^1-$norm (sum of absolute values), and $\|\cdot\|_2$ represents the $\ell^2-$norm (Euclidean distance). The final generator loss is formulated as follows:
	\begin{equation}
		\mathcal{L}_{\small\textrm{G}}=\gamma_1 \,\mathcal{L}_{\small\textrm{TF}} + \gamma_2 \,\mathcal{L}_{\small\textrm{GAN}} + \gamma_3 \,\mathcal{L}_{\small\textrm{Time}},
	\end{equation}
	where $\gamma_1, \gamma_2$ and $\gamma_3$ are the weights of the corresponding losses and they are chosen to reflect equal importance.
	\vspace{-1mm}
	\section{Experiments}
		\vspace{-1mm}
	\subsection{Datasets}\label{sec:denoising_data}
	\subsubsection{Denoising} 
	We investigate our proposed approach on the commonly used publicly available Voice Bank+DEMAND dataset \cite{valentini2016voicebank}. The clean tracks are selected from the Voice Bank corpus \cite{veaux2013voicebank} which includes 11,572 utterances from 28 speakers in the training set and 824 utterances from 2 unseen speakers in the test set. In the training set, the clean utterances are mixed with background noise (8 noise types from DEMAND database \cite{thiemann2013demand} and 2 artificial noise types) at SNRs of 0 dB, 5 dB, 10 dB and 15 dB. In the test set, the clean utterances are mixed with 5 unseen noise types from the DEMAND database at SNRs of 2.5 dB, 7.5dB, 12.5 dB and 17.5 dB. The noise types are mostly challenging, e.g., public space noises (cafeteria, restaurant and office), domestic noises (kitchen and living room) and transportation/street noises (car, metro, bus, busy traffic, public square and subway station). All utterances are resampled to 16 kHz in our experiments.
	
	\subsubsection{Dereverberation}
	We choose the REVERB challenge dataset \cite{kinoshita2016reverb}, the utterances are divided into simulated and real recordings. The simulated data is based on the wall street journal corpus (WSJCAM0) \cite{robinson1995wsj0} distorted by measured RIRs and a stationary ambient noise of SNR $=$ 20 dB. The measured RIRs represent three different room sizes: small -- room 1, medium -- room 2 and large -- room 3, with a 60 dB reverberation time ($\tau$) of 0.3, 0.6 and 0.7 seconds, respectively. For each room, the microphone is placed at a near condition (0.5 m) and a far condition (2 m). The real data is based on the multi-channel wall street journal audio-visual (MC-WSJ-AV) corpus \cite{lincoln2005mcwsj}, where the speakers are recording in a large room of $\tau$ = 0.7 seconds at a near (1 m) and a far (2.5 m) microphone conditions. The training set includes 7861 paired utterances from the simulated data. The test set contains both simulated paired utterances (2176) and real reverberant utterances (372). Different room recordings are used for the training and test sets. The datasets were originally captured in a single-channel, two-channel and eight-channel configuration with a 16 kHz sampling frequency. However, for the scope of this study, we only use the single-channel configuration.
	
	\subsubsection{Super-resolution}
	For comparative analysis, we utilize the English multi-speaker corpus (VCTK) \cite{yamagishi2019vctk}. The VCTK dataset contains 44 hours recordings from 108 speakers with various English accents. For the super-resolution experiment, we follow the design choice of \cite{kuleshov2017sr}, where the low-resolution audio signal is generated from the 16 kHz original tracks by subsampling the signal with the desired upscaling ratio ($s$). The first task uses a single VCTK speaker (p225), the first 223 recordings are used for training and the last 8 recordings are used for testing. The second task takes the first 100 VCTK speakers as the training set and tests on the last 8 speakers. The upscaling ratios for both the single-speaker and the multi-speaker tasks are set to \{2, 4, 8\}, representing a reconstruction from 8 kHz, 4 kHz, 2 kHz to 16 kHz.
	
	\begin{table*}[t!]
		\centering
		\caption{Performance comparison on the Voice Bank+DEMAND dataset \cite{valentini2016voicebank}. “-” denotes the result is not provided in the original paper. Model size represents the number of trainable parameters in million.}\vspace{-1mm}
		\resizebox{0.99\textwidth}{!}{
			\small
			\begin{tabular}{lllcccccccc}
				\toprule
				Method  & Year & Input & Model Size (M) &  PESQ & CSIG & CBAK & COVL & SSNR & STOI  \\
				\midrule
				Noisy  & \hspace{2.5mm}- & \hspace{2.5mm}- & -&  1.97 & 3.35 & 2.44 & 2.63 & 1.68 & 0.91  \\
				\midrule
				SEGAN \cite{pascual2017segan} & 2017 & Time & 97.47& 2.16 & 3.48 & 2.94 & 2.80 & 7.73 & 0.92  \\
				MetricGAN \cite{fu2019metricgan} & 2019 & Magnitude & - & 2.86 & 3.99 & 3.18 & 3.42 & - & - \\
				PHASEN \cite{yin2020phasen} & 2020 & Magnitude+Phase & - & 2.99 & 4.21 & 3.55 & 3.62 & 10.08 & -  \\
				TSTNN \cite{wang2021tstnn} & 2021 & Time & 0.92& 2.96 & 4.10 & 3.77 & 3.52 & 9.70 & 0.95 \\
				DEMUCS \cite{defossez2020demucs} & 2021 & Time & 128& 3.07 & 4.31 & 3.40 & 3.63 & - & 0.95  \\
				PFPL \cite{hsieh2020pfpl} & 2021 & Complex & - & 3.15 & 4.18 & 3.60 & 3.67 & - & 0.95  \\
				MetricGAN+ \cite{fu2021metricgan+} & 2021 & Magnitude & - & 3.15 & 4.14 & 3.16 & 3.64 & - & -  \\
				SE-Conformer \cite{kim2021seconformer} & 2021 & Time & -& 3.13 & 4.45 & 3.55 & 3.82 & - & 0.95  \\
				DB-AIAT \cite{yu2022dualbranch} & 2021 & Complex+Magnitude & 2.81& 3.31 & 4.61 & 3.75 & 3.96 & 10.79 & \textbf{0.96}  \\
				DPT-FSNet \cite{dang2021dpt} & 2021 & Complex & 0.91& 3.33 & 4.58 & 3.72 & 4.00 & - &  \textbf{0.96}  \\
				\textbf{CMGAN} & 2022 & Complex+Magnitude & 1.83& \textbf{3.41} & \textbf{4.63} & \textbf{3.94} & \textbf{4.12} & \textbf{11.10} & \textbf{0.96}  \\
				\bottomrule
			\end{tabular}
		}
		\label{tab:results_demand}
		\vspace{-4.5mm}
	\end{table*}
	
	\vspace{-3mm}
	\subsection{Experimental setup}
	The utterances in the training set are sliced into 2 seconds, while in the test set, no slicing is utilized and the length is kept variable. A Hamming window with 25 ms window length (400-point FFT) and hop size of 6.25 ms (75\% overlap) is employed. Thus, the resultant spectrogram will have 200 frequency bins $F$, while the time dimension $T$ depends on the variable track duration.
	The number of two-stage conformer blocks $N$, the batch size $B$ and the 
	channel number $C$ in the generator are set to 4, 4 and 64, respectively. The 
	channel numbers in the metric discriminator are set to \{16, 32, 64, 128\}. In 
	the training stage, AdamW optimizer \cite{loshchilov2017decoupled} is used for 
	both the generator and the discriminator to train for 50 epochs. The learning 
	rate is set to 5$\times$10$^{-4}$ for the generator and 1$\times$10$^{-3}$ for 
	the discriminator. A learning rate scheduler is applied with a decay factor of 
	0.5 every 12 epochs.  In the generator loss $\mathcal{L}_{\small\textrm{G}}$, 
	the weights are set to $\{\gamma_1=1, \gamma_2=0.01, \gamma_3=1\}$.
	The detailed parameter setup of both generator and discriminator is presented in 
	Appendix.
	\vspace{-5mm}
	\section{Results and discussion} 
	For the three tasks discussed in this section, all methods, including baselines and proposed approaches, are evaluated using standard benchmarks. The reported numbers adhere to the consistent train/test splits that these benchmarks specify.
	\vspace{-4mm}
	\subsection{Denoising} \label{sec:denoising_results}
	\subsubsection*{Objective scores}
	We choose a set of commonly used metrics to evaluate the denoised speech 
	quality, i.e., PESQ with a score range from -0.5 to 4.5, segmental 
	signal-to-noise ratio (SSNR) and composite mean opinion score (MOS) 
	\cite{hu2007mos} based metrics: MOS prediction of the signal distortion (CSIG), 
	MOS prediction of the intrusiveness of background noise (CBAK) and MOS 
	prediction of the overall effect (COVL), all of them are within a score range 
	of 1 to 5. Additionally, we utilize STOI with a score range from 0 to 1 to 
	judge speech intelligibility. Higher values indicate better performance for all 
	given metrics.
	
	\subsubsection*{Results analysis}
	Our proposed CMGAN is objectively compared with other state-of-the-art (SOTA) 
	denoising baselines, as shown in Table~\ref{tab:results_demand}. For the 
	time-domain methods, we included the standard SEGAN \cite{pascual2017segan} and 
	three recent methods: TSTNN \cite{wang2021tstnn}, DEMUCS 
	\cite{defossez2020demucs} and SE-Conformer \cite{kim2021seconformer}. It is worth noting that DEMUCS is trained not only on the standard Voice Bank+DEMAND dataset \cite{valentini2016voicebank}, but also incorporates additional data from the DNS-Challenge \cite{dubey2022dns}. For the 
	TF-domain methods, we evaluate six recent SOTA methods, i.e., MetricGAN 
	\cite{fu2019metricgan}, PHASEN \cite{yin2020phasen}, PFPL \cite{hsieh2020pfpl}, 
	MetricGAN+ \cite{fu2021metricgan+}, DB-AIAT \cite{yu2022dualbranch} and 
	DPT-FSNet \cite{dang2021dpt}. It can be observed that most of the TF-domain 
	methods outperform the time-domain counterparts over all utilized metrics. 
	Moreover, our proposed TF conformer-based approach shows a major improvement 
	over the time-domain SE-Conformer. Compared to frameworks involving metric 
	discriminators (MetricGAN+), we have 0.26, 0.49, 0.78 and 0.48 improvements on 
	the PESQ, CSIG, CBAK and COVL scores, respectively. Finally, our framework also 
	outperforms recent improved transformer-based methods, such as DB-AIAT and 
	DPT-FSNet in all of the evaluation scores with a relatively low model size of 
	only 1.83~M parameters.
	
	\begin{table}[b!]
		\vspace{-4.5mm}
		\centering
		\midsepremove
		\caption{Results of the denoising ablation study.}\vspace{-1.5mm}
		\def\arraystretch{1.15}
		\resizebox{0.975\columnwidth}{!}{
			\setlength{\tabcolsep}{0.8mm}
			\small
			\begin{tabular}{lcccccc}
				\toprule  
				Method  & PESQ & CSIG & CBAK & COVL & SSNR & STOI \\
				\midrule
				\rowcolor{tblcolortransparent}
				\textbf{CMGAN} & 3.41 & 4.63 & 3.94 & 4.12 & 11.10 & 0.96 \\
				\midrule
				\rowcolor[gray]{0.8}
				\multicolumn{7}{c}{Inputs \& Decoders}\\
				\midrule
				Single-Mask & 3.23 & 4.60 & 3.76 & 4.00 & 9.82 & 0.95 \\
				Single-Complex & 3.35 & 4.56 & 3.79 & 4.05 & 9.19 & \textbf{0.96} \\
				Dual-Mask & 3.23 & 4.62 & 3.79 & 4.01 & 10.20 & \textbf{0.96} \\
				Dual-Complex & \textbf{3.39} & \textbf{4.63} & 3.82 & \textbf{4.10} & 9.41 & 0.95 \\
				Single-Path & 3.38 & 4.54 & \textbf{3.86} & 4.05 & 10.19 & \textbf{0.96} \\
				Mask + cIRM & 3.28 & 4.60 & 3.83 & 4.03 & \textbf{10.40} & \textbf{0.96}\\
				\midrule
				\rowcolor[gray]{0.8}
				\multicolumn{7}{c}{Loss functions \& Discriminator}\\
				\midrule
				w/o Time Loss & \textbf{3.45} & 4.56 & 3.86 & 4.11 & 9.71 & \textbf{0.96} \\
				w/o Disc. & 3.24 & 4.46 & 3.82 & 3.93 & 10.56 & \textbf{0.96} \\
				Patch Disc. & 3.28 & 4.48 & 3.85 & 3.96 & 10.75 & \textbf{0.96} \\
				Pretrained Disc. & 3.23 & 4.48 & 3.83 & 3.93 & \textbf{10.76} & \textbf{0.96}\\
				MetricGAN+ Disc. & 3.44 & 4.61 & \textbf{3.87} & \textbf{4.13} & 9.83 & \textbf{0.96}\\
				PESQ + STOI Disc. & 3.41 & \textbf{4.63} & 3.89 & 4.12 & 10.34 & \textbf{0.96} \\
				\midrule
				\rowcolor[gray]{0.8}
				\multicolumn{7}{c}{TS-Conformer design}\\
				\midrule
				Parallel-Conf. & 3.35 & 4.54 & 3.87 & 4.03 & 10.63 & \textbf{0.96} \\
				Freq. $\rightarrow$ Time & \textbf{3.39} & \textbf{4.56} & \textbf{3.91} & \textbf{4.07} & \textbf{10.84} & \textbf{0.96} \\
				\midrule
				\rowcolor[gray]{0.8}
				\multicolumn{7}{c}{Magnitude mask activation function}\\
				\midrule
				Sigmoid & 3.34 & 4.52 & 3.80 & 4.02 & 10.70 & \textbf{0.96} \\
				ReLU & 3.32 & 4.54 & 3.80 & \textbf{4.04} & 10.69 & \textbf{0.96} \\
				Softplus & \textbf{3.43} & \textbf{4.58} & \textbf{3.83} & 4.02 & \textbf{10.75} & \textbf{0.96} \\
				\bottomrule 
			\end{tabular}
		}
		\label{tab:ablation}
	\end{table}
	
	\subsubsection*{Ablation study}
	To verify our design choices, an ablation study is conducted, as shown in Table~\ref{tab:ablation}. We first investigate the influence of different inputs and decoders. For the Single-Mask configuration, only magnitude is used as the input. The enhanced magnitude is then combined with the noisy phase for the ISTFT operation, omitting the complex decoder. Conversely, the Single-Complex approach solely uses the complex spectrogram as input, and the mask decoder is removed. The drop in the PESQ score from 3.35 in Single-Complex to 3.23 in Single-Mask emphasizes the crucial influence of phase information on the enhanced speech quality. Turning our attention to SSNR, the Single-Mask setup using magnitude-based masking reaches 9.82 dB, while the Single-Complex with its complete input representation attains only 9.19 dB. This suggests that, for segment-specific metrics like SSNR, explicit masking remains vital.
	
	To ascertain whether the advancements of CMGAN stemmed from the synergy between both mask and complex decoders rather than just an increased parameter count, we introduced two additional configurations: Dual-Mask and Dual-Complex. The Dual-Mask design used magnitude-only as input and featured two mask decoders. Both decoders received the same input, with the output from the first mask decoder being subsequently masked by the second before being combined with the noisy phase. With an identical parameter count to CMGAN, this configuration displayed minor improvements in most metrics when compared to its Single-Mask counterpart. Notably, there was a rise of 0.38 dB in SSNR, reinforcing the link between SSNR and effective masking techniques. In contrast, the Dual-Complex setup solely relied on the complex representation and incorporated an additional complex decoder, with both decoders summing their outputs to produce the final result. While this approach displayed metric improvements, it did not achieve the performance levels exhibited by CMGAN, underscoring the importance of integrating both mask and complex decoders in our design.
	
	To further evaluate the efficacy of our design decisions, we implemented two additional modifications. First, we explored the necessity of decoder decoupling in complex refinement through the Single-Path decoder configuration. In this setup, we employed a single decoder that produces three channels: the first channel undergoes a PReLU activation for magnitude, while the subsequent two channels, which represent the complex component, have no activation. A comparison with our primary mask/complex decoders indicates a decline across all metrics, most significantly in the SSNR metric with a decrease of 0.91 dB. In another separate alteration, Mask + cIRM, we maintained the mask decoder while adjusting the complex decoder to integrate a cIRM inspired by \cite{williamson2016cirm}. This change resulted in a notable reduction in both PESQ and SSNR when compared to the original CMGAN setup.
	\begin{figure}[b!]
		\vspace{-5mm}
		\centering
		%		\resizebox{\columnwidth}{!}{\input{Figs/ablation_curves/mask_hist.tex}}
		\includegraphics[width=0.99\columnwidth]{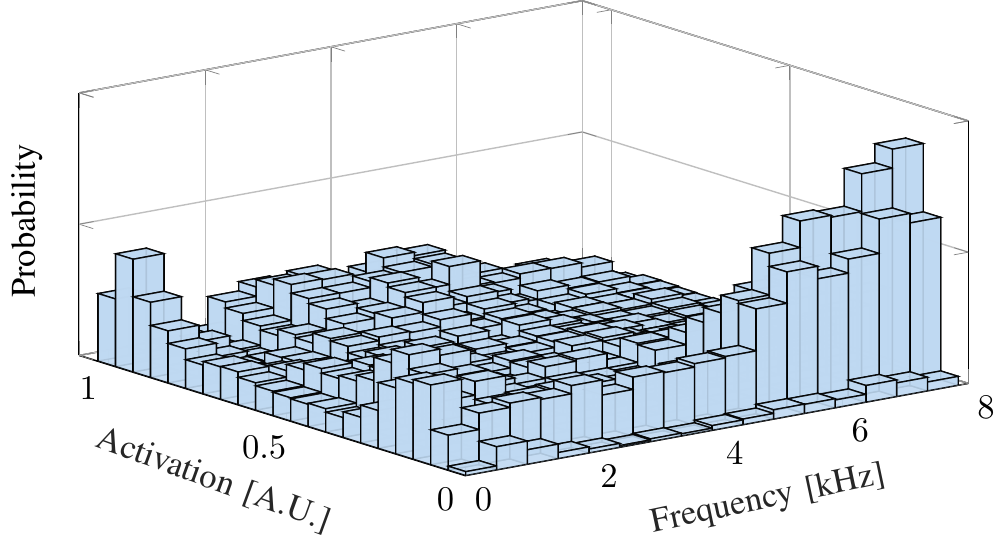}
		\caption{Histogram of masking PReLU activations.}
		\label{fig:mask_hist}
		\vspace{-0.5mm}
	\end{figure}
	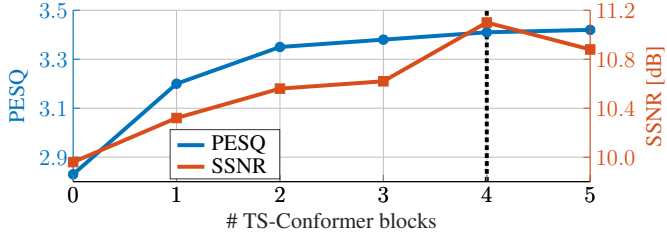
\begin{figure}[t!]
		\centering
		\resizebox{\columnwidth}{!}{\begin{tikzpicture}

\begin{axis}[%
width=\figW,
height=\figH,
at={(0in,0in)},
scale only axis,
xmin=0,
xmax=5,
xlabel style={font=\color{white!15!black}},
xlabel={\Large \# TS-Conformer blocks},
every outer y axis line/.append style={pesqcolor},
every y tick label/.append style={font=\Large\color{pesqcolor},/pgf/number format/fixed,
	/pgf/number format/fixed zerofill,
	/pgf/number format/precision=1},
every x tick label/.append style={font=\Large},
every y tick/.append style={pesqcolor},
ymin=2.8,
ymax=3.5,
ytick={2.9,3.1,3.3,3.5},
xtick={0,1,2,3,4,5},
ylabel style={font=\color{pesqcolor}},
ylabel={\Large  \textcolor{pesqcolor}{PESQ}},
axis background/.style={fill=white},
axis x line*=bottom,
axis y line*=left,
xmajorgrids,
ymajorgrids,
legend style={legend cell align=left, align=left, draw=none}
]

\addplot [color=pesqcolor, line width=2.5pt, solid,draw=pesqcolor]
table[row sep=crcr]{%
0	2.83\\
1	3.2\\
2	3.35\\
3	3.38\\
4	3.41\\
5	3.42\\
};
\label{lab:pesq_conf}

\addplot [only marks, color=pesqcolor, solid, mark=*, mark size=3.000pt, mark indices = {1,2,3,4,5,6}]
table[row sep=crcr]{%
	0	2.83\\
	1	3.2\\
	2	3.35\\
	3	3.38\\
	4	3.41\\
	5	3.42\\
};
\addplot[mark=none, densely dashed, line width=2.5pt] coordinates {(4,2.8) (4,3.5)};
\end{axis}

\begin{axis}[%
width=\figW,
height=\figH,
at={(0in,0in)},
scale only axis,
xmin=0,
xmax=5,
every outer y axis line/.append style={ssnrcolor},
every y tick label/.append style={font=\Large\color{ssnrcolor},/pgf/number format/fixed,
	/pgf/number format/fixed zerofill,
	/pgf/number format/precision=1},
every y tick/.append style={ssnrcolor},
ymin=9.8,
ymax=11.2,
every x tick label/.append style={font=\Large},
xtick={0,1,2,3,4,5},
ytick={10,10.4,10.8,11.2},
ylabel style={font=\Large\color{ssnrcolor}},
ylabel={\Large SSNR [dB]},
axis x line*=bottom,
axis y line*=right,
legend style={legend cell align=left, align=left, draw=none,yshift = -37, xshift = -5,nodes={scale=0.99, transform shape}}
]

\addplot [color=ssnrcolor, line width=2.5pt, solid, draw=ssnrcolor]
table[row sep=crcr]{%
0	9.96\\
1	10.32\\
2	10.56\\
3	10.62\\
4	11.1\\
5	10.88\\
};
\label{lab:ssnr_conf}

\addplot [only marks, color=ssnrcolor, solid, mark=square*, mark size=3.0000pt, mark indices = {1,2,3,4,5,6}]
table[row sep=crcr]{%
	0	9.96\\
	1	10.32\\
	2	10.56\\
	3	10.62\\
	4	11.1\\
	5	10.88\\
};
\coordinate (legendPos) at (axis description cs:0.4,0.16);
\end{axis}
\node[anchor=east,text width=2.2cm,align=left,fill=white,draw=black!80,line width=0.1mm] at (legendPos) {\ref{lab:pesq_conf} \Large PESQ \vspace{1mm}\\ \ref{lab:ssnr_conf} \Large SSNR};
\end{tikzpicture}%
}
		\caption{Influence of TS-Conformer blocks on objective scores.}
		\label{fig:n_conformers}
		\vspace{-6mm}
	\end{figure}
	
	In our exploration of loss functions, several variations revealed insightful impacts on performance metrics. 
	The omission of time loss (w/o Time Loss) led to an improvement in the PESQ score, reaching 3.45, but with a noticeable effect on SSNR. This indicates the effectiveness of the time loss in balancing the performance for both PESQ and SSNR scores. Further, two discriminator-based assessments were conducted: one that entirely removed the discriminator (w/o Disc.) and another that substituted the metric discriminator with a patch discriminator, commonly employed in image generation tasks \cite{isola2017pix2pix}. It can be realized that the absence of the discriminator negatively impacted all the given scores. Similarly, adding a patch discriminator only showed a marginal improvement, which reflects that the generator is fully capable of enhancing the tracks without the aid of a normal patch discriminator. However, a metric discriminator to directly improve the evaluation scores is proven to be beneficial.
	
	For the Pretrained Disc. experiment, the discriminator was pretrained to predict normalized PESQ. During its training, the discriminator lacked access to enhanced tracks, with noisy tracks taking their place. The generator then integrated this loss, which was derived from the pretrained discriminator, with other reconstruction losses. Throughout the generator training, the discriminator remained frozen, receiving no updates from the generator. This approach resulted in reduced performance across all metrics compared to CMGAN. These findings highlight the importance of adversarial training, which can be largely attributed to the dynamic feedback loop it establishes for better model generalization. With the MetricGAN+ Disc. setup, noisy tracks were included in the discriminator loss. While there was a minor uplift in PESQ, other metrics remained unchanged or even registered a decline in SSNR. Such outcomes suggest potential instabilities when directing the discriminator with a broader objective set. Our hands-on observations led us to persist with the original MetricGAN loss structure. Exploring the PESQ + STOI Disc. method, we incorporated an extra discriminator to estimate the STOI in conjunction with PESQ. The results closely paralleled those of CMGAN, suggesting a possible ceiling in STOI optimization. The aforementioned discriminator adjustments illustrate the intricate interplay between metric discriminators and SSNR. Introducing extra tailored discriminators or additional objectives can negatively impact SSNR, highlighting the significance of careful design decisions.
	
	Furthermore, we investigate the influence of the two-stage conformer design. Given an input feature map, the two-stage conformer will separately focus on the time and frequency dimensions. To this end, two different configurations can be proposed, either sequential or parallel. Accordingly, we compare our sequential CMGAN to a parallel connection counterpart without any further modifications (Parallel-Conf.). The results illustrate that the parallel approach lags behind the proposed sequential design, with the PESQ and SSNR scores reduced by 0.06 and 0.47 dB, respectively. A possible explanation is that parallel conformers might learn redundant or conflicting patterns, lacking the synergistic benefits seen in sequential models. Additionally, we flipped the order of the sequential conformer blocks (Freq. $\rightarrow$ Time) and found the scores to be similar with a slight improvement in favor of the standard CMGAN (Time $\rightarrow$ Freq.). Note that designing a single conformer to attend over both time and frequency is theoretically possible. However, in this case, the complexity will grow exponentially \cite{liu2021swin}. 
	
	Preliminary literature mostly assumes the predicted magnitude mask to be between 0 and 1 \cite{yin2020phasen, fu2019metricgan, yu2022dualbranch, hu2020dccrn, fu2021metricgan+}. Hence, sigmoid activation is usually the preferred activation to reflect this interval. Although this is true, a bounded sigmoid function would restrict the model to allocate values between 0 and 0.5 to all aggregated negative activations from the previous layer. On the other hand, an unbounded activation function such as PReLU could automatically learn this interval while mitigating the negative activations issue by learning a relevant slope to each frequency band as explained in Sec.~\ref{sec:generator}. To confirm this assumption, we construct a histogram of several magnitude masks from different noisy tracks. As shown in Fig.~\ref{fig:mask_hist}, the PReLU activations would always lie in the 0 to 1 interval. Moreover, the majority of low activations are assigned to frequency bands above 5 kHz (beyond human speech) \cite{virag1995critical}. We also extend our ablation study to involve different bounded and unbounded activations for the mask decoder, namely sigmoid, ReLU and the soft version of ReLU (softplus) \cite{zheng2015softplus}. According to Table~\ref{tab:ablation}, both sigmoid and ReLU activations are comparable and they report lower scores than CMGAN with PReLU activation. Softplus achieves slightly higher PESQ, but at the expense of other metrics. 
	\begin{table*}[t!]
		\centering
		\caption{Results of simulated and real data on near microphone case.}\vspace{-1.5mm}
		\large
		\def\arraystretch{1.15}	
		\resizebox{0.98\textwidth}{!}{
			\begin{tabular}{l|cccc|cccc|cccc|cccc|c}
				\toprule  
				& \multicolumn{4}{c}{CD $\downarrow$} \vline & \multicolumn{4}{c}{LLR $\downarrow$} \vline & \multicolumn{4}{c}{FWSegSNR $\uparrow$} \vline & \multicolumn{4}{c}{SRMR $\uparrow$} \vline & SRMR-real $\uparrow$ \\
				Room & 1 & 2 & \multicolumn{1}{c|}{3} & Avg. & 1 & 2 & \multicolumn{1}{c|}{3} & Avg. & 1 & 2 & \multicolumn{1}{c|}{3} & Avg. & 1 & 2 & \multicolumn{1}{c|}{3} & Avg. & - \\
				\hline
				& & & \multicolumn{1}{c|}{} & & & & \multicolumn{1}{c|}{} & & & & \multicolumn{1}{c|}{} & & & & \multicolumn{1}{c|}{} & & \\[-2.5ex]
				Reverberant speech & 1.99 & 4.63 & \multicolumn{1}{c|}{4.38} & 3.67 & 0.35 & 0.49 & \multicolumn{1}{c|}{0.65} & 0.50 & 8.12 & 3.35 & \multicolumn{1}{c|}{2.27} & 4.58 & 4.50 & 3.74 & \multicolumn{1}{c|}{3.57} & 3.94 & 3.17 \\
				Xiao et al. \cite{xiao2014reverb} & 1.58 & 2.65 & \multicolumn{1}{c|}{2.68} & 2.30 & 0.37 & 0.50 & \multicolumn{1}{c|}{0.52} & 0.46 & 9.79 & 7.27 & \multicolumn{1}{c|}{6.83} & 7.96 & \textbf{5.74} & \textbf{6.49} & \multicolumn{1}{c|}{5.86} & \textbf{6.03} & 4.29 \\
				WRN \cite{ribas2019reverb}& 2.02 & 4.61 & \multicolumn{1}{c|}{4.15} & 3.59 & 0.36 & 0.46 & \multicolumn{1}{c|}{0.60} & 0.47 & 8.28 & 3.57 & \multicolumn{1}{c|}{2.54} & 4.80 & 4.04 & 3.46 & \multicolumn{1}{c|}{3.27} & 3.59 & - \\
				U-Net \cite{ernst2018reverb} & 1.75 & 2.58 & \multicolumn{1}{c|}{2.53} & 2.28 & 0.20 & 0.41 & \multicolumn{1}{c|}{0.45} & 0.35 & 13.32 & 10.87 & \multicolumn{1}{c|}{10.40} & 11.53 & 4.51 & 5.09 & \multicolumn{1}{c|}{4.94} & 4.85 & 5.47 \\
				SkipConvNet \cite{kothapally2020reverb} & 1.86 & 2.57 & \multicolumn{1}{c|}{2.45} & 2.29 & 0.19 & 0.30 & \multicolumn{1}{c|}{0.35} & 0.28 & 13.07 & 10.96 & \multicolumn{1}{c|}{10.22} & 11.42 & 4.99 & 4.75 & \multicolumn{1}{c|}{4.56} & 4.77 & 7.27 \\
				\bottomrule
				\toprule
				CMGAN & \textbf{1.46} & \textbf{2.14} & \multicolumn{1}{c|}{\textbf{2.27}} & \textbf{1.96} & \textbf{0.14} & \textbf{0.25} & \multicolumn{1}{c|}{0.34} & 0.24 & 14.36 & \textbf{13.49} & \multicolumn{1}{c|}{\textbf{11.69}} & \textbf{13.18} & 5.42 & 5.74 & \multicolumn{1}{c|}{5.29} & 5.48 & 6.49 \\
				CMGAN-LLR & 1.69 & 2.56 & \multicolumn{1}{c|}{2.43} & 2.23 & 0.15 & 0.25 & \multicolumn{1}{c|}{\textbf{0.25}} & \textbf{0.22} & \textbf{14.48} & 12.49 & \multicolumn{1}{c|}{11.03} & 12.67 & 5.48 & 5.80 & \multicolumn{1}{c|}{\textbf{6.02}} & 5.77 &  \textbf{7.71} \\			
				\bottomrule 
			\end{tabular}
		}
		\label{tab:reverb_near}
		\vspace{-1.5mm}
	\end{table*}
	
	\begin{table*}[t!]
		\centering
		\caption{Results of simulated and real data on far microphone case.}\vspace{-1.5mm}
		\large
		\def\arraystretch{1.15}	
		\resizebox{0.98\textwidth}{!}{
			\begin{tabular}{l|cccc|cccc|cccc|cccc|c}
				\toprule  
				& \multicolumn{4}{c}{CD $\downarrow$} \vline & \multicolumn{4}{c}{LLR $\downarrow$} \vline & \multicolumn{4}{c}{FWSegSNR $\uparrow$} \vline & \multicolumn{4}{c}{SRMR $\uparrow$} \vline & SRMR-real $\uparrow$ \\
				Room & 1 & 2 & \multicolumn{1}{c|}{3} & Avg. & 1 & 2 & \multicolumn{1}{c|}{3} & Avg. & 1 & 2 & \multicolumn{1}{c|}{3} & Avg. & 1 & 2 & \multicolumn{1}{c|}{3} & Avg. & - \\
				\hline
				& & & \multicolumn{1}{c|}{} & & & & \multicolumn{1}{c|}{} & & & & \multicolumn{1}{c|}{} & & & & \multicolumn{1}{c|}{} & & \\[-2.5ex]
				Reverberant speech & 2.67 & 5.21 & \multicolumn{1}{c|}{4.96} & 4.28 & 0.38 & 0.75 & \multicolumn{1}{c|}{0.84} & 0.66 & 6.68 & 1.04 & \multicolumn{1}{c|}{0.24} & 2.65 & 4.58 & 2.97 & \multicolumn{1}{c|}{2.73} & 3.43 & 3.19 \\
				Xiao et al. \cite{xiao2014reverb} & 1.92 & 3.17 & \multicolumn{1}{c|}{2.99} & 2.69 & 0.41 & 0.61 & \multicolumn{1}{c|}{0.58} & 0.53 & 9.12 & 6.31 & \multicolumn{1}{c|}{5.97} & 7.13 & 5.67 & 5.80 & \multicolumn{1}{c|}{5.03} & 5.50 & 4.42 \\
				WRN \cite{ribas2019reverb}& 2.43 & 4.99 & \multicolumn{1}{c|}{4.56} & 3.99 & 0.35 & 0.59 & \multicolumn{1}{c|}{0.67} & 0.54 & 7.54 & 1.79 & \multicolumn{1}{c|}{0.88} & 3.40 & 4.48 & 3.32 & \multicolumn{1}{c|}{2.84} & 3.55 & - \\
				U-Net \cite{ernst2018reverb} & 2.05 & 3.19 & \multicolumn{1}{c|}{2.92} & 2.72 & 0.26 & 0.57 & \multicolumn{1}{c|}{0.56} & 0.46 & \textbf{12.08} & 9.00 & \multicolumn{1}{c|}{9.05} & 10.04 & 4.76 & 5.27 & \multicolumn{1}{c|}{4.71} & 4.91 & 5.68 \\
				SkipConvNet \cite{kothapally2020reverb} & 2.12 & 3.06 & \multicolumn{1}{c|}{\textbf{2.82}} & 2.67 & \textbf{0.22} & 0.46 & \multicolumn{1}{c|}{0.46} & 0.38 & 11.80 & 8.88 & \multicolumn{1}{c|}{8.16} & 9.61 & 5.10 & 4.76 & \multicolumn{1}{c|}{4.25} & 4.70 & 6.87 \\
				\bottomrule
				\toprule
				CMGAN & \textbf{1.88} & \textbf{2.90} & \multicolumn{1}{c|}{2.85} & \textbf{2.54} & 0.24 & \textbf{0.43} & \multicolumn{1}{c|}{0.47} & 0.38 & 11.65 & \textbf{10.34} & \multicolumn{1}{c|}{8.91} & \textbf{10.30} & 5.78 & \textbf{5.87} & \multicolumn{1}{c|}{4.69} & 5.45 & 6.61 \\
				CMGAN-LLR & 2.07 & 3.32 & \multicolumn{1}{c|}{3.05} & 2.81 & 0.24 & 0.46 & \multicolumn{1}{c|}{\textbf{0.40}} & \textbf{0.37} & 11.21 & 9.22 & \multicolumn{1}{c|}{\textbf{9.48}} & 9.97 & \textbf{5.93} & 5.54 & \multicolumn{1}{c|}{\textbf{5.19}} & \textbf{5.55} &  \textbf{7.62} \\	
				\bottomrule 
			\end{tabular}
		}
		\label{tab:reverb_far}
		\vspace{-5mm}
	\end{table*}
	
	Finally, we experiment with the number of TS-Conformer blocks. As shown in Fig.~\ref{fig:n_conformers}, the performance of CMGAN without any conformer blocks is acceptable and even comparable with other SOTA methods, such as MetricGAN. However, only one conformer block effectively improves the PESQ by 0.4. The performance gradually increases with more blocks until no further improvement is observed after four blocks. Due to space constraints, the original CMGAN will be considered for upcoming tasks with few relevant ablation studies.
	
	\vspace{-3mm}
	\subsection{Dereverberation}
	\subsubsection*{Objective scores} For dereverberation, we utilize the recommended measures in the REVERB challenge paper \cite{kinoshita2016reverb}: cepstrum distance (CD) \cite{kitawaki1988cd}, log-likelihood ratio (LLR) \cite{hansen1998llr}, frequency weighted segmental SNR (FWSegSNR) \cite{tribolet1978fwssnr} and speech-to-dereverberation modulation energy ratio (SRMR) \cite{falk2010srmr}. The paper also recommended PESQ as an optional measure, although most of the latest dereverberation literature did not take it into account.  For outliers reduction, authors in \cite{hu2007mos} suggested limiting the ranges of CD to [0,10] and LLR to [0,2]. Lower values indicate better scores for CD and LLR, while higher values indicate better speech quality for FWSegSSNR, PESQ and SRMR. The CD, LLR, FWSegSNR and PESQ are chosen as they correlate to listening tests, albeit they are all intrusive scores, i.e., enhanced speech and clean reference are required. Accordingly, SRMR is employed as a non-intrusive score to operate on enhanced speech without a clean reference. Thus, it is quite important to measure the quality and intelligibility of enhanced unpaired real recordings.
	
	\subsubsection*{Results} For quantitative analysis, the CMGAN is compared with recent dereverberation methods. As discussed in Sec.~\ref{sec:reverb}, using time-domain approaches in dereverberation is limited and these methods did not use the REVERB challenge data. Thus, the chosen methods would all consider TF-domain analysis. For fair comparison, only papers recording individual room scores are considered. Based on this criteria, we compare against four recent methods:  Xiao \emph{et~al.} \cite{xiao2014reverb}, U-Net \cite{ernst2018reverb}, wide residual network (WRN) \cite{ribas2019reverb} and SkipConvNet \cite{kothapally2020reverb}. Unfortunately, none of these papers reported the PESQ scores, so it is excluded from the comparative analysis. However, PESQ is still used as the objective score to be maximized by the metric discriminator in CMGAN. 
	
	The results for both near and far microphone cases are shown in 
	Table~\ref{tab:reverb_near} and \ref{tab:reverb_far}, respectively. The first 
	four columns represent the simulated data results for the three different room 
	sizes (small -- room 1, medium -- room 2, large -- room 3 and average score). 
	The last column represents the SRMR of the real recordings. As expected, larger 
	rooms and further microphone placements result in lower scores, as these 
	scenarios would introduce more distortions to the speech. 
	
	In the simulated near 
	microphone case, the proposed CMGAN shows superior performance compared to 
	other methods in the majority of metrics, particularly FWSegSNR. For SRMR, Xiao 
	\emph{et~al.} reports a higher SRMR score on simulated near data, but a 
	significant drop is observed in near real recordings. SkipConvNet achieves 
	better real SRMR scores in the near case but worse on the simulated data. U-Net 
	and SkipConvNet report overall competitive scores, although CMGAN outperforms 
	in average CD and FWSegSNR with 0.3 and 1.65 dB, respectively. For the far 
	microphone, CMGAN is still able to show a gain in overall scores, especially 
	FWSegSNR. Xiao \emph{et~al.} is still slightly better in SRMR for simulated 
	data, but the gap is much closer than the near microphone case, only 0.05 on 
	average. The same holds for SkipConvNet with slightly better real SRMR scores 
	than the proposed CMGAN. 
	
	\subsubsection*{Ablation study} To validate the PESQ choice for metric 
	discriminator, we introduce a CMGAN variant operating on LLR as the objective 
	metric discriminator score (CMGAN-LLR). LLR is chosen as it reflects a bounded 
	metric and based on the LS-GAN formulation \cite{mao2017lsgans}, the metric 
	discriminator is more robust when the optimization space is bounded by a 
	normalized score. Accordingly, we modify Eq.~\ref{eq:gan_loss} to involve the 
	normalized LLR scores $Q_{LLR}$ instead of $Q_{PESQ}$ and the term 1 is changed 
	to 0 in both $\mathcal{L}_{\small\textrm{GAN}}$ and 
	$\mathcal{L}_{\small\textrm{D}}$. Thus, the score is minimized to 0 instead of 
	maximized to 1. It can be shown in Table~\ref{tab:reverb_near} and 
	\ref{tab:reverb_far} that the LLR score is marginally better than the original 
	CMGAN trained with PESQ. However, a considerable improvement is observed in 
	SRMR scores for both simulated and real recordings, especially in the near 
	microphone case. Moreover, the CMGAN-LLR variant outperforms the SkipConvNet in 
	real recordings for near and far microphone cases by 0.44 and 0.75, 
	respectively. Comparing both CMGAN and CMGAN-LLR shows a balanced performance 
	over most of the given metrics in favor of the standard proposed CMGAN, which 
	indicates that the PESQ is a robust metric to optimize and is highly correlated 
	with most of the given quality metrics.
	
	\vspace{-3mm}
	\subsection{Super-resolution}\label{sec:sr_results}
		\begin{figure*}[b!]
		\vspace{-4.5mm}
		\captionsetup[subfigure]{justification=centering}
		\centering
		\centerline{
			\begin{subfigure}[b]{.23\textwidth}
				\centering
%				{\includegraphics[width=\columnwidth]{Figs/sr_figs/lr_input26.pdf}}
				{\resizebox{\columnwidth}{!}{\begin{tikzpicture}
\begin{axis}[%
width=\figWSubMag,
height=\figHSubMag,
at={(0in,0in)},
scale only axis,
axis on top,
xmin=0.02875,
xmax=3.2539375,
xlabel style={font=\color{white!15!black}, yshift = -4pt},
xlabel={\mtlargeMSub Time [s]},
ylabel={\mtlargeMSub Frequency [kHz]},
ytick={0,2,4,6,8},
xtick={0.5,1,1.5,2,2.5,3},
ytick style={draw=none},
xtick style={draw=none},
yticklabel style = {font=\mtlargeMSubTick},
xticklabel style = {font=\mtlargeMSubTick, yshift = -5pt},
ymin=-0.00800819625565892,
ymax=8.00018805940326,
ylabel style={font=\color{white!15!black}, yshift = 8pt},
axis background/.style={fill=white},
legend style={legend cell align=left, align=left, draw=white!15!black}
]
\addplot [forget plot] graphics [xmin=0.02875, xmax=3.2539375, ymin=-0.00800819625565892, ymax=8.00018805940326] {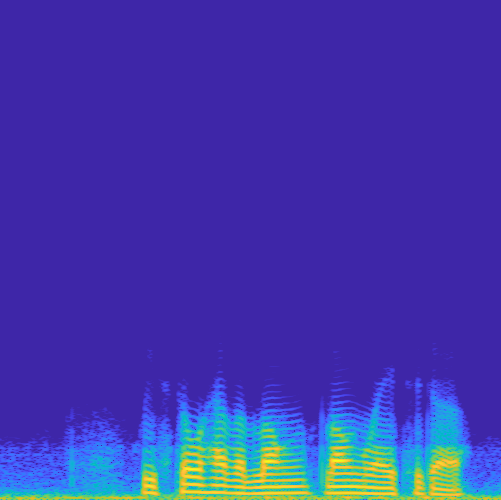};
\end{axis}
\end{tikzpicture}%
}}
				\caption{Low-resolution input}
				\label{fig:lr_input}
			\end{subfigure}%
			\hspace{3.7mm}
			\begin{subfigure}[b]{.23\textwidth}
				\centering
%				{\includegraphics[width=\columnwidth]{Figs/sr_figs/sr_mask26.pdf}}
				{\resizebox{\columnwidth}{!}{\begin{tikzpicture}
\begin{axis}[%
width=\figWSubMag,
height=\figHSubMag,
at={(0in,0in)},
scale only axis,
axis on top,
xmin=0.02875,
xmax=3.2539375,
xlabel style={font=\color{white!15!black}, yshift = -4pt},
xlabel={\mtlargeMSub Time [s]},
ylabel={\mtlargeMSub Frequency [kHz]},
ytick={0,2,4,6,8},
xtick={0.5,1,1.5,2,2.5,3},
ytick style={draw=none},
xtick style={draw=none},
yticklabel style = {font=\mtlargeMSubTick},
xticklabel style = {font=\mtlargeMSubTick, yshift = -5pt},
ymin=-0.00800819625565892,
ymax=8.00018805940326,
ylabel style={font=\color{white!15!black}, yshift = 8pt},
axis background/.style={fill=white},
legend style={legend cell align=left, align=left, draw=white!15!black}
]
\addplot [forget plot] graphics [xmin=0.02875, xmax=3.2539375, ymin=-0.00800819625565892, ymax=8.00018805940326] {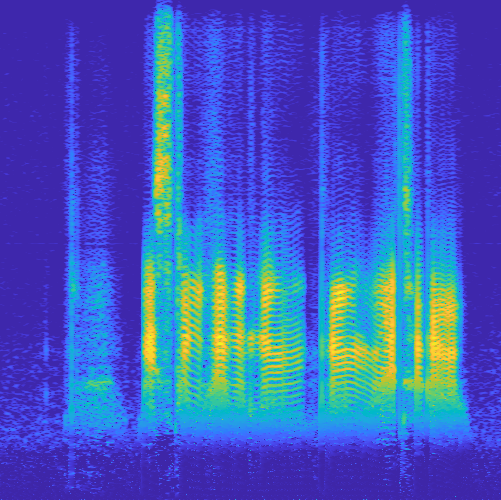};
\end{axis}
\end{tikzpicture}%
}}
				\caption{Predicted mask $M'$}
				\label{fig:sr_mask}
			\end{subfigure}
			\hspace{2.7mm}
			\begin{subfigure}[b]{.23\textwidth}
				\centering
%				{\includegraphics[width=\columnwidth]{Figs/sr_figs/pr_output26.pdf}}
				{\resizebox{\columnwidth}{!}{\begin{tikzpicture}
\begin{axis}[%
width=\figWSubMag,
height=\figHSubMag,
at={(0in,0in)},
scale only axis,
axis on top,
xmin=0.02875,
xmax=3.2539375,
xlabel style={font=\color{white!15!black}, yshift = -4pt},
xlabel={\mtlargeMSub Time [s]},
ylabel={\mtlargeMSub Frequency [kHz]},
ytick={0,2,4,6,8},
xtick={0.5,1,1.5,2,2.5,3},
ytick style={draw=none},
xtick style={draw=none},
yticklabel style = {font=\mtlargeMSubTick},
xticklabel style = {font=\mtlargeMSubTick, yshift = -5pt},
ymin=-0.00800819625565892,
ymax=8.00018805940326,
ylabel style={font=\color{white!15!black}, yshift = 8pt},
axis background/.style={fill=white},
legend style={legend cell align=left, align=left, draw=white!15!black}
]
\addplot [forget plot] graphics [xmin=0.02875, xmax=3.2539375, ymin=-0.00800819625565892, ymax=8.00018805940326] {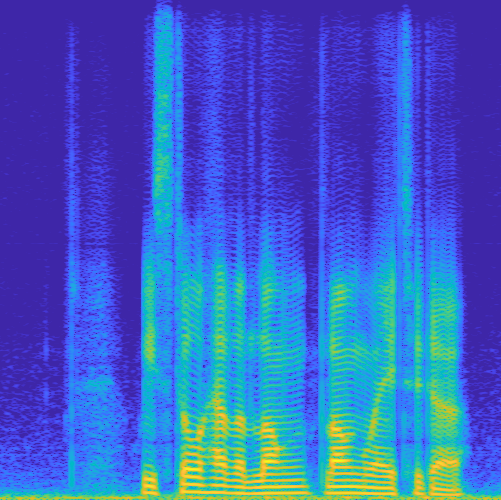};
\end{axis}
\end{tikzpicture}%
}}
				\caption{Predicted output}
				\label{fig:pr_output}
			\end{subfigure}
			\hspace{2.7mm}
			\begin{subfigure}[b]{.23\textwidth}
				\centering
%				{\includegraphics[width=\columnwidth]{Figs/sr_figs/hr_reference26.pdf}}
				{\resizebox{\columnwidth}{!}{\begin{tikzpicture}
\begin{axis}[%
width=\figWSubMag,
height=\figHSubMag,
at={(0in,0in)},
scale only axis,
axis on top,
xmin=0.02875,
xmax=3.2539375,
xlabel style={font=\color{white!15!black}, yshift = -4pt},
xlabel={\mtlargeMSub Time [s]},
ylabel={\mtlargeMSub Frequency [kHz]},
ytick={0,2,4,6,8},
xtick={0.5,1,1.5,2,2.5,3},
ytick style={draw=none},
xtick style={draw=none},
yticklabel style = {font=\mtlargeMSubTick},
xticklabel style = {font=\mtlargeMSubTick, yshift = -5pt},
ymin=-0.00800819625565892,
ymax=8.00018805940326,
ylabel style={font=\color{white!15!black}, yshift = 8pt},
axis background/.style={fill=white},
legend style={legend cell align=left, align=left, draw=white!15!black}
]
\addplot [forget plot] graphics [xmin=0.02875, xmax=3.2539375, ymin=-0.00800819625565892, ymax=8.00018805940326] {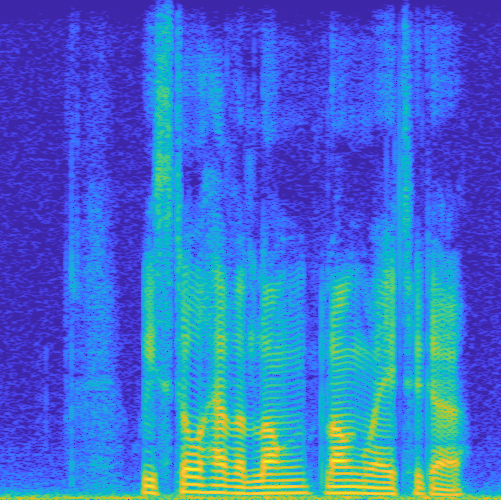};
\end{axis}
\end{tikzpicture}%
}}
				\caption{High-resolution reference}
				\label{fig:hr_reference}
			\end{subfigure}
		}
		\vspace{1.5mm}
		\centering
		\centerline{
			\begin{subfigure}[b]{.25\textwidth}
				\centering
%				{\includegraphics[width=\columnwidth]{Figs/sr_figs/lr_input_time17_zoom.pdf}}
				{\resizebox{\textwidth}{!}{\begin{tikzpicture}
\begin{axis}[%
width=\figWtime,
height=\figHtime,
at={(0in,0in)},
scale only axis,
xmin=1.82,
xmax=1.84,
xlabel style={font=\color{white!15!black}},
xlabel={\mtlargeTSub Time [ms]},
ytick={-1,0,1},
xticklabels={},
xtick style={draw=none},
ymin=-1.2,
ymax=1.2,
ylabel style={font=\color{white!15!black},yshift = -10pt},
ylabel={\mtlargeTSub Value [A.U.]},
yticklabel style = {font=\mtlargeTSubTick},
xticklabel style = {font=\mtlargeTSubTick},
axis background/.style={fill=white},
legend style={legend cell align=left, align=left, draw=white!15!black}
]
\addplot [color=ssnrcolor, line width=\SRLineWidthT]
  table[row sep=crcr]{%
1.82011463122756	-0.526829268292683\\
1.82036464697361	-0.343774069319641\\
1.82061466271966	-0.387676508344031\\
1.82086467846571	-0.117073170731707\\
1.82111469421175	-0.0955070603337612\\
1.8213647099578	-0.0238767650834403\\
1.82161472570385	0.0613607188703466\\
1.8218647414499	-0.0256739409499358\\
1.82211475719594	0.0741976893453145\\
1.82236477294199	0.0225930680359435\\
1.82261478868804	0.231065468549422\\
1.82286480443409	0.237997432605905\\
1.82311482018013	0.481899871630295\\
1.82336483592618	0.560718870346598\\
1.82361485167223	0.736585365853659\\
1.82386486741828	0.911424903722721\\
1.82411488316433	0.187163029525032\\
1.82436489891037	0.726572528883184\\
1.82461491465642	-0.634403080872914\\
1.82486493040247	-0.710141206675225\\
1.82511494614852	-0.766367137355584\\
1.82536496189456	-0.78408215661104\\
1.82561497764061	-0.0225930680359435\\
1.82586499338666	-0.290629011553273\\
1.82611500913271	0.403337612323492\\
1.82636502487876	0.287548138639281\\
1.8266150406248	0.246469833119384\\
1.82686505637085	0.0523748395378691\\
1.8271150721169	-0.237740693196406\\
1.82736508786295	-0.275224646983312\\
1.82761510360899	-0.441078305519897\\
1.82786511935504	-0.171245186136072\\
1.82811513510109	-0.250320924261874\\
1.82836515084714	0.252631578947368\\
1.82861516659319	0.0546854942233633\\
1.82886518233923	0.166623876765083\\
1.82911519808528	-0.0567394094993582\\
1.82936521383133	-0.311168164313222\\
1.82961522957738	-0.419768934531451\\
1.82986524532342	-0.562002567394095\\
1.83011526106947	-0.372272143774069\\
1.83036527681552	-0.361745827984596\\
1.83061529256157	-0.127342747111682\\
1.83086530830761	-0.138125802310655\\
1.83111532405366	-0.0110397946084724\\
1.83136533979971	-0.00539152759948652\\
1.83161535554576	-0.0518613607188703\\
1.83186537129181	0.0474967907573813\\
1.83211538703785	0.0213093709884467\\
1.8323654027839	0.234659820282413\\
1.83261541852995	0.256482670089859\\
1.832865434276	0.484210526315789\\
1.83311545002204	0.563543003851091\\
1.83336546576809	0.726059050064185\\
1.83361548151414	0.914249037227214\\
1.83386549726019	1\\
1.83411551300624	-0.133504492939666\\
1.83436552875228	0.476508344030809\\
1.83461554449833	-0.372528883183569\\
1.83486556024438	-0.650577663671374\\
1.83511557599043	-0.698074454428755\\
1.83536559173647	-0.563286264441592\\
1.83561560748252	-0.0644415917843389\\
1.83586562322857	0.0629011553273428\\
1.83611563897462	0.360462130937099\\
1.83636565472067	0.372528883183569\\
1.83661567046671	0.308600770218229\\
1.83686568621276	-0.119640564826701\\
1.83711570195881	-0.154300385109114\\
1.83736571770486	-0.431578947368421\\
1.8376157334509	-0.267265725288832\\
1.83786574919695	-0.115532734274711\\
1.838115764943	-0.089602053915276\\
1.83836578068905	0.309114249037227\\
1.83861579643509	0.220795892169448\\
1.83886581218114	0.22798459563543\\
1.83911582792719	-0.114762516046213\\
1.83936584367324	-0.113735558408216\\
1.83961585941929	-0.659050064184852\\
1.83986587516533	-0.574069319640565\\
1.84011589091138	-0.41181001283697\\
};
\end{axis}
\end{tikzpicture}%
}}
				\caption{4 kHz input segment}
				\label{fig:lr_input_time}
			\end{subfigure}%
			\hspace{12mm}
			\begin{subfigure}[b]{.25\textwidth}
				\centering
%				{\includegraphics[width=\columnwidth]{Figs/sr_figs/pr_output_time17_zoom.pdf}}
				{\resizebox{\textwidth}{!}{\input{Figs/sr_figs/pr_output_time17_zoom.tex}}}
				\caption{16 kHz predicted segment}
				\label{fig:pr_output_time}
			\end{subfigure}
			\hspace{10mm}
			\begin{subfigure}[b]{.25\textwidth}
				\centering
%				{\includegraphics[width=\columnwidth]{Figs/sr_figs/hr_reference_time17_zoom.pdf}}
				{\resizebox{\textwidth}{!}{\input{Figs/sr_figs/hr_reference_time17_zoom.tex}}}
				\caption{16 kHz reference segment}
				\label{fig:hr_reference_time}
			\end{subfigure}
		}
		\caption{Example of scale 4 super-resolution (4 kHz $\rightarrow$ 16 kHz). The upper row represents the TF-magnitude representations of the relevant spectrograms. The bottom row shows a 20 ms segment of the corresponding time-domain signals. \label{fig:sr_task}}
	\end{figure*}
	\subsubsection*{Objective scores} Two metrics, log-spectral distance (LSD) and signal-to-noise ratio (SNR), are used to evaluate super-resolution. Based on our literature review, the LSD definition is not the same for all papers. Mathematically, LSD measures the log distance between the magnitude spectrogram component of the enhanced speech with respect to the clean reference. Some papers would use the log to the base $e$, while others would evaluate the log to base 10. In both definitions, the STFT is evaluated with a Hanning window of 2048 samples and a hop size of 512. To ensure a fair comparison, the same STFT parameterization is used and the LSD results based on the two different definitions in the literature are presented. A lower LSD and a higher SNR represent better speech quality.
	\subsubsection*{Results} Since masking-based methods are not relevant for the super-resolution task, as previously stated in Sec.~\ref{sec:sr}. Therefore, the CMGAN mask decoder part is modified by involving an element-wise addition instead of element-wise multiplication. This is reflected in Eq.~\ref{eq:comb} as follows:
	\begin{equation}
		\vspace{-0.5mm}
		\begin{aligned}
			&\hat{X}_r=(M'+Y_m) \cos{Y_p}+\hat{X}'_r \\
			&\hat{X}_i=(M'+Y_m) \sin{Y_p}+\hat{X}'_i,
		\end{aligned}
		\vspace{-0.5mm}
	\end{equation}
	where $M'$ represents the modified output of the mask decoder. Unlike the prior cases of denoising and dereverberation, the network is not learning mask activations between 0 and 1 to suppress the noise and preserve the speech, but rather activations that can complete the missing high-frequency bands while preserving the given low-frequency bands. 
	
	\begin{table}[t!]
		\vspace{1.5mm}
		\centering
		\caption{Performance comparison for super-resolution, “-” denotes the result is not provided in the original paper.}\vspace{-1.5mm}
		\def\arraystretch{1.2}	
		\resizebox{0.98\columnwidth}{!}{
			\huge
			\begin{tabular}{l@{\hskip 0.2in}c|ccc|ccc}
				\toprule  
				&  & \multicolumn{3}{c}{VCTK-Single} \vline & \multicolumn{3}{c}{VCTK-Multi.} \\
				\hline 
				Method& $s$ & $\textrm{LSD}_e \downarrow$ & $\textrm{LSD}_{10} \downarrow$ & SNR $\uparrow$ & $\textrm{LSD}_e \downarrow$ & $\textrm{LSD}_{10} \downarrow$ & SNR $\uparrow$ \\
				\hline
				U-Net \cite{kuleshov2017sr} & 2 & 3.2 & - & 21.1 & 3.1 & - & 20.7 \\
				TFiLM \cite{birnbaum2019sr} & 2 & 2.5 & - & 19.5 & 1.8 & - & 19.8 \\
				AFILM \cite{rakotonirina2021sr} & 2 & 2.3 & - & 19.3 & 1.7 & - & 20.0 \\
				AE \cite{wang2021sr} & 2 & - & 0.9 & 22.4 & - & 0.9 & 22.1  \\
				NVSR \cite{liu2022nvsr} & 2 & - & - & - & - & 0.8 & -  \\
				CMGAN & 2 & 1.7 & 0.7 & \textbf{24.7} & 1.6 & 0.7 & \textbf{24.4} \\
				CMGAN-Mag. & 2 & \textbf{1.4} & \textbf{0.6} & 22.2 & \textbf{1.3} & \textbf{0.6} & 23.4 \\		
				\hline
				U-Net \cite{kuleshov2017sr} & 4 & 3.6 & - & 17.1 & 3.5 & - & 16.1 \\
				TFiLM \cite{birnbaum2019sr} & 4 & 3.5 & - & 16.8 & 2.7 & - & 15.0 \\
				AFILM \cite{rakotonirina2021sr} & 4 & 3.1 & - & 17.2 & 2.3 & - & 17.2 \\
				TFNet \cite{lim2018sr} & 4 & - & 1.3 & 18.5 & - & 1.3 & 17.5 \\
				AE \cite{wang2021sr} & 4 & - & 0.9 & \textbf{18.9} & - & 1.0 & 18.1  \\
				NVSR \cite{liu2022nvsr} & 4 & - & - & - & - & 0.9 & -  \\
				CMGAN & 4 & 2.3 & 1.0 & 18.6 & 2.2 & 1.0 & \textbf{19.1} \\
				CMGAN-Mag. & 4 & \textbf{1.7} & \textbf{0.7} & 16.9 & \textbf{1.8} & \textbf{0.8} & 16.1 \\		
				\hline
				TFiLM \cite{birnbaum2019sr} & 8 & 4.3 & - & 12.9 & 2.9 & - & 12.0 \\
				AFILM \cite{rakotonirina2021sr} & 8 & 3.7 & - & 12.9 & 2.7 & - & 12.0 \\
				TFNet \cite{lim2018sr} & 8 & - & 1.9 & \textbf{15.0} & - & 1.9 & 12.0 \\
				NVSR \cite{liu2022nvsr} & 8 & - & - & - & - & 1.1 & -  \\
				CMGAN & 8 & 2.6 & 1.1 & 12.9 & 2.7 & 1.2 & \textbf{14.1} \\
				CMGAN-Mag. & 8 & \textbf{1.9} &  \textbf{0.8} & 10.9 & \textbf{2.0} & \textbf{0.9} & 10.9 \\
				\bottomrule 
			\end{tabular}
		}
		\label{tab:sr}
		\vspace{-6mm}
	\end{table}
	As shown in Table~\ref{tab:sr}, we compare our approach with five other 
	methods: the U-Net architecture proposed by Kuleshov \textit{et al.} 
	\cite{kuleshov2017sr}, TFiLM \cite{birnbaum2019sr}, AFiLM 
	\cite{rakotonirina2021sr}, hybrid TFNet \cite{lim2018sr}, hybrid AE 
	\cite{wang2021sr} and NVSR \cite{liu2022nvsr}. All the scores are from the 
	corresponding original papers. The value $s=$ 2/4/8 implies upsampling scale 
	from 8 kHz/4 kHz/2 kHz to 16 kHz speech. In the VCTK-Single experiment, our 
	method achieved the best score in all three metrics on scale 2 when converting 
	the audio signal from 8 kHz to 16 kHz, especially in SNR, a 2.3 dB improvement 
	compared to the SOTA AE method. As for scale 4, the AE method shows a marginal 
	improvement of 0.3 dB and 0.1 in SNR and $\textrm{LSD}_{10}$, respectively. In 
	the scale 8 task, our method exceeds other methods in terms of 
	$\textrm{LSD}_{e}$ and $\textrm{LSD}_{10}$. However, the SNR is lower than 
	TFNet and similar to TFiLM and AFiLM approaches. We hypothesize that this is
	accounted for the limited training samples in the VCTK-Single dataset, which 
	can lead to model overfitting. On the other hand, in the VCTK-Multi. 
	evaluation, our method outperforms other approaches in all upscaling ratios on 
	all metrics. Specifically, our method has an improvement of 2.3 dB, 1.0 dB and 
	2.1 dB on SNR on scales 2/4/8. Note that CMGAN has a much better performance on 
	scale 8 compared to the same scale in VCTK-Single evaluation, which verifies 
	the overfitting assumption.
	\subsubsection*{Ablation study} To demonstrate the effectiveness of complex TF-domain super-resolution. The CMGAN is modified to eliminate both complex decoder and metric discriminator, leaving only the magnitude loss (CMGAN-Mag.). A substantial improvement in both $\textrm{LSD}_{e}$ and $\textrm{LSD}_{10}$ is observed when the complex branch is removed and this is expected as the LSD is defined in magnitude component only. This LSD gain comes at the expense of a significant drop in the SNR scores, which considers the reconstructed time-domain signal. Thus, removing the complex branch would give a push in the LSD as the network would focus only on enhancing the magnitude component but with a degradation in the overall signal quality. 
	
	An illustration of the input, predicted and reference tracks from a scale 4 example is depicted in Fig.~\ref{fig:sr_task}. Excitation of high-frequency bands are clear in the output mask $M'$. Comparing Fig.~\ref{fig:pr_output} and \ref{fig:hr_reference} shows the potential of the CMGAN in constructing missing high-frequency bands just from observing different speech phonetics in the training data. This performance is also reflected as an accurate interpolation of intermediate samples in the time-domain Fig.~\ref{fig:lr_input_time}, \ref{fig:pr_output_time} and \ref{fig:hr_reference_time}.
	
	\vspace{-2mm}
	\section{Opinion score evaluation}
	Till now, the proposed architecture is quantitatively compared to different 
	SOTA methods using objective metrics scores. Although these scores can serve as 
	an indication of how well is the proposed method, they still cannot fully 
	replace the subjective quality measure. Since subjective listening tests are 
	costly and time consuming as it requires many participants and ideal listening 
	conditions. Therefore, finding an objective measure that can highly correlate 
	with the subjective quality score is still an open research topic 
	\cite{streijl2016mos}. The most noticeable work in this area is introduced in 
	\cite{hu2007mos}, where the authors proposed a composite mean opinion score 
	(MOS) based on traditional regression analysis methods \cite{friedman1991mars}. 
	Note that these scores are used in Sec.~\ref{sec:denoising_results} to evaluate 
	speech denoising performance. The study involved 1792 speech samples rated 
	according to ITU-T P.835 standards \cite{p835standards} and well-established 
	objective measures such as PESQ, segmental SNR, LLR and weighted spectral slope 
	(WSS) \cite{klatt1982wss} are utilized as basis functions for construction of 
	three different composite scores reflecting the signal distortion, background 
	noise and overall quality. The proposed composite measure reported a 
	correlation of 0.9 to 0.91 with the subjective ratings and the authors 
	emphasized the importance of PESQ as it shows the highest correlation (0.89). 
	However, this study is limited to only four background noise types under two 
	SNR conditions (5 and 10 dB) and most importantly, the proposed scores are 
	intrusive (requiring both paired clean and enhanced speech). 
	
	Recently, DNNs have been utilized for finding a subjective alternative score 
	\cite{fu2018subjective,avila2019subjective,catellier2020subjective,serra2021subjective,reddy2021dnsmos,reddy2022dnsmos}.
	Unlike the previous composite measure, most of these methods will take the 
	track as an input and the network is trained to mimic the subjective ratings. 
	Thus, the scores will not depend on non-optimal objective scores, but rather on 
	the whole track. Additionally, these scores are non-intrusive, hence evaluating 
	enhanced tracks without the need for clean reference is possible. The standard 
	score used as a subjective baseline for many recent studies is the DNSMOS 
	proposed by Microsoft in \cite{reddy2021dnsmos,reddy2022dnsmos}. The DNSMOS is 
	trained on 75 hours of rated speech. In accordance to ITU-T P.835, listeners 
	assign a score between 1 and 5 (higher is better) for signal distortion, 
	background noise and overall quality. A significant correlation of 0.94 to 
	0.98 is reported over the three given quality assessment scores.
	
	Due to non-availability of open-source 
	implementations, especially in dereverberation, the MOS evaluation will 
	focus on the denoising aspect of the SE problem. Accordingly, four different 
	denoising use cases are included in this study to indicate the generalization 
	capability of the network to unseen noise conditions, real noise samples and 
	additional distortions not included in training. To this end, the frameworks 
	will be all trained on a single use case (Voice Bank+DEMAND), then the models 
	will be evaluated on four different datasets:
	\begin{enumerate}[(a),wide = 0pt]
		\item Voice Bank+DEMAND test set \cite{valentini2016voicebank}: including 
		35 minutes (824 tracks) of noisy speech from two unseen speakers using 
		noise types from DEMAND dataset \cite{thiemann2013demand} which are not 
		included in the training as explained in Sec.~\ref{sec:denoising_data}.
		\item CHiME-3 \cite{barker2015chime}: including 7.8 hours (4560 tracks) of 
		real noisy speech recordings from 12 speakers at four different 
		environments: bus, cafe, pedestrian area and street junction. In this data, 
		no clean reference tracks are available.
		\item DNS-Challenge \cite{dubey2022dns}: the original data includes 1934 
		English speaker reading speech samples from 
		Librivox\footnote{\textit{\href{https://librivox.org/}{https://librivox.org/}}}
		and 181 hours of 150 different noise types from Audio Set 
		\cite{gemmeke2017audioset} and 
		Freesound\footnote{\textit{\href{https://freesound.org/}{https://freesound.org/}}}.
		Based on this dataset, we construct 9 hours (3240 tracks) of noisy speech 
		with SNRs from 0 to 10 dB. It is worth noting that the DNS-Challenge dataset features noises with unique spectral characteristics and harmonics, such as doorbells, church bells, squeaky chairs, and musical distortions. This stands in contrast to the predominantly wideband noises found in DEMAND.
		\item DNS-Challenge+Reverb: we use the same 9 hours, but we simulate 
		reverberant conditions on the speech, then we add the same noise in the DNS 
		challenge part. The RIRs are chosen from openSLR26/28 \cite{ko2017openslr}, 
		including 248 real and 60k synthetic conditions. The RIRs are recorded in 
		three different room sizes with a 60 dB reverberation time of 0.3-1.3 seconds.
	\end{enumerate}\vspace{-1mm}
	
	All tracks are resampled to 16 kHz and the ratio of male-to-female speakers is 
	50\%. From Table~\ref{tab:results_demand}, we choose a representative for each 
	denoising paradigm. The methods were chosen based on the availability of 
	open-source implementations and the reproducibility of the reported results in 
	the corresponding papers. As a representative for metric discriminator, we used 
	the MetricGAN+ \cite{fu2021metricgan+}. For time-domain methods, DEMUCS 
	\cite{defossez2020demucs} is selected. For TF-domain complex denoising, PHASEN 
	\cite{yin2020phasen} is chosen as it attempts to correct magnitude and phase 
	components. In addition to, PFPL \cite{hsieh2020pfpl} utilizing a deep 
	complex-valued network to enhance both real and imaginary parts. Most of 
	the papers provided an official implementation with pretrained models. PHASEN 
	is the only exception, as a non-official code is used and we trained the model 
	to reproduce the results in the paper. For DEMUCS, the available model is 
	pretrained on both Voice Bank+DEMAND and DNS-Challenge data. Thus, we retrain 
	DEMUCS using the recommended configuration on Voice Bank+DEMAND data only 
	to ensure a fair comparison between all presented models. In this study, a DNN-based approach is used to evaluate a MOS on the aforementioned four different datasets ($\approx$ 26 hours). Then a subset is selected from these tracks to construct a listening test experiment.  
	\begin{figure*}[t!]
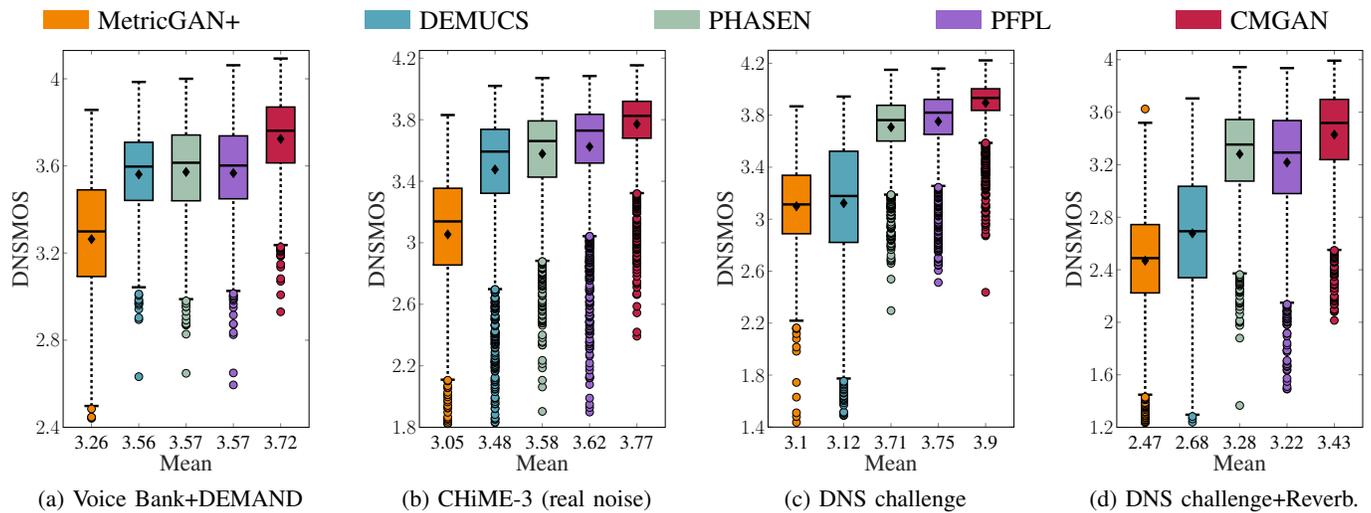

		\hspace{4mm}\metricganrect MetricGAN+ \hfill \demucsrect DEMUCS \hfill 
		\phasenrect PHASEN \hfill \pfplrect PFPL \hfill \cmganrect CMGAN 
		\hspace{4mm}
		
		\captionsetup[subfigure]{oneside,margin={0.3cm,0cm}} 
		\centering
		\vspace{0.25cm}
		\centerline{
			\begin{subfigure}[b]{.22\textwidth}
				\centering
%				{\includegraphics[width=\columnwidth]{Figs/boxplots/bp_vctk_0.pdf}}
				{\resizebox{\columnwidth}{!}{\begin{tikzpicture}
\pgfplotsset{every tick label/.append style={font=\huge}}
\begin{axis}[%
width=\figWbp,
height=\figHbp,
at={(0in,0in)},
scale only axis,
xmin=0.4,
xmax=5.6,
xtick={1,2,3,4,5},
xticklabels={3.26,3.56,3.57,3.57,3.72},
xticklabel style = {font=\huge, yshift = -5pt},
xlabel style={font=\color{white!15!black}},
xlabel={\mtlarge Mean},
ymin=2.4,
ymax=4.13,
ytick={2.4,2.8, ...,4.4},
ylabel style={font=\color{white!15!black}, yshift = 1pt},
ylabel={\mtlarge DNSMOS},
axis background/.style={fill=white},
legend style={legend cell align=left, align=left, draw=white!15!black}
]

%% Whiskers
\addplot [color=black, dashed, line width=\bplinewidth, forget plot]
  table[row sep=crcr]{%
1	3.49005043291932\\
1	3.85716453860028\\
};
\addplot [color=black, dashed, line width=\bplinewidth, forget plot]
  table[row sep=crcr]{%
2	3.70864015348259\\
2	3.98554833755049\\
};
\addplot [color=black, dashed, line width=\bplinewidth, forget plot]
  table[row sep=crcr]{%
3	3.74196588371\\
3	4.00016385431763\\
};
\addplot [color=black, dashed, line width=\bplinewidth, forget plot]
  table[row sep=crcr]{%
4	3.73800181520844\\
4	4.0622093558367\\
};
\addplot [color=black, dashed, line width=\bplinewidth, forget plot]
  table[row sep=crcr]{%
5	3.87046639648466\\
5	4.09304547575044\\
};
\addplot [color=black, dashed, line width=\bplinewidth, forget plot]
  table[row sep=crcr]{%
1	2.49798524276692\\
1	3.09230567970063\\
};
\addplot [color=black, dashed, line width=\bplinewidth, forget plot]
  table[row sep=crcr]{%
2	3.04293680190274\\
2	3.44155082941629\\
};
\addplot [color=black, dashed, line width=\bplinewidth, forget plot]
  table[row sep=crcr]{%
3	2.98844652386544\\
3	3.43997716244777\\
};
\addplot [color=black, dashed, line width=\bplinewidth, forget plot]
  table[row sep=crcr]{%
4	3.02613190733867\\
4	3.44904428027506\\
};
\addplot [color=black, dashed, line width=\bplinewidth, forget plot]
  table[row sep=crcr]{%
5	3.23666164298879\\
5	3.61375169642012\\
};

%% Whiskers edges
\addplot [color=black, line width=\bplinewidth, forget plot]
  table[row sep=crcr]{%
0.85	3.85716453860028\\
1.15	3.85716453860028\\
};
\addplot [color=black, line width=\bplinewidth, forget plot]
  table[row sep=crcr]{%
1.85	3.98554833755049\\
2.15	3.98554833755049\\
};
\addplot [color=black, line width=\bplinewidth, forget plot]
  table[row sep=crcr]{%
2.85	4.00016385431763\\
3.15	4.00016385431763\\
};
\addplot [color=black, line width=\bplinewidth, forget plot]
  table[row sep=crcr]{%
3.85	4.0622093558367\\
4.15	4.0622093558367\\
};
\addplot [color=black, line width=\bplinewidth, forget plot]
  table[row sep=crcr]{%
4.85	4.09304547575044\\
5.15	4.09304547575044\\
};
\addplot [color=black, line width=\bplinewidth, forget plot]
  table[row sep=crcr]{%
0.85	2.49798524276692\\
1.15	2.49798524276692\\
};
\addplot [color=black, line width=\bplinewidth, forget plot]
  table[row sep=crcr]{%
1.85	3.04293680190274\\
2.15	3.04293680190274\\
};
\addplot [color=black, line width=\bplinewidth, forget plot]
  table[row sep=crcr]{%
2.85	2.98844652386544\\
3.15	2.98844652386544\\
};
\addplot [color=black, line width=\bplinewidth, forget plot]
  table[row sep=crcr]{%
3.85	3.02613190733867\\
4.15	3.02613190733867\\
};
\addplot [color=black, line width=\bplinewidth, forget plot]
  table[row sep=crcr]{%
4.85	3.23666164298879\\
5.15	3.23666164298879\\
};

%% Boxes
\addplot [ultra thick, color=black, fill = metricgan, forget plot]
  table[row sep=crcr]{%
0.7	3.09230567970063\\
0.7	3.49005043291932\\
1.3	3.49005043291932\\
1.3	3.09230567970063\\
0.6835	3.09230567970063\\
};
\addplot [ultra thick, color=black, fill = demucs, forget plot]
  table[row sep=crcr]{%
1.7	3.44155082941629\\
1.7	3.70864015348259\\
2.3	3.70864015348259\\
2.3	3.44155082941629\\
1.6835	3.44155082941629\\
};
\addplot [ultra thick, color=black, fill = phasen, forget plot]
  table[row sep=crcr]{%
2.7	3.43997716244777\\
2.7	3.74196588371\\
3.3	3.74196588371\\
3.3	3.43997716244777\\
2.6835	3.43997716244777\\
};
\addplot [ultra thick, color=black, fill = pfpl, forget plot]
  table[row sep=crcr]{%
3.7	3.44904428027506\\
3.7	3.73800181520844\\
4.3	3.73800181520844\\
4.3	3.44904428027506\\
3.6835	3.44904428027506\\
};
\addplot [ultra thick, color=black, fill = cmgan, forget plot]
  table[row sep=crcr]{%
4.7	3.61375169642012\\
4.7	3.87046639648466\\
5.3	3.87046639648466\\
5.3	3.61375169642012\\
4.6835	3.61375169642012\\
};

%% Median
\addplot [color=black, line width=\bplinewidth, forget plot]
  table[row sep=crcr]{%
0.7	3.29916050221308\\
1.3	3.29916050221308\\
};
\addplot [color=black, line width=\bplinewidth, forget plot]
  table[row sep=crcr]{%
1.7	3.59703875087386\\
2.3	3.59703875087386\\
};
\addplot [color=black, line width=\bplinewidth, forget plot]
  table[row sep=crcr]{%
2.7	3.61467690464597\\
3.3	3.61467690464597\\
};
\addplot [color=black, line width=\bplinewidth, forget plot]
  table[row sep=crcr]{%
3.7	3.60184745546717\\
4.3	3.60184745546717\\
};
\addplot [color=black, line width=\bplinewidth, forget plot]
  table[row sep=crcr]{%
4.7	3.76180523557592\\
5.3	3.76180523557592\\
}; \label{plot:median}

%% Outliers
\addplot [only marks, color=black, draw=none, mark=*, mark size=\msize-0.5pt, mark options={solid, fill=metricgan}, forget plot]
  table[row sep=crcr]{%
1	2.01918781675801\\
1	2.14534604477358\\
1	2.17419128341978\\
1	2.2954428919435\\
1	2.34386637438878\\
1	2.36358425299569\\
1	2.36987217625874\\
1	2.39422365853357\\
1	2.4413223830274\\
1	2.44552255302408\\
1	2.44696251964817\\
1	2.48455718167563\\
}; \label{plot:metricgan}
\addplot [only marks, color=black, draw=none, mark=*, mark size=\msize-0.5pt, mark options={solid, fill=demucs}, forget plot]
  table[row sep=crcr]{%
2	2.14806997511035\\
2	2.63251652355439\\
2	2.89445373435639\\
2	2.90337731226683\\
2	2.94272116855312\\
2	2.96234548124925\\
2	2.96959008587529\\
2	2.97144492145353\\
2	2.9800118517142\\
2	2.98658812907524\\
2	3.00551719041022\\
2	3.01176365922574\\
}; \label{plot:demucs}
\addplot [only marks, color=black, draw=none, mark=*, mark size=\msize-0.5pt, mark options={solid, fill=phasen}, forget plot]
  table[row sep=crcr]{%
3	2.64792618281648\\
3	2.82772759886091\\
3	2.86961645045424\\
3	2.87137989997352\\
3	2.87362956600396\\
3	2.87630946623401\\
3	2.89219663006096\\
3	2.91060515344956\\
3	2.93254463811877\\
3	2.95373464509693\\
3	2.95375406256091\\
3	2.96875438710249\\
3	2.98010460385692\\
}; \label{plot:phasen}
\addplot [only marks, color=black, draw=none, mark=*, mark size=\msize-0.5pt, mark options={solid, fill=pfpl}, forget plot]
  table[row sep=crcr]{%
4	1.93953924887124\\
4	2.18962264516539\\
4	2.59461884280071\\
4	2.64958148974948\\
4	2.82524657770893\\
4	2.83506284840302\\
4	2.87386924410221\\
4	2.87493116336296\\
4	2.91573284756012\\
4	2.95018224493027\\
4	2.96530358582577\\
4	2.98260381811229\\
4	2.98973978904713\\
4	2.99301108346709\\
4	3.00144596100104\\
4	3.00260808221539\\
4	3.00456538104505\\
4	3.01375997417055\\
}; \label{plot:pfpl}
\addplot [only marks, color=black, draw=none, mark=*, mark size=\msize-0.5pt, mark options={solid, fill=cmgan}, forget plot]
  table[row sep=crcr]{%
5	2.93027032762398\\
5	3.00811332796957\\
5	3.07058831742781\\
5	3.08247568590847\\
5	3.13544109821664\\
5	3.14896255222145\\
5	3.18608071124185\\
5	3.18772390022398\\
5	3.19246521314864\\
5	3.20548523231214\\
5	3.21211655415347\\
5	3.21407056300907\\
5	3.22120646666983\\
5	3.22567210185158\\
5	3.22744327916223\\
}; \label{plot:cmgan}

%% Mean
\addplot[only marks, mark=diamond*, mark options={}, mark size=\msize, draw=black, fill=black] table[row sep=crcr]{%
x	y\\
1	3.26382135375611\\
2	3.56116162558782\\
3	3.57214124833496\\
4	3.56691923448183\\
5	3.7238110601902\\
};
\end{axis} \label{plot:mean}
\end{tikzpicture}%
}}
				\caption{Voice Bank+DEMAND}
				\label{fig:bp_vctk}
			\end{subfigure}
			\hspace{5mm}
			\begin{subfigure}[b]{.22\textwidth}
				\centering
%				{\includegraphics[width=\columnwidth]{Figs/boxplots/bp_chime_0.pdf}}
				{\resizebox{\columnwidth}{!}{\input{Figs/boxplots/bp_chime_0.tex}}}
				\caption{CHiME-3 (real noise)}
				\label{fig:bp_chime}
			\end{subfigure}
			\hspace{4mm}
			\begin{subfigure}[b]{.22\textwidth}
				\centering
%				{\includegraphics[width=\columnwidth]{Figs/boxplots/bp_DNS_5_15_0.pdf}}
				{\resizebox{\columnwidth}{!}{\input{Figs/boxplots/bp_DNS_5_15_0.tex}}}
				\caption{DNS-Challenge}
				\label{fig:bp_dns}
			\end{subfigure}
			\hspace{4mm}
			\begin{subfigure}[b]{.22\textwidth}
				\centering
%				{\includegraphics[width=\columnwidth]{Figs/boxplots/bp_DNS_5_15_reverb_0.pdf}}
				{\resizebox{\columnwidth}{!}{\begin{tikzpicture}
\pgfplotsset{every tick label/.append style={font=\huge}}
\begin{axis}[%
width=\figWbp,
height=\figHbp,
at={(0in,0in)},
scale only axis,
xmin=0.4,
xmax=5.6,
xtick={1,2,3,4,5},
xticklabels={2.47,2.68,3.28,3.22,3.43},
xticklabel style = {font=\huge, yshift = -5pt},
xlabel style={font=\color{white!15!black}},
xlabel={\mtlarge Mean},
ymin=1.2,
ymax=4.07,
ytick={1.2,1.6, ...,4.2},
ylabel style={font=\color{white!15!black}, yshift = 1pt},
ylabel={\mtlarge DNSMOS},
axis background/.style={fill=white},
legend style={legend cell align=left, align=left, draw=white!15!black}
]

%% Whiskers
\addplot [color=black, dashed, line width=\bplinewidth, forget plot]
  table[row sep=crcr]{%
1	2.74421457471752\\
1	3.5193695706498\\
};
\addplot [color=black, dashed, line width=\bplinewidth, forget plot]
  table[row sep=crcr]{%
2	3.03571722699906\\
2	3.70500429381691\\
};
\addplot [color=black, dashed, line width=\bplinewidth, forget plot]
  table[row sep=crcr]{%
3	3.54424121594116\\
3	3.94291014335925\\
};
\addplot [color=black, dashed, line width=\bplinewidth, forget plot]
  table[row sep=crcr]{%
4	3.5365843461707\\
4	3.93575953398291\\
};
\addplot [color=black, dashed, line width=\bplinewidth, forget plot]
  table[row sep=crcr]{%
5	3.6973985480494\\
5	3.99253612062501\\
};
\addplot [color=black, dashed, line width=\bplinewidth, forget plot]
  table[row sep=crcr]{%
1	1.44860704956782\\
1	2.22352023185627\\
};
\addplot [color=black, dashed, line width=\bplinewidth, forget plot]
  table[row sep=crcr]{%
2	1.29593785720672\\
2	2.33944796126464\\
};
\addplot [color=black, dashed, line width=\bplinewidth, forget plot]
  table[row sep=crcr]{%
3	2.37226043257323\\
3	3.07519464481097\\
};
\addplot [color=black, dashed, line width=\bplinewidth, forget plot]
  table[row sep=crcr]{%
4	2.14911783000979\\
4	2.98014518815949\\
};
\addplot [color=black, dashed, line width=\bplinewidth, forget plot]
  table[row sep=crcr]{%
5	2.55109150087724\\
5	3.23868978255966\\
};

%% Whiskers edges
\addplot [color=black, line width=\bplinewidth, forget plot]
  table[row sep=crcr]{%
0.85	3.5193695706498\\
1.15	3.5193695706498\\
};
\addplot [color=black, line width=\bplinewidth, forget plot]
  table[row sep=crcr]{%
1.85	3.70500429381691\\
2.15	3.70500429381691\\
};
\addplot [color=black, line width=\bplinewidth, forget plot]
  table[row sep=crcr]{%
2.85	3.94291014335925\\
3.15	3.94291014335925\\
};
\addplot [color=black, line width=\bplinewidth, forget plot]
  table[row sep=crcr]{%
3.85	3.93575953398291\\
4.15	3.93575953398291\\
};
\addplot [color=black, line width=\bplinewidth, forget plot]
  table[row sep=crcr]{%
4.85	3.99253612062501\\
5.15	3.99253612062501\\
};
\addplot [color=black, line width=\bplinewidth, forget plot]
  table[row sep=crcr]{%
0.85	1.44860704956782\\
1.15	1.44860704956782\\
};
\addplot [color=black, line width=\bplinewidth, forget plot]
  table[row sep=crcr]{%
1.85	1.29593785720672\\
2.15	1.29593785720672\\
};
\addplot [color=black, line width=\bplinewidth, forget plot]
  table[row sep=crcr]{%
2.85	2.37226043257323\\
3.15	2.37226043257323\\
};
\addplot [color=black, line width=\bplinewidth, forget plot]
  table[row sep=crcr]{%
3.85	2.14911783000979\\
4.15	2.14911783000979\\
};
\addplot [color=black, line width=\bplinewidth, forget plot]
  table[row sep=crcr]{%
4.85	2.55109150087724\\
5.15	2.55109150087724\\
};

%% Boxes
\addplot [ultra thick, color=black, fill = metricgan, forget plot]
  table[row sep=crcr]{%
0.7	2.22352023185627\\
0.7	2.74421457471752\\
1.3	2.74421457471752\\
1.3	2.22352023185627\\
0.6835	2.22352023185627\\
};
\addplot [ultra thick, color=black, fill = demucs, forget plot]
  table[row sep=crcr]{%
1.7	2.33944796126464\\
1.7	3.03571722699906\\
2.3	3.03571722699906\\
2.3	2.33944796126464\\
1.6835	2.33944796126464\\
};
\addplot [ultra thick, color=black, fill = phasen, forget plot]
table[row sep=crcr]{%
2.7	3.07519464481097\\
2.7	3.54424121594116\\
3.3	3.54424121594116\\
3.3	3.07519464481097\\
2.6835	3.07519464481097\\
};
\addplot [ultra thick, color=black, fill = pfpl, forget plot]
  table[row sep=crcr]{%
3.7	2.98014518815949\\
3.7	3.5365843461707\\
4.3	3.5365843461707\\
4.3	2.98014518815949\\
3.6835	2.98014518815949\\
};
\addplot [ultra thick, color=black, fill = cmgan, forget plot]
  table[row sep=crcr]{%
4.7	3.23868978255966\\
4.7	3.6973985480494\\
5.3	3.6973985480494\\
5.3	3.23868978255966\\
4.6835	3.23868978255966\\
};

%% Median
\addplot [color=black, line width=\bplinewidth, forget plot]
  table[row sep=crcr]{%
0.7	2.48837675512485\\
1.3	2.48837675512485\\
};
\addplot [color=black, line width=\bplinewidth, forget plot]
  table[row sep=crcr]{%
1.7	2.69332082986478\\
2.3	2.69332082986478\\
};
\addplot [color=black, line width=\bplinewidth, forget plot]
  table[row sep=crcr]{%
2.7	3.35367213227353\\
3.3	3.35367213227353\\
};
\addplot [color=black, line width=\bplinewidth, forget plot]
  table[row sep=crcr]{%
3.7	3.29278632381991\\
4.3	3.29278632381991\\
};
\addplot [color=black, line width=\bplinewidth, forget plot]
  table[row sep=crcr]{%
4.7	3.51802889533401\\
5.3	3.51802889533401\\
};

%% Outliers
\addplot [color=black, draw=none, mark=*, mark size=\msize-0.5pt, mark options={solid, fill=metricgan}, forget plot]
  table[row sep=crcr]{%
1	1.09725900263428\\
1	1.12554346539229\\
1	1.13828427707992\\
1	1.16651884102578\\
1	1.17077843623389\\
1	1.17169439232826\\
1	1.17284342000218\\
1	1.17654284109695\\
1	1.1952898930176\\
1	1.23553815149775\\
1	1.23939774962602\\
1	1.24069193433361\\
1	1.24218859578198\\
1	1.24635613886532\\
1	1.25497368874917\\
1	1.25921267151623\\
1	1.27179734169167\\
1	1.2719950331106\\
1	1.27520352485662\\
1	1.28689093005395\\
1	1.29064011794574\\
1	1.29142616576826\\
1	1.30239567049304\\
1	1.3147959288635\\
1	1.3200850322499\\
1	1.32033056413058\\
1	1.33504704707821\\
1	1.34578373069688\\
1	1.35376872620436\\
1	1.35457506529067\\
1	1.35906565562839\\
1	1.36074232976474\\
1	1.36593636476242\\
1	1.37059742488391\\
1	1.38005455037386\\
1	1.38894734573434\\
1	1.41060549600283\\
1	1.41083926891472\\
1	1.41100658415255\\
1	1.41780600944164\\
1	1.4190095966692\\
1	1.42802797528749\\
1	1.4297703304683\\
1	3.62430607741921\\
};
\addplot [color=black, draw=none, mark=*, mark size=\msize-0.5pt, mark options={solid, fill=demucs}, forget plot]
  table[row sep=crcr]{%
2	1.23957768916209\\
2	1.26371279856937\\
2	1.28241360415812\\
};
\addplot [color=black, draw=none, mark=*, mark size=\msize-0.5pt, mark options={solid, fill=phasen}, forget plot]
  table[row sep=crcr]{%
3	1.36645099677116\\
3	1.87947893368985\\
3	1.97572200385233\\
3	2.00045861472473\\
3	2.00570565963283\\
3	2.01203921359846\\
3	2.06210612012588\\
3	2.0959036556907\\
3	2.12099925035681\\
3	2.14349459372202\\
3	2.14579269296774\\
3	2.15113831549951\\
3	2.15848810194993\\
3	2.16853760921906\\
3	2.20249567220887\\
3	2.22606631251256\\
3	2.23137764873839\\
3	2.23889725707936\\
3	2.2421025092466\\
3	2.24361989133772\\
3	2.24887850154439\\
3	2.27828440088421\\
3	2.28832474504246\\
3	2.29073622869949\\
3	2.29690703400573\\
3	2.305071772646\\
3	2.30995632289625\\
3	2.32127760274378\\
3	2.33327187375942\\
3	2.33569116070076\\
3	2.34229493254444\\
3	2.3498037098466\\
3	2.35122149569252\\
3	2.35137405868703\\
3	2.36395486124133\\
};
\addplot [color=black, draw=none, mark=*, mark size=\msize-0.5pt, mark options={solid, fill=pfpl}, forget plot]
  table[row sep=crcr]{%
4	0.977754658375953\\
4	1.49101526327761\\
4	1.49461676461262\\
4	1.510203003845\\
4	1.54915911742544\\
4	1.57313141003403\\
4	1.6333238603571\\
4	1.64071052455535\\
4	1.66316542694598\\
4	1.6634727622585\\
4	1.669840605877\\
4	1.68318662709411\\
4	1.70279591381938\\
4	1.70856356526935\\
4	1.77546658748564\\
4	1.7870839825987\\
4	1.78789246500205\\
4	1.82565834812556\\
4	1.84238987257354\\
4	1.89614900491168\\
4	1.91556706330543\\
4	1.91600071128717\\
4	1.98104406151074\\
4	1.9822129751768\\
4	1.9832924827723\\
4	1.9981691364085\\
4	2.00587335578414\\
4	2.00655283466473\\
4	2.00680707385312\\
4	2.02192716005463\\
4	2.02819252331846\\
4	2.03046623524406\\
4	2.03234021337208\\
4	2.04290375781089\\
4	2.04826431969723\\
4	2.06150555186295\\
4	2.06496268021484\\
4	2.0783785786533\\
4	2.0828431389613\\
4	2.08328985530501\\
4	2.08777681809126\\
4	2.09601003896895\\
4	2.11155239630046\\
4	2.11867408227099\\
4	2.12716887268351\\
4	2.13246973351779\\
4	2.13639364240612\\
};
\addplot [color=black, draw=none, mark=*, mark size=\msize-0.5pt, mark options={solid, fill=cmgan}, forget plot]
  table[row sep=crcr]{%
5	2.01410737941641\\
5	2.0799382938255\\
5	2.08654041910779\\
5	2.09736338856064\\
5	2.09868031125884\\
5	2.13552185589919\\
5	2.14010159202306\\
5	2.16090361246847\\
5	2.1683306848928\\
5	2.18184545902411\\
5	2.18629525687602\\
5	2.19518425693988\\
5	2.19943271310383\\
5	2.23857450003307\\
5	2.26052822346998\\
5	2.28309117745138\\
5	2.28735378025217\\
5	2.29192356518884\\
5	2.31253887749481\\
5	2.31329408852941\\
5	2.34925465513679\\
5	2.36481793892147\\
5	2.3662234397673\\
5	2.36831550471102\\
5	2.3687327894737\\
5	2.36902464729465\\
5	2.37163229207149\\
5	2.37498176486392\\
5	2.38872447429049\\
5	2.39096264831996\\
5	2.39445074454018\\
5	2.40178832670149\\
5	2.40585348197656\\
5	2.40591799545911\\
5	2.42091710636259\\
5	2.42327370999046\\
5	2.45764239071472\\
5	2.45803888763197\\
5	2.45959105874819\\
5	2.46572829116979\\
5	2.46670137420192\\
5	2.48511717478218\\
5	2.49351442200169\\
5	2.4993475384472\\
5	2.5034901734789\\
5	2.50449965177916\\
5	2.51587485071976\\
5	2.52099648764615\\
5	2.53150607234807\\
5	2.53391348164991\\
5	2.53940633918085\\
5	2.54036545124748\\
5	2.54341951609505\\
5	2.5462273096449\\
5	2.5471962059189\\
};

%% Mean
\addplot[only marks, mark=diamond*, mark options={}, mark size=\msize, draw=black, fill=black] table[row sep=crcr]{%
x	y\\
1	2.46896497638885\\
2	2.67703588540171\\
3	3.28107780787068\\
4	3.21741844278704\\
5	3.42926105295193\\
};
\end{axis}
\end{tikzpicture}%
}}
				\caption{DNS-Challenge+Reverb}
				\label{fig:bp_dns_reverb}
			\end{subfigure}
		}
		\caption{DNSMOS of subjective evaluation methods tested on four different 
			datasets. In the boxplots, the mean is represented by ($\mydiamond$), 
			median ($\boldsymbol{-}$) and the width of each box indicates the 
			interquartile range (25\textsuperscript{th} and 75\textsuperscript{th} 
			percentile). The whiskers show the maximum and minimum values excluding the 
			outliers (\ref{plot:metricgan}). The mean value for each method is 
			presented on the x-axis. \label{fig:bp_all}}
		\vspace{-5mm}
	\end{figure*}
	\vspace{-3mm}
	\subsection{DNN-Based MOS}
	Following the literature \cite{yu2022dbt,li2022filter}, the DNSMOS of the overall speech quality will be evaluated as an objective measure to replicate the subjective evaluation metric, as shown in Fig.~\ref{fig:bp_all}. From the boxplots, CMGAN is outperforming all methods in the four use cases. For instance, CMGAN shows an 
	average improvement of 0.15 in comparison to the most competitive approach 
	(PFPL) in the first three use cases. Moreover, the interquartile range of the 
	CMGAN is much narrower than all other methods, which indicates a low variance 
	and thus a confident prediction, especially in the DNS-Challenge 
	(Fig.~\ref{fig:bp_dns}). On the other hand, MetricGAN+ is showing the worst 
	performance in all use cases. We hypothesize that although the PESQ score is 
	relatively high (3.15), the SSNR score that we calculate is below 1 dB, 
	indicating that the metric discriminator in MetricGAN+ case, is only focusing on 
	enhancing the PESQ at the expense of other metrics. Note that the SSNR score is 
	not reported in the original paper. DEMUCS representing the time-domain 
	paradigm is showing a robust performance over Voice Bank+DEMAND and real 
	CHiME-3 use cases. However, it is not generalizing to the DNS-Challenge 
	dataset. This generalization issue is clearly mitigated in the TF-domain 
	complex denoising methods (PHASEN, PFPL and CMGAN). From 
	Fig.~\ref{fig:bp_dns_reverb}, the overall DNSMOS of DN-Challenge with 
	additional reverberation dropped by 0.5 on average in comparison to DNS 
	challenge (Fig.~\ref{fig:bp_dns}). This is expected as generalizing to unseen 
	effects such as reverberation is more challenging than unseen noise types. 
	Despite this drop, CMGAN is still showing superior performance over other 
	competitive TF-domain approaches (PHASEN and PFPL). Audio samples\footnote{\textit{\href{https://sherifabdulatif.github.io/}{https://sherifabdulatif.github.io/}}} from all subjective evaluation methods are available online for interested readers.
	
	\vspace{-3mm}
	\subsection{Listening test}
	We organize a medium-scale listening test experiment with 25 participants and 10 samples, which were carefully selected from the Voice Bank+DEMAND, DNS-Challenge, and DNS-Challenge+Reverb datasets. Each sample has an average duration of 10 seconds, containing a complete sentence. We ensure that the tracks are of different noise types and mostly challenging conditions. During the experiment, each participant listened to both the noisy track and its clean reference. Subsequently, they were asked to assign a rating between 1 and 5 (where a higher score indicates better quality) to represent the absolute overall quality of the sample for the five methods: MetricGAN+, DEMUCS, PHASEN, PFPL and CMGAN. In compliance with the ITU-T P.835 standards \cite{p835standards}, the methods are unknown to the participants and the order of the methods is randomly shuffled for each sample.
	\begin{figure}[b!]
		\vspace{-2mm}
		\metricganrect MetricGAN+ \hfill \demucsrect DEMUCS \hfill \phasenrect PHASEN
		\vspace{2mm}
		\hspace{7mm}\pfplrect PFPL \hspace{10mm} \cmganrect CMGAN 
		\vspace{1mm}
		\centering
%		{\includegraphics[width=0.98\columnwidth]{Figs/boxplots/box_plots_lt_24pts.pdf}}
		\resizebox{0.98\columnwidth}{!}{\begin{tikzpicture}
\pgfplotsset{every tick label/.append style={font=\huge}}
\begin{axis}[%
width=\figWbplt,
height=\figHbplt,
at={(0in,0in)},
scale only axis,
unbounded coords=jump,
xmin=0.4,
xmax=5.6,
xtick={1,2,3,4,5},
xticklabels={1.92,2.23,3.67,2.93,4.48},
xticklabel style = {font=\large},
xlabel style={font=\color{white!15!black}},
xlabel={\large Mean},
ymin=0.795,
ymax=5.30641021743705,
ylabel style={font=\color{white!15!black}, yshift = 1pt},
ylabel={\large Overall quality rating},
ytick={1,2,3,4,5},
yticklabel style = {font=\large},
axis background/.style={fill=white},
legend style={legend cell align=left, align=left, draw=white!15!black}
]

%% Whiskers
\addplot [color=black, dashed, line width=\bpltlinewidth, forget plot]
  table[row sep=crcr]{%
1	2\\
1	3\\
};
\addplot [color=black, dashed, line width=\bpltlinewidth, forget plot]
  table[row sep=crcr]{%
2	3\\
2	5\\
};
\addplot [color=black, dashed, line width=\bpltlinewidth, forget plot]
  table[row sep=crcr]{%
3	4\\
3	5\\
};
\addplot [color=black, dashed, line width=\bpltlinewidth, forget plot]
  table[row sep=crcr]{%
4	3.5\\
4	5\\
};
\addplot [color=black, dashed, line width=\bpltlinewidth, forget plot]
  table[row sep=crcr]{%
5	5\\
5	5\\
};
\addplot [color=black, dashed, line width=\bpltlinewidth, forget plot]
  table[row sep=crcr]{%
1	1\\
1	1\\
};
\addplot [color=black, dashed, line width=\bpltlinewidth, forget plot]
  table[row sep=crcr]{%
2	1\\
2	1\\
};
\addplot [color=black, dashed, line width=\bpltlinewidth, forget plot]
  table[row sep=crcr]{%
3	2\\
3	3\\
};
\addplot [color=black, dashed, line width=\bpltlinewidth, forget plot]
  table[row sep=crcr]{%
4	1\\
4	2\\
};
\addplot [color=black, dashed, line width=\bpltlinewidth, forget plot]
  table[row sep=crcr]{%
5	3\\
5	4\\
};

%% Whislers edges
\addplot [color=black, line width=\bpltlinewidth, forget plot]
  table[row sep=crcr]{%
0.9	3\\
1.1	3\\
};
\addplot [color=black, line width=\bpltlinewidth, forget plot]
  table[row sep=crcr]{%
1.9	5\\
2.1	5\\
};
\addplot [color=black, line width=\bpltlinewidth, forget plot]
  table[row sep=crcr]{%
2.9	5\\
3.1	5\\
};
\addplot [color=black, line width=\bpltlinewidth, forget plot]
  table[row sep=crcr]{%
3.9	5\\
4.1	5\\
};
\addplot [color=black, line width=\bpltlinewidth, forget plot]
  table[row sep=crcr]{%
4.9	5\\
5.1	5\\
};
\addplot [color=black, line width=\bpltlinewidth, forget plot]
  table[row sep=crcr]{%
0.9	1\\
1.1	1\\
};
\addplot [color=black, line width=\bpltlinewidth, forget plot]
  table[row sep=crcr]{%
1.9	1\\
2.1	1\\
};
\addplot [color=black, line width=\bpltlinewidth, forget plot]
  table[row sep=crcr]{%
2.9	2\\
3.1	2\\
};
\addplot [color=black, line width=\bpltlinewidth, forget plot]
  table[row sep=crcr]{%
3.9	1\\
4.1	1\\
};
\addplot [color=black, line width=\bpltlinewidth, forget plot]
  table[row sep=crcr]{%
4.9	3\\
5.1	3\\
};

%% Boxes
\addplot [ultra thick, color=black, fill = metricgan, forget plot]
  table[row sep=crcr]{%
0.8	1\\
0.8	2.01\\
1.2	2.01\\
1.2	1\\
0.7875	1\\
};
\addplot [ultra thick, color=black, fill = demucs, forget plot]
  table[row sep=crcr]{%
1.8	1\\
1.8	3\\
2.2	3\\
2.2	1\\
1.7875	1\\
};
\addplot [ultra thick, color=black, fill = phasen, forget plot]
  table[row sep=crcr]{%
2.8	3\\
2.8	4.01\\
3.2	4.01\\
3.2	3\\
2.7875	3\\
};
\addplot [ultra thick, color=black, fill = pfpl, forget plot]
  table[row sep=crcr]{%
3.8	2\\
3.8	3.5\\
4.2	3.5\\
4.2	2\\
3.7875	2\\
};
\addplot [ultra thick, color=black, fill = cmgan, forget plot]
  table[row sep=crcr]{%
4.8	4\\
4.8	5.005\\
5.2	5.005\\
5.2	4\\
4.7875	4\\
};

%% Median
\addplot [color=black, line width=\bpltlinewidth, forget plot]
  table[row sep=crcr]{%
0.8	2\\
1.2	2\\
};
\addplot [color=black, line width=\bpltlinewidth, forget plot]
  table[row sep=crcr]{%
1.8	2\\
2.2	2\\
};
\addplot [color=black, line width=\bpltlinewidth, forget plot]
  table[row sep=crcr]{%
2.8	4\\
3.2	4\\
};
\addplot [color=black, line width=\bpltlinewidth, forget plot]
  table[row sep=crcr]{%
3.8	3\\
4.2	3\\
};
\addplot [color=black, line width=\bpltlinewidth, forget plot]
  table[row sep=crcr]{%
4.8	5\\
5.2	5\\
};

%% Outliers
\addplot [only marks, color=black, draw=none, mark=*, mark size=\msize-1pt, mark options={solid, fill=metricgan}, forget plot]
  table[row sep=crcr]{%
1	4\\
1	4\\
1	4\\
1	4\\
1	4\\
};
\addplot [only marks, color=black, draw=none, mark=*, mark size=\msize-1pt, mark options={solid, fill=demucs}, forget plot]
  table[row sep=crcr]{%
nan	nan\\
};
\addplot [only marks, color=black, draw=none, mark=*, mark size=\msize-1pt, mark options={solid, fill=phasen}, forget plot]
  table[row sep=crcr]{%
3	1\\
3	1\\
3	1\\
};
\addplot [only marks, color=black, draw=none, mark=*, mark size=\msize-1pt, mark options={solid, fill=pfpl}, forget plot]
  table[row sep=crcr]{%
nan	nan\\
};
\addplot [only marks, color=black, draw=none, mark=*, mark size=\msize-1pt, mark options={solid, fill=cmgan}, forget plot]
  table[row sep=crcr]{%
5	2\\
5	2\\
5	2\\
5	2\\
};

%% Mean
\addplot[only marks, mark=diamond*, mark options={}, mark size=\msize-0.5pt, draw=black, fill=black] table[row sep=crcr]{%
x	y\\
1	1.92272727272727\\
2	2.225\\
3	3.67083333333333\\
4	2.93333333333333\\
5	4.47916666666667\\
};
\end{axis}
\end{tikzpicture}%
}
		\vspace{-2mm}
		\caption{Overall speech quality ratings of listening test.}
		\label{fig:bp_lt}
	\end{figure}
	
	As illustrated in Fig.~\ref{fig:bp_lt}, the participants rated the CMGAN as the best method in the subjective overall quality. On average, CMGAN outperformed the second-best method (PHASEN) by 0.81. The findings obtained from the non-intrusive DNSMOS in Fig.~\ref{fig:bp_all} generally align with the listening test experiment, with the exception that PHASEN exhibited better performance than PFPL. This discrepancy could be attributed to the limited tracks included in the listening test experiment or the relatively non-optimal performance of the DNSMOS ratings. The performance of MetricGAN+ and time-domain DEMUCS is relatively poor, indicating that TF-domain complex denoising methods (PHASEN, PFPL and CMGAN) continue to demonstrate superior generalization capabilities and subjective quality ratings.
	
	\vspace{-5mm}
	\section{Limitations}
	Despite the above results, this study is not without limitations. For instance, CMGAN has not been tested for real-time SE, i.e., CMGAN can access the whole track. In the future, CMGAN should be modified to only access few TF bins from the old samples and not the entire track, together with an extensive study on the exact amount of floating point operations in the real-time scenario. Another point to consider is the focus of this investigation on individual task experimentation. The superimposed effect (denoising and dereverberation) is only briefly addressed in the subjective evaluation part, so training and evaluating CMGAN for this use case would be an important extension of our work. Lastly, the current evaluation does not encompass ASR performance, a dimension that would mark a promising avenue for subsequent research.
	
	\section{Conclusion}
	This paper introduces CMGAN as a unified framework operating on both magnitude and complex spectrogram components for various SE tasks, including denoising, dereverberation and super-resolution. Our approach combines recent conformers that can capture long-term dependencies as well as local features in both time and frequency dimensions, together with a metric discriminator that resolves metric mismatch by directly enhancing non-differentiable evaluation scores. Experimental results demonstrate that the proposed method achieves superior or competitive performance against SOTA methods in each task with relatively few parameters (1.83~M). Additionally, we conduct an ablation study to verify the fragmented benefits of each utilized component and loss in the proposed CMGAN framework. Finally, subjective evaluation illustrates that CMGAN outperforms other methods with a robust generalization to unseen noise types and distortions.
	\begin{table}[t!]
		\centering
		\large
		\caption{CMGAN architectural details.} \vspace{-1.5mm}
		\resizebox{\columnwidth}{!}{\begin{tabular}{l|llll}
				\toprule 
				& Layer  & Input size & Hyperparameters  & Output size  \\
				\midrule
				\multirow{6}*{\rotatebox[origin=c]{90}{Encoder}} & 2D-Conv. & $B\!\times\!T\!\times\!201\!\times\!3$ & $1\!\times\!1, (1,1), 64$ & $B\!\times\!T\!\times\!201\!\times\!64$ \\
				~ & Dil. Dense-1  &  $B\!\times\!T\!\times\!201\!\times\!64$ & $2\!\times\!3, (1,1), 64, 1$ & $B\!\times\!T\!\times\!201\!\times\!64$ \\
				~ & Dil. Dense-2 &  $B\!\times\!T\!\times\!201\!\times\!64$ & $2\!\times\!3, (1,1), 64, 2$ & $B\!\times\!T\!\times\!201\!\times\!64$ \\
				~ & Dil. Dense-3  &  $B\!\times\!T\!\times\!201\!\times\!64$ & $2\!\times\!3, (1,1), 64, 4$ & 
				$B\!\times\!T\!\times\!201\!\times\!64$ \\
				~ & Dil. Dense-4  &  $B\!\times\!T\!\times\!201\!\times\!64$ & $2\!\times\!3, (1,1), 64, 8$ & $B\!\times\!T\!\times\!201\!\times\!64$ \\
				~ & 2D-Conv. & $B\!\times\!T\!\times\!201\!\times\!64$ & $1\!\times\!3, (1,2), 64$ & 
				$B\!\times\!T\!\times\!101\!\times\!64$ \\
				\midrule
				\multirow{5}*{ \hspace{-1mm}\rotatebox[origin=c]{90}{TS-Conf. \hspace{-3mm} $\times 4$}} & Reshape & $B\!\times\!T\!\times\!101\!\times\!64$ & -- & $101B\!\times\!T\!\times\!64$ \\
				~ & Time-Conf. & $101B\!\times\!T\!\times\!64$ & -- & $101B\!\times\!T\!\times\!64$ \\
				~ & Reshape & $101B\!\times\!T\!\times\!64 $ & -- & 
				$BT\!\times\!101\!\times\!64$ \\
				~ & Freq.-Conf. & $BT\!\times\!101\!\times\!64 $ & -- & $BT\!\times\!101\!\times\!64$ \\
				~ & Reshape & $BT\!\times\!101\!\times\!64$ & -- & $B\!\times\!T\!\times\!101\!\times\!64$ \\
				\midrule
				\multirow{7}*{\rotatebox[origin=c]{90}{Mask Dec.}} & Dil. Dense-1  &  $B\!\times\!T\!\times\!101\!\times\!64$ & $2\!\times\!3, (1,1), 64, 1$ & $B\!\times\!T\!\times\!101\!\times\!64$ \\
				~ & Dil. Dense-2 &  $B\!\times\!T\!\times\!101\!\times\!64$ & $2\!\times\!3, (1,1), 64, 2$ & $B\!\times\!T\!\times\!101\!\times\!64$ \\
				~ & Dil. Dense-3  &  $B\!\times\!T\!\times\!101\!\times\!64$ & $2\!\times\!3, (1,1), 64, 4$ & $B\!\times\!T\!\times\!101\!\times\!64$ \\
				~ & Dil. Dense-4  &  $B\!\times\!T\!\times\!101\!\times\!64$ & $2\!\times\!3, (1,1), 64, 8$ & $B\!\times\!T\!\times\!101\!\times\!64$ \\
				~ & Sub-pixel & $B\!\times\!T\!\times\!101\!\times\!64$ & 
				$1\!\times\!3, (1,1), 256$ & $B\!\times\!T\!\times\!202\!\times\!256$ \\
				~ & 2D-Conv. & $B\!\times\!T\!\times\!202\!\times\!256$ & 
				$1\!\times\!2,\!(1,1),\!1$ & $B\!\times\!T\!\times\!201\!\times\!1$ \\
				%		~ & 2D-Conv. & $B\!\times\!T\!\times\!201\!\times\!64$ & $1\!\times\!1, (1,1), 1$ & $B\!\times\!T\!\times\!201\!\times\!1$ \\
				~ & PReLU & -- & $201$ & -- \\
				\midrule
				\multirow{6}*{\rotatebox[origin=c]{90}{Complex Dec.}} & Dil. Dense-1  &  $B\!\times\!T\!\times\!101\!\times\!64$ & $2\!\times\!3, (1,1), 64, 1$ & $B\!\times\!T\!\times\!101\!\times\!64$ \\
				~ & Dil. Dense-2 &  $B\!\times\!T\!\times\!101\!\times\!64$ & $2\!\times\!3, (1,1), 64, 2$ & $B\!\times\!T\!\times\!101\!\times\!64$ \\
				~ & Dil. Dense-3  &  $B\!\times\!T\!\times\!101\!\times\!64$ & $2\!\times\!3, (1,1), 64, 4$ & $B\!\times\!T\!\times\!101\!\times\!64$ \\
				~ & Dil. Dense-4  &  $B\!\times\!T\!\times\!101\!\times\!64$ & $2\!\times\!3, (1,1), 64, 8$ & $B\!\times\!T\!\times\!101\!\times\!64$ \\
				~ & Sub-pixel & $B\!\times\!T\!\times\!101\!\times\!64$ & 
				$1\!\times\!3, (1,1), 256$ & $B\!\times\!T\!\times\!202\!\times\!256$ \\
				~ & 2D-Conv. & $B\!\times\!T\!\times\!202\!\times\!256$ & 
				$1\!\times\!2, (1,1), 2$ & $B\!\times\!T\!\times\!201\!\times\!2$ \\
				\bottomrule 
				\toprule
				\multirow{8}*{\rotatebox[origin=c]{90}{Metric Disc.}} & 2D-Conv.-1  &  $B\!\times\!T\!\times\!201\!\times\!2$ & $4\!\times\!4, (2,2), 16$ & $B\!\times\!T/2\!\times\!100\!\times\!16$ \\
				~ & 2D-Conv.-2  &  $B\!\times\!T/2\!\times\!100\!\times\!16$ & $4\!\times\!4, (2,2), 32$ & $B\!\times\!T/4\!\times\!50\!\times\!32$ \\
				~ & 2D-Conv.-3  &  $B\!\times\!T/4\!\times\!50\!\times\!32$ & $4\!\times\!4, (2,2), 64$ &
				$B\!\times\!T/8\!\times\!25\!\times\!64$ \\
				~ & 2D-Conv.-4  &  $B\!\times\!T/8\!\times\!25\!\times\!64$ & $4\!\times\!4, (2,2), 128$ &
				$B\!\times\!T/16\!\times\!12\!\times\!128$ \\
				~ & Avg. Pooling  &  $B\!\times\!T/16\!\times\!12\!\times\!128$ & -- &
				$B\!\times\!128$ \\
				~ & Linear-1  &  $B\!\times\!128$ & $64$ &
				$B\!\times\!64$ \\
				~ & Linear-2  &  $B\!\times\!64$ & $1$ &
				$B\!\times\!1$ \\
				~ & Sigmoid  &  -- & -- &
				-- \\
				\bottomrule 
		\end{tabular}}
		\label{tab:parameters}
		\vspace{-5mm}
	\end{table}
	\vspace{-4mm}
	\section*{Acknowledgments}
	We would like to thank the Institute of Natural Language Processing, University of Stuttgart for providing useful datasets to support this research.
	\vspace{-1.5mm}
	\section*{Appendix} \label{sec:appendix}
	The complete architectural details of both generator and discriminator are outlined in Table~\ref{tab:parameters}. The hyperparameters for 2D-Conv. layers represent kernel sizes, strides, and number of channels. For dilated dense blocks, the dilation factor is appended at the end. In linear layers only the number of channels is presented. For the conformer blocks, we follow the same baseline in \cite{gulati2020conformer}. 
	
\begin{figure*}[t!]
	\captionsetup[subfigure]{justification=centering}
	\centering
	\centerline{
		\begin{subfigure}[b]{.23\textwidth}
			\centering			
			\includegraphics[width=\columnwidth]{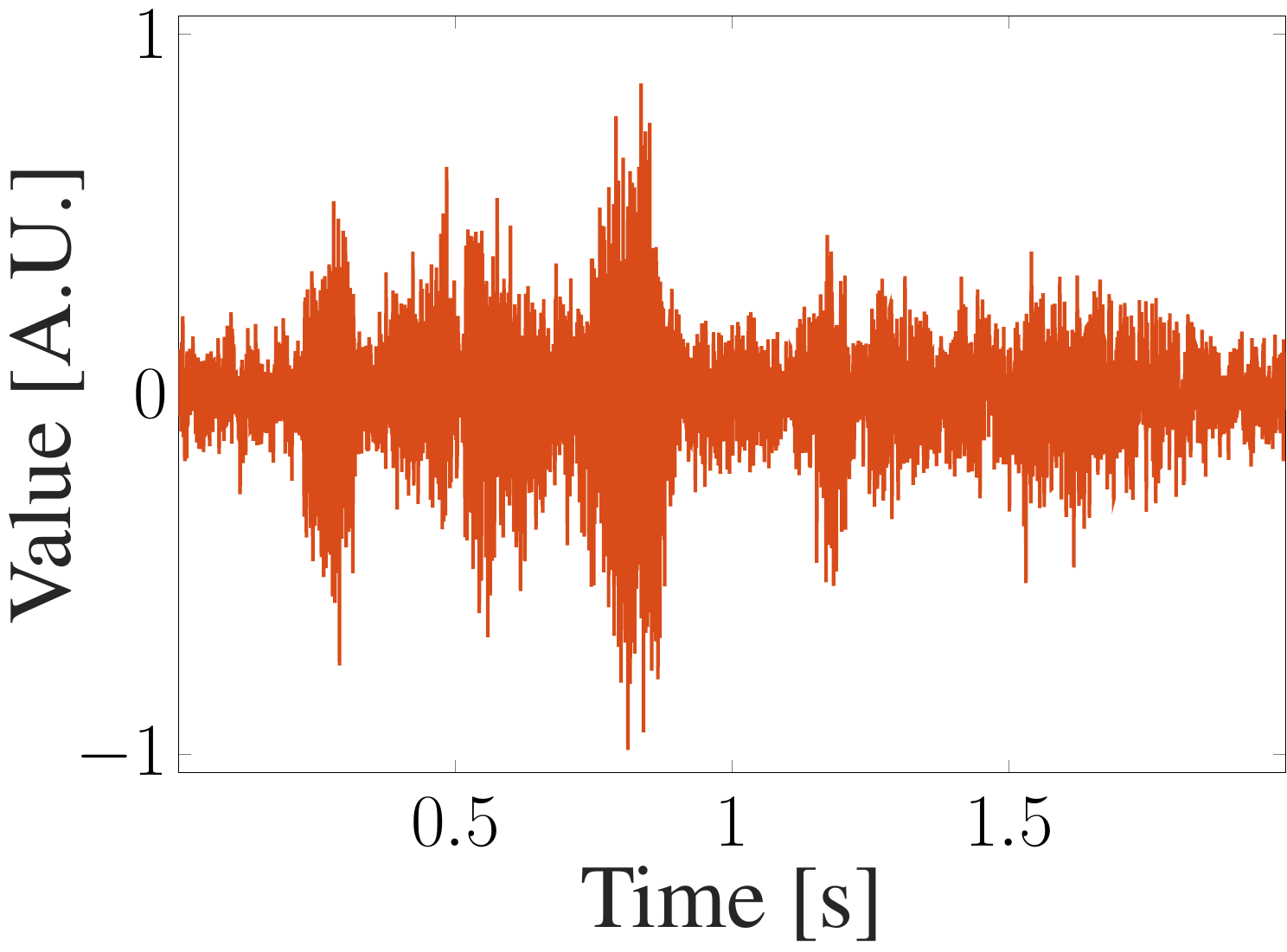}
			\caption{Noisy}
			\label{fig:wb_noisy_time}
		\end{subfigure}
		\hspace{3.7mm}
		\begin{subfigure}[b]{.23\textwidth}
			\centering
			\includegraphics[width=\columnwidth]{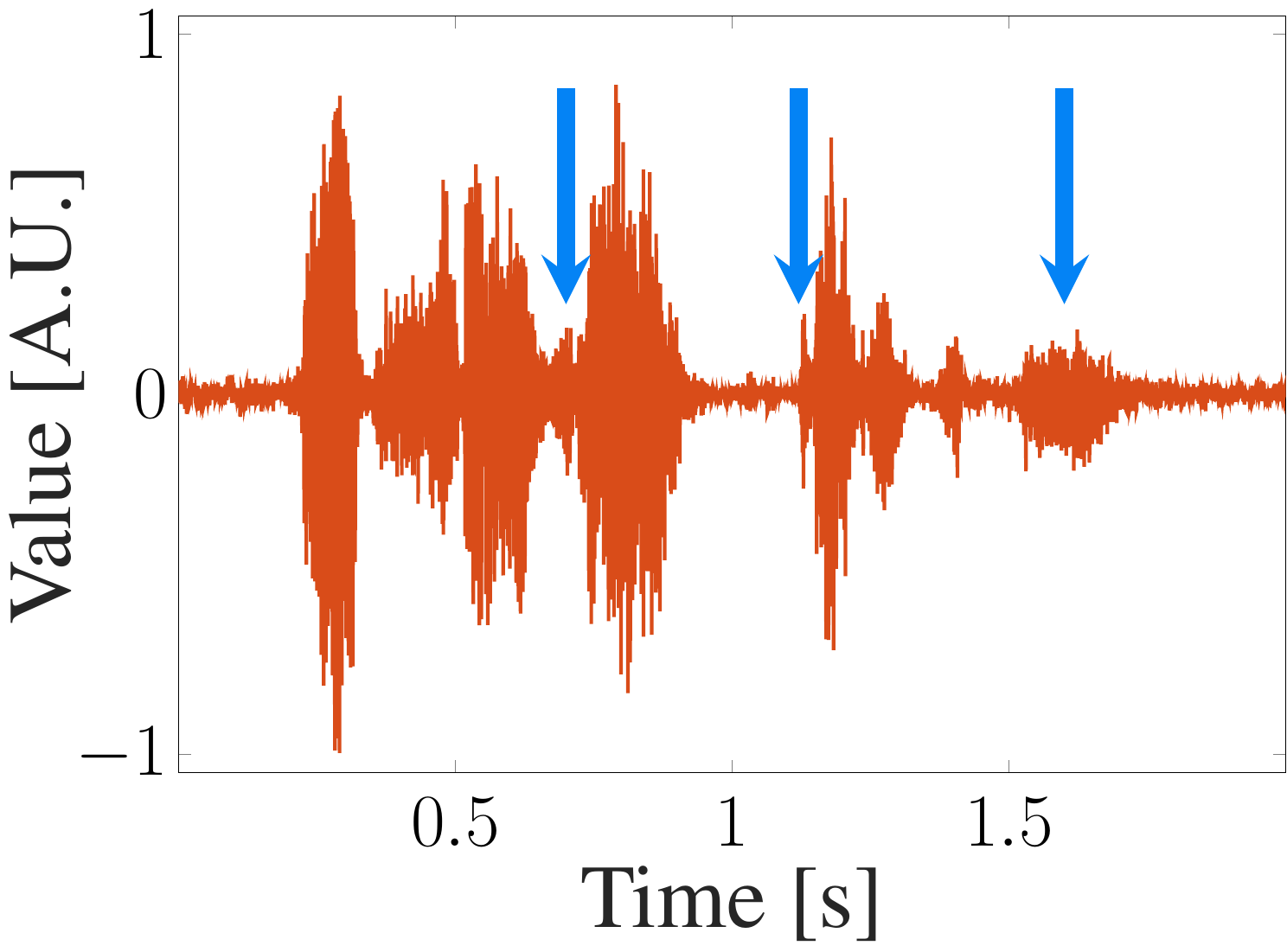}
			\caption{MetricGAN+}
			\label{fig:wb_metricgan_time}
		\end{subfigure}
		\hspace{2.7mm}
		\begin{subfigure}[b]{.23\textwidth}
			\centering		
			\includegraphics[width=\columnwidth]{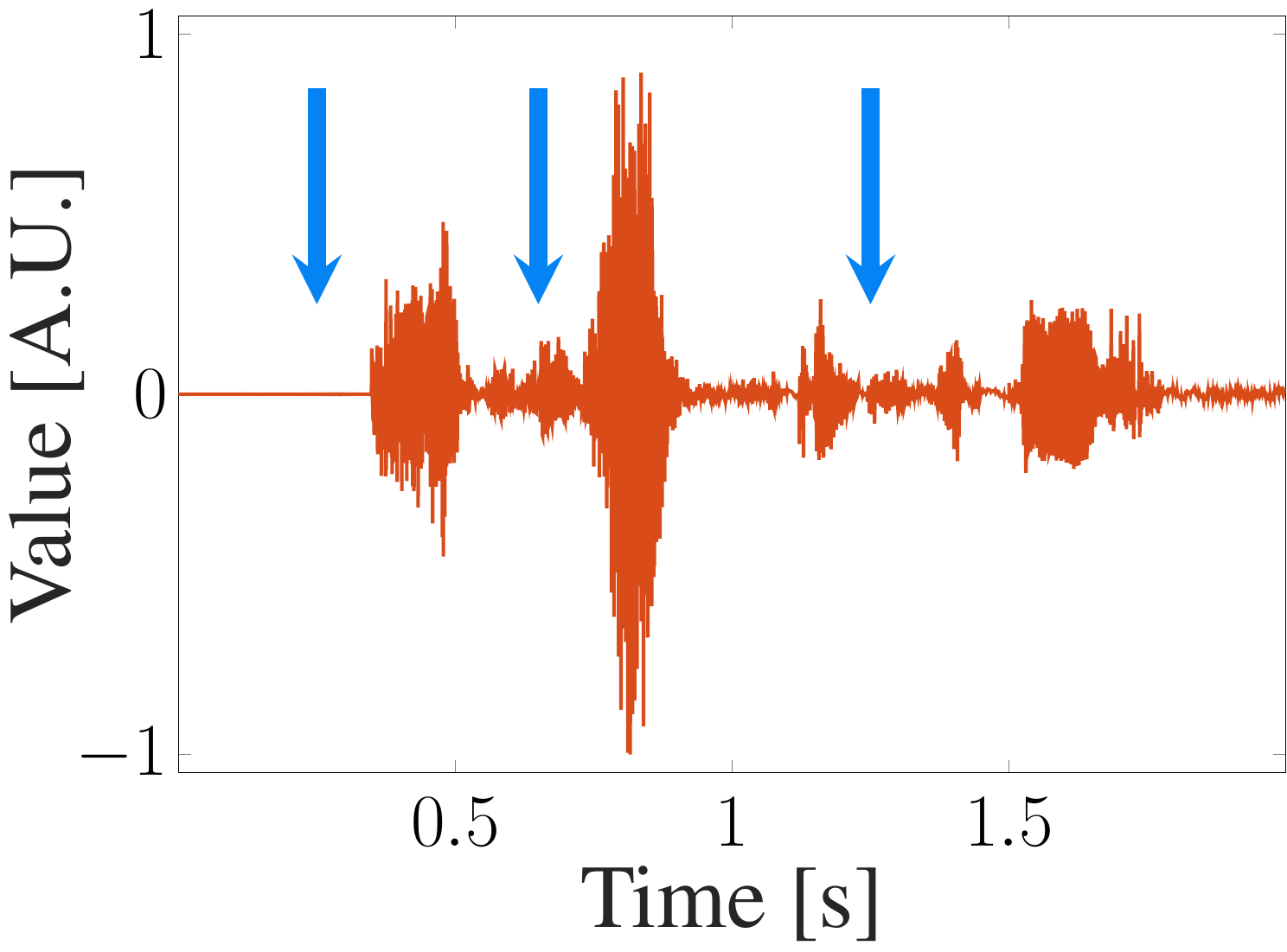}
			\caption{DEMUCS}
			\label{fig:wb_demucs_time}
		\end{subfigure}
		\hspace{2.7mm}
		\begin{subfigure}[b]{.23\textwidth}
			\centering
			\includegraphics[width=\columnwidth]{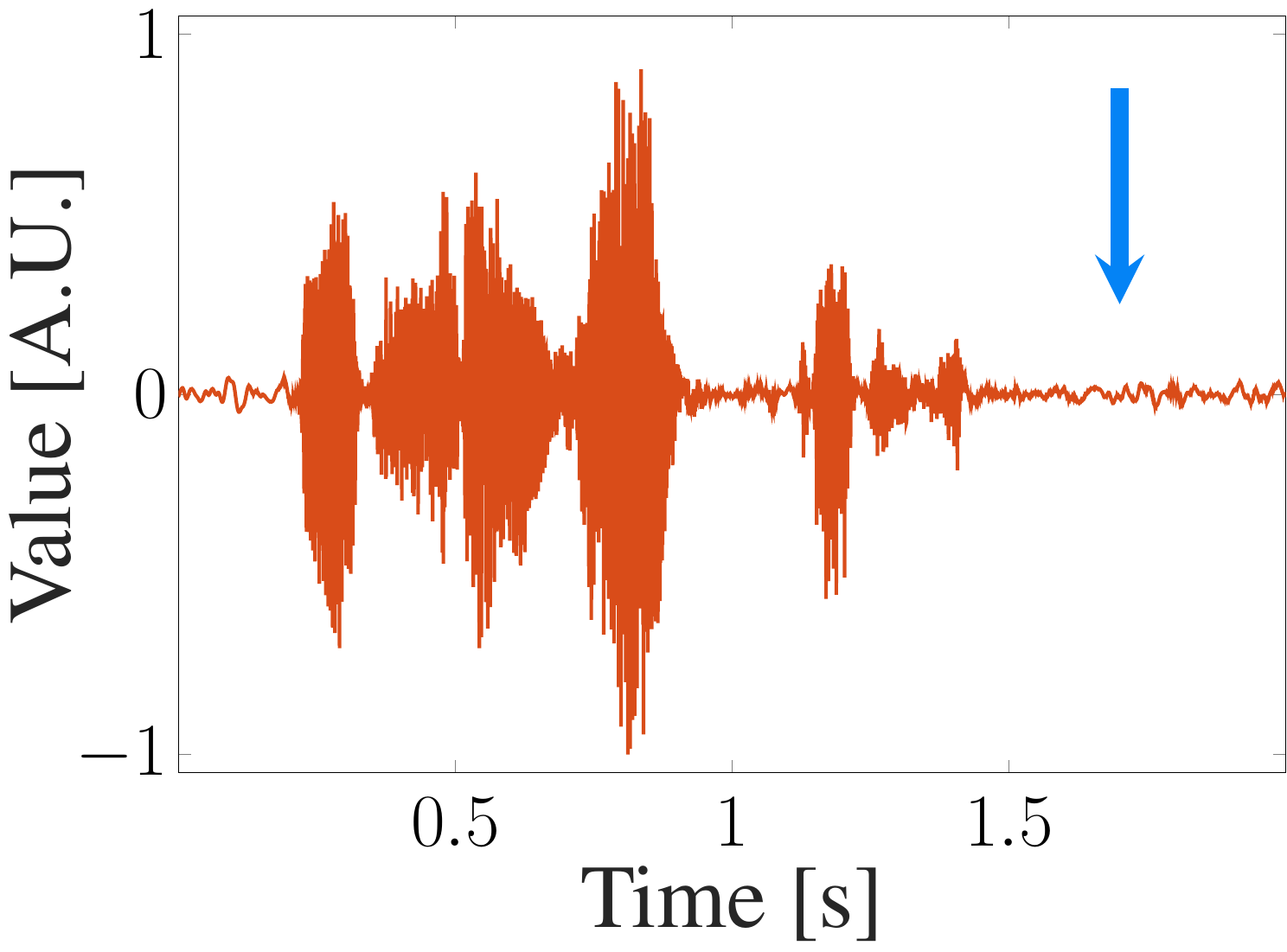}
			\caption{PHASEN}
			\label{fig:wb_phasen_time}
		\end{subfigure}
	}
	\vspace{1mm}
	\centering
	\centerline{
		\begin{subfigure}[b]{.23\textwidth}
			\centering
			\includegraphics[width=\columnwidth]{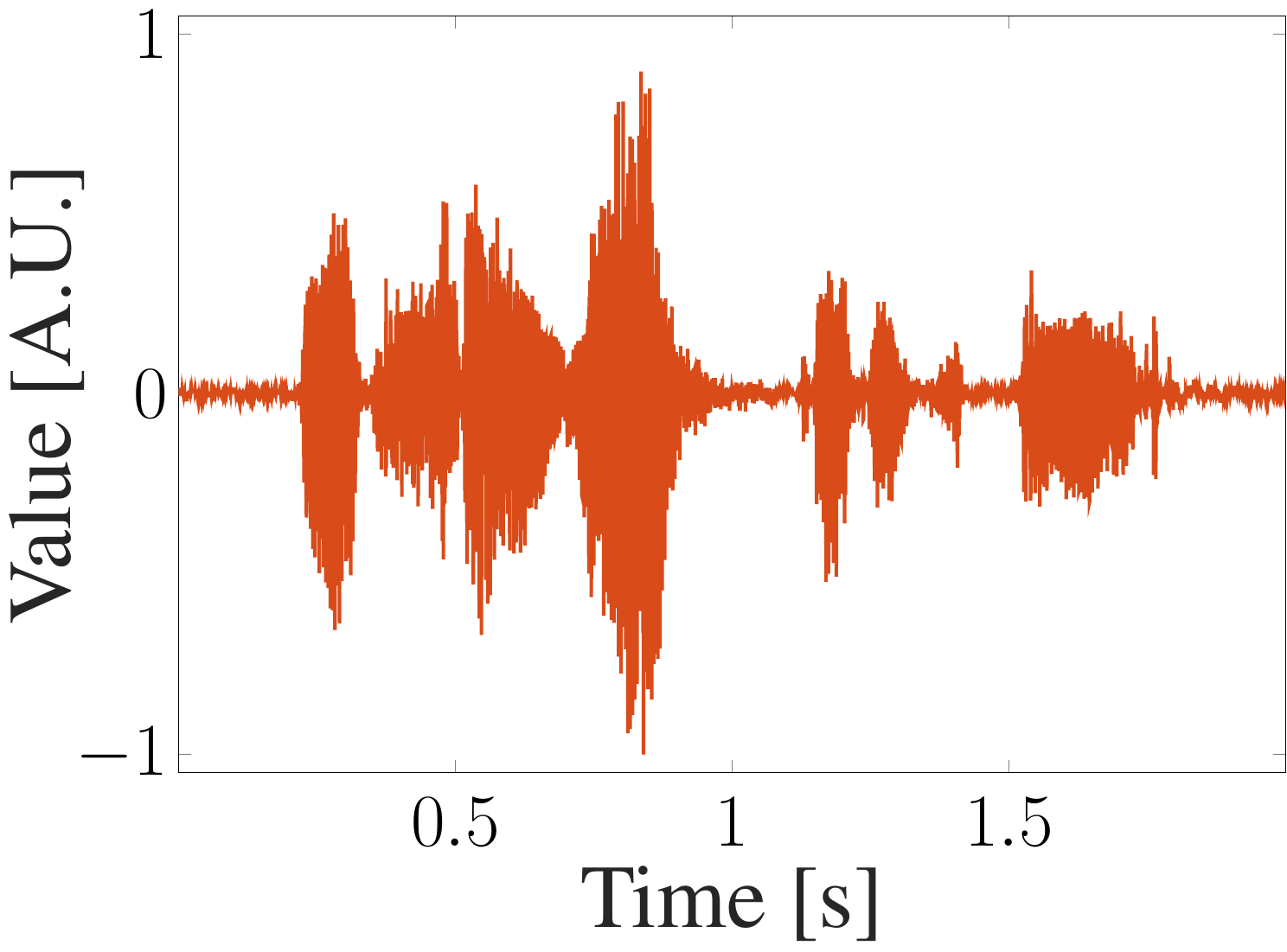}
			\caption{PFPL}
			\label{fig:wb_pfpl_time}
		\end{subfigure}
		\hspace{12mm}
		\begin{subfigure}[b]{.23\textwidth}
			\centering
			\includegraphics[width=\columnwidth]{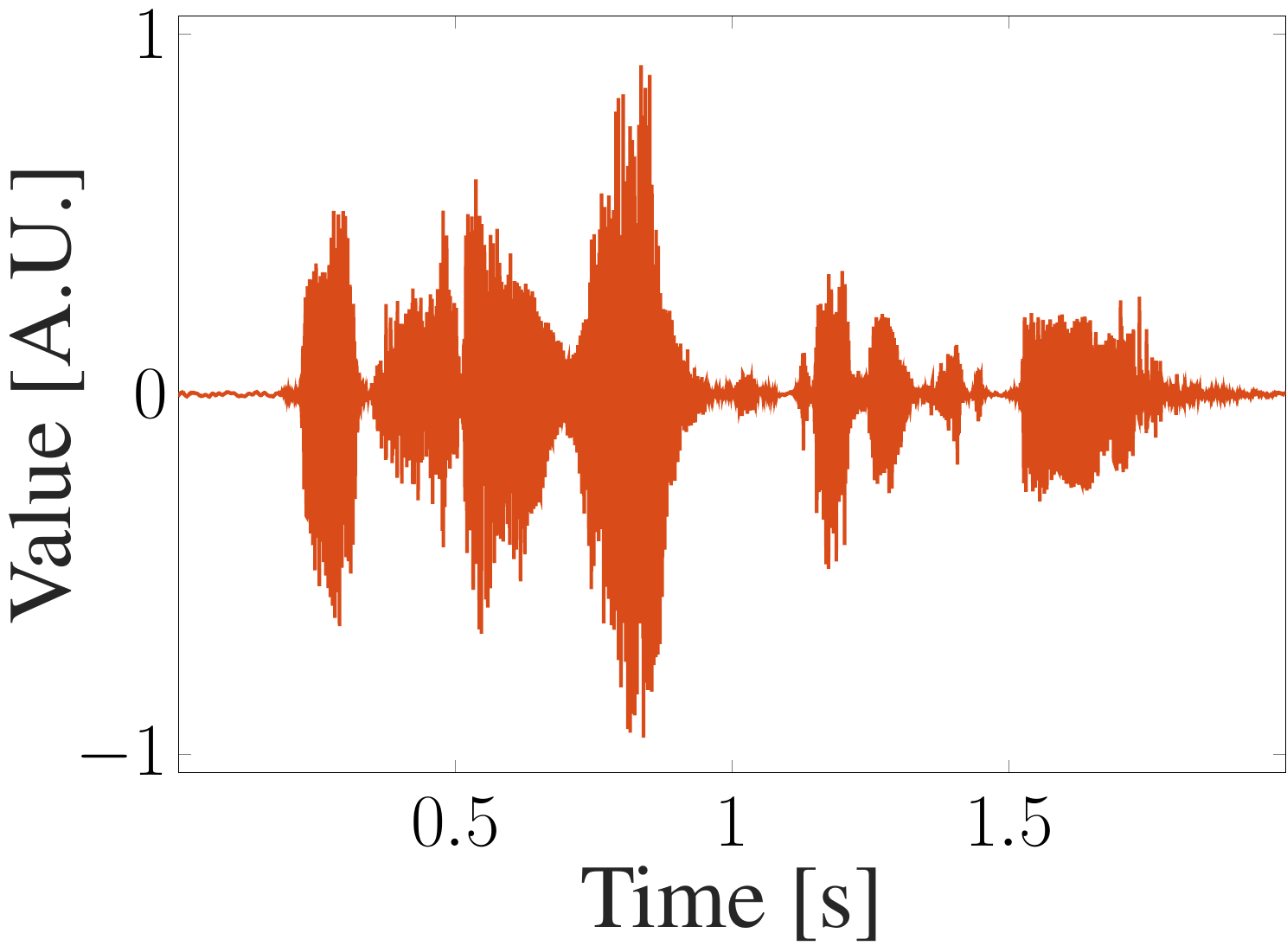}
			\caption{CMGAN}
			\label{fig:wb_cmgan_time}
		\end{subfigure}
		\hspace{10mm}
		\begin{subfigure}[b]{.23\textwidth}
			\centering
			\includegraphics[width=\columnwidth]{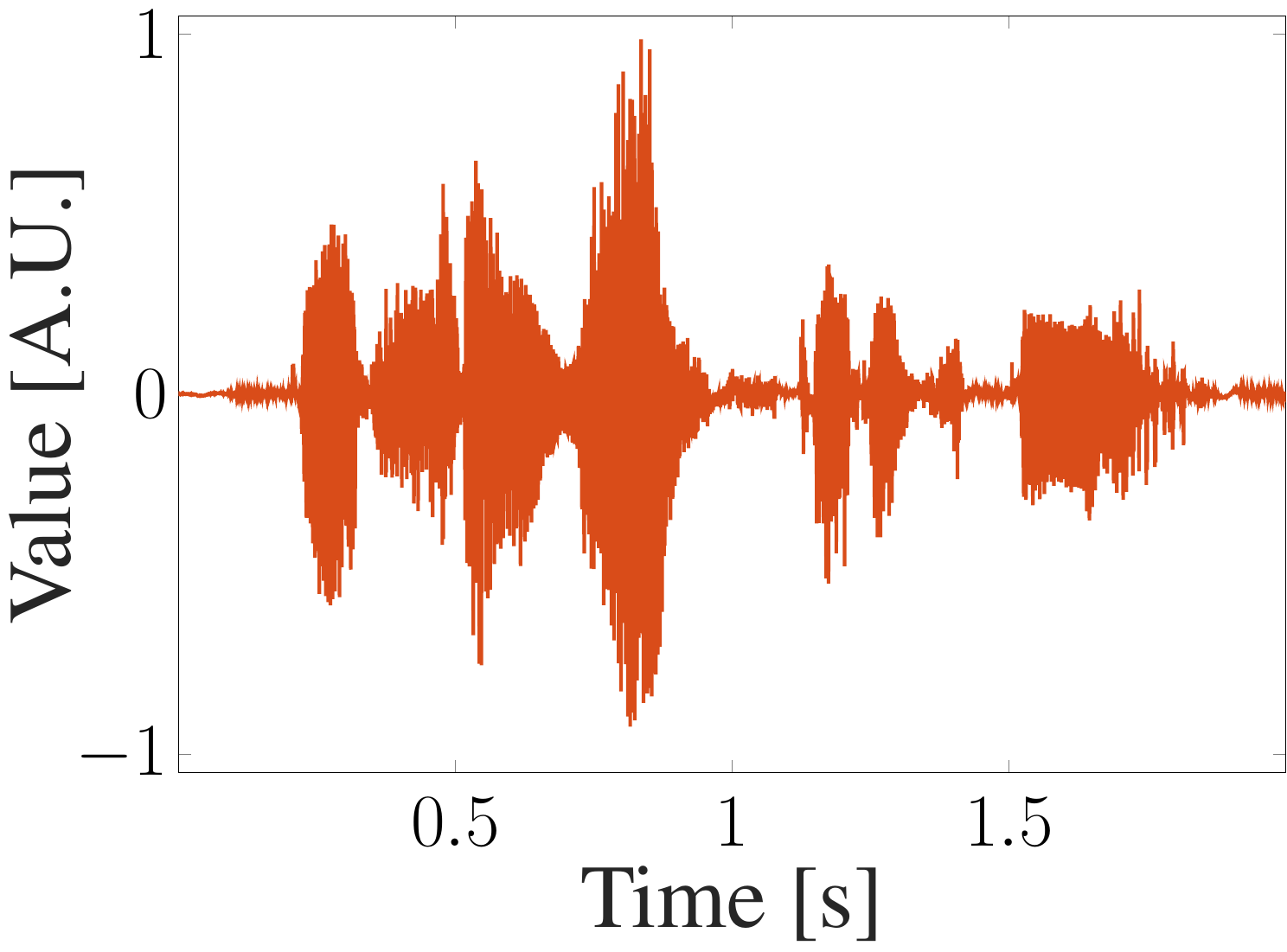}
			\caption{Clean}
			\label{fig:wb_clean_time}
		\end{subfigure}
	}
	\vspace{1mm}
	\centering
	\centerline{
		\begin{subfigure}[b]{.23\textwidth}
			\centering			
			\includegraphics[width=\columnwidth]{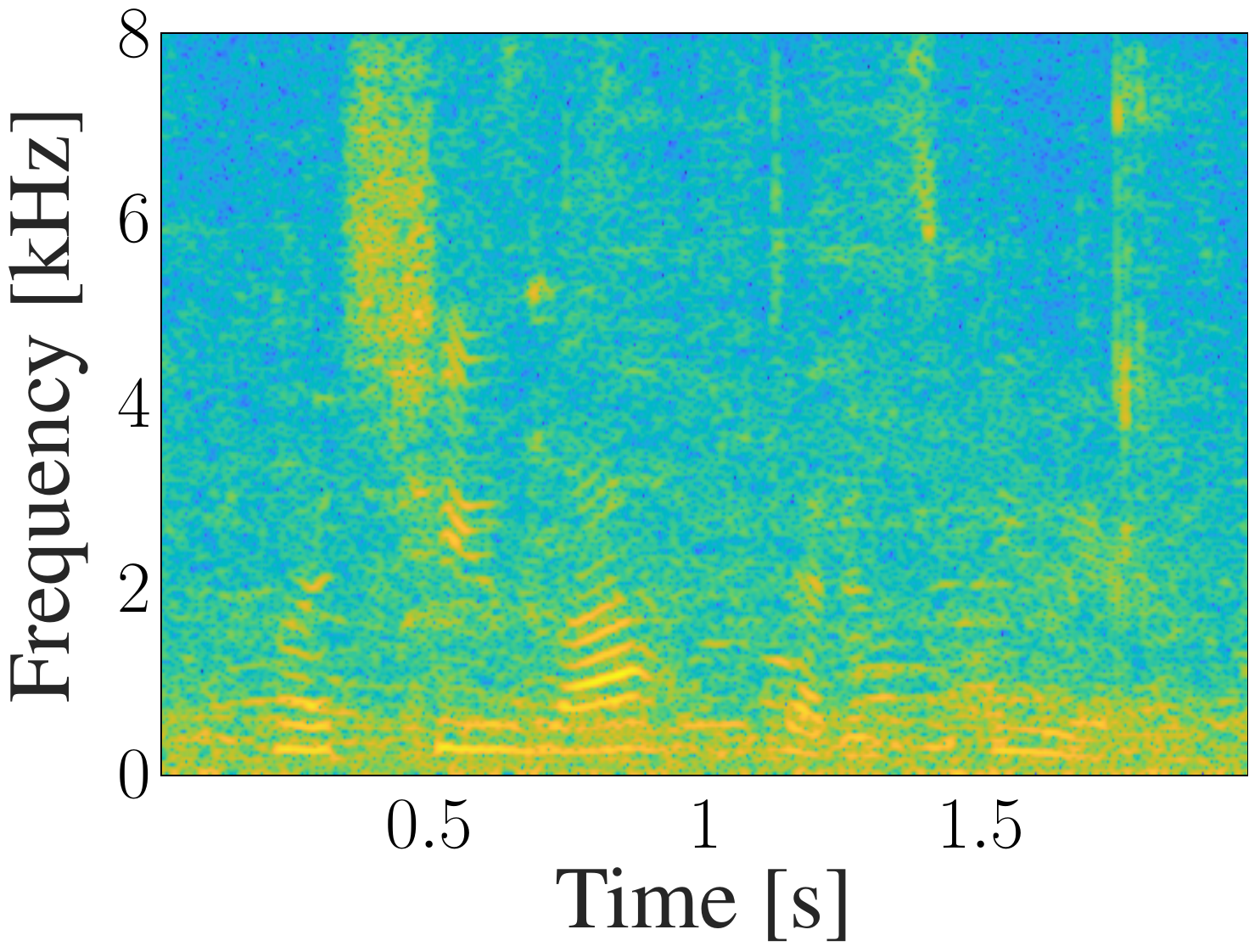}
			\caption{Noisy}
			\label{fig:wb_noisy_mag}
		\end{subfigure}
		\hspace{3.7mm}
		\begin{subfigure}[b]{.23\textwidth}
			\centering
			\includegraphics[width=\columnwidth]{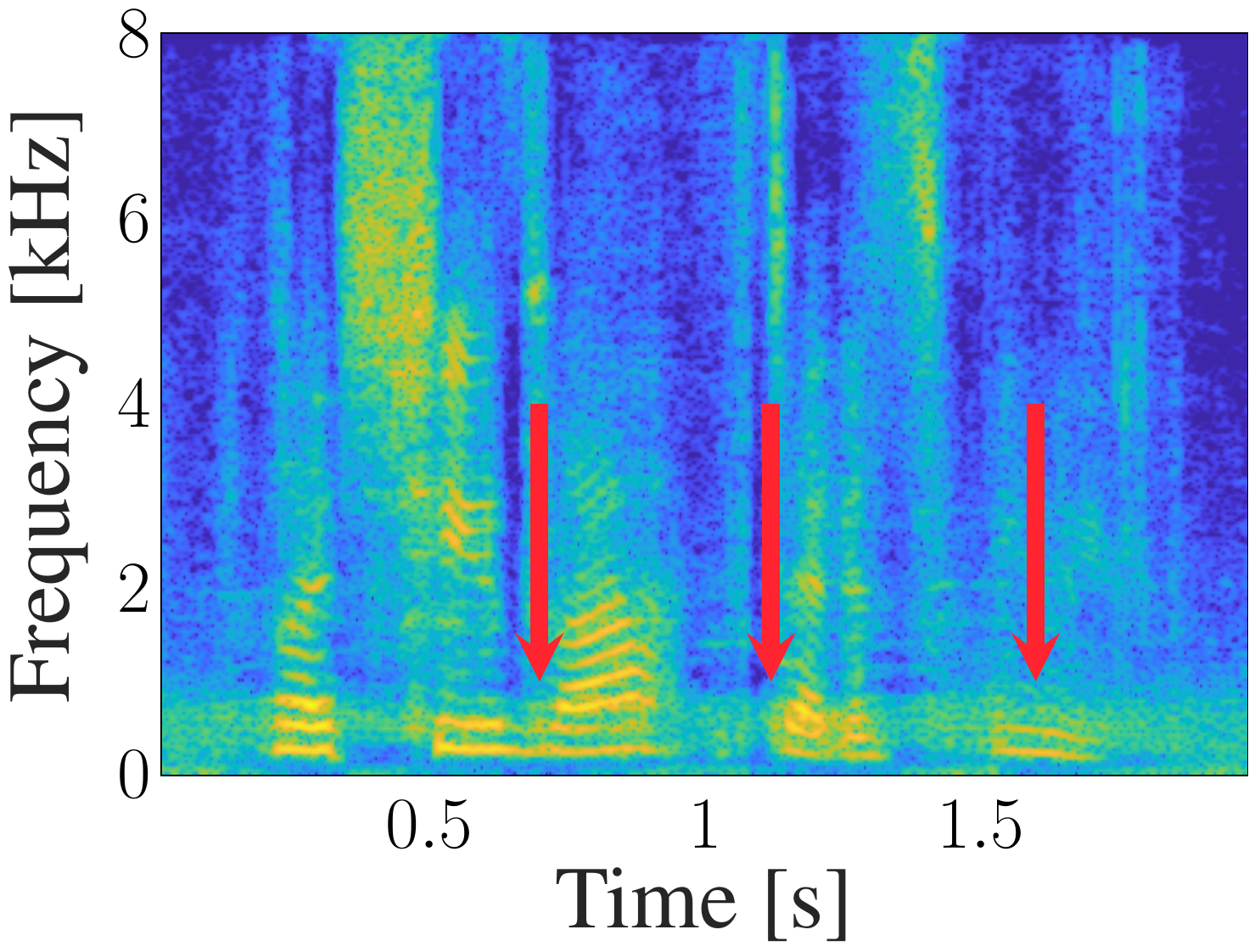}
			\caption{MetricGAN+}
			\label{fig:wb_metricgan_mag}
		\end{subfigure}
		\hspace{2.7mm}
		\begin{subfigure}[b]{.23\textwidth}
			\centering		
			\includegraphics[width=\columnwidth]{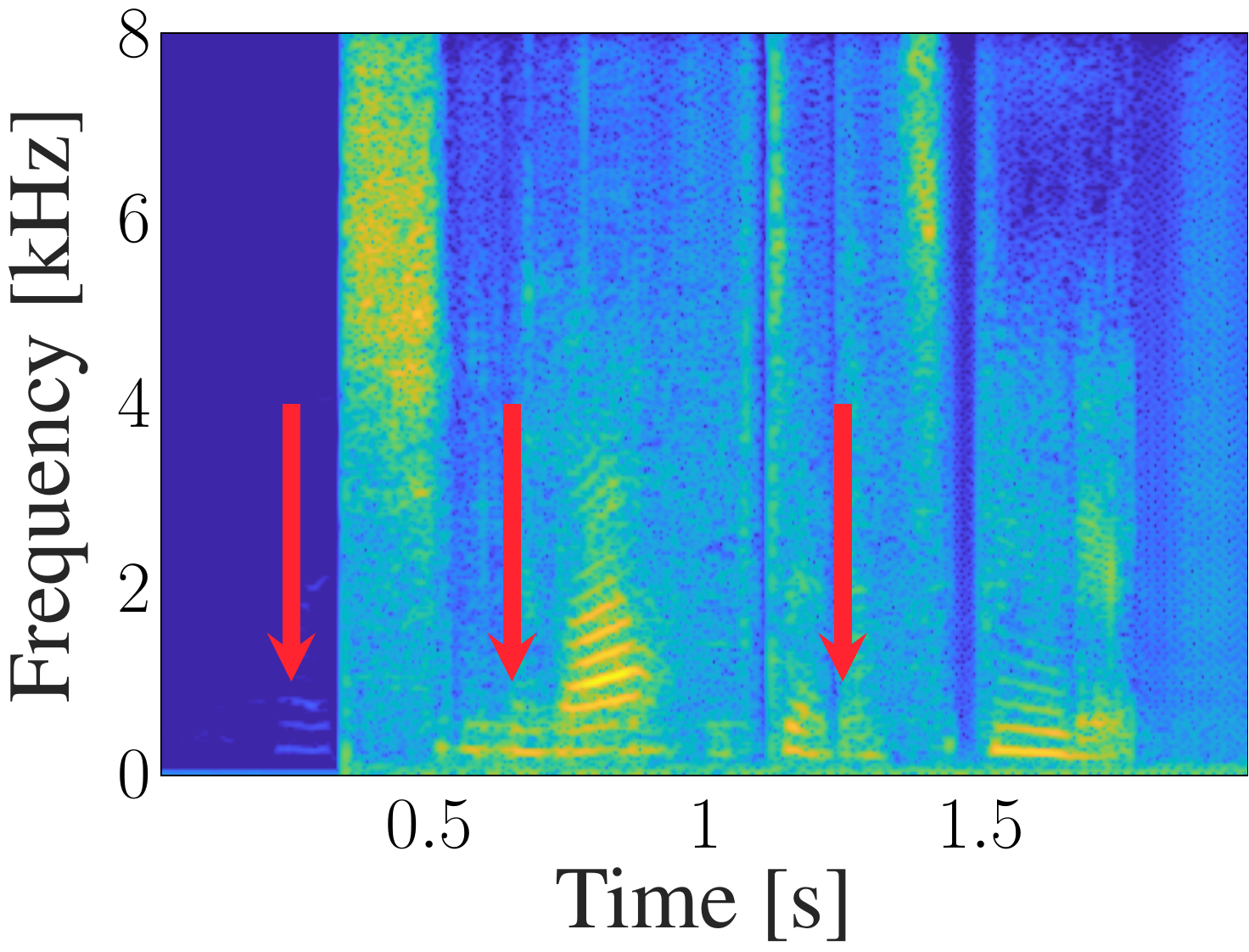}
			\caption{DEMUCS}
			\label{fig:wb_demucs_mag}
		\end{subfigure}
		\hspace{2.7mm}
		\begin{subfigure}[b]{.23\textwidth}
			\centering
			\includegraphics[width=\columnwidth]{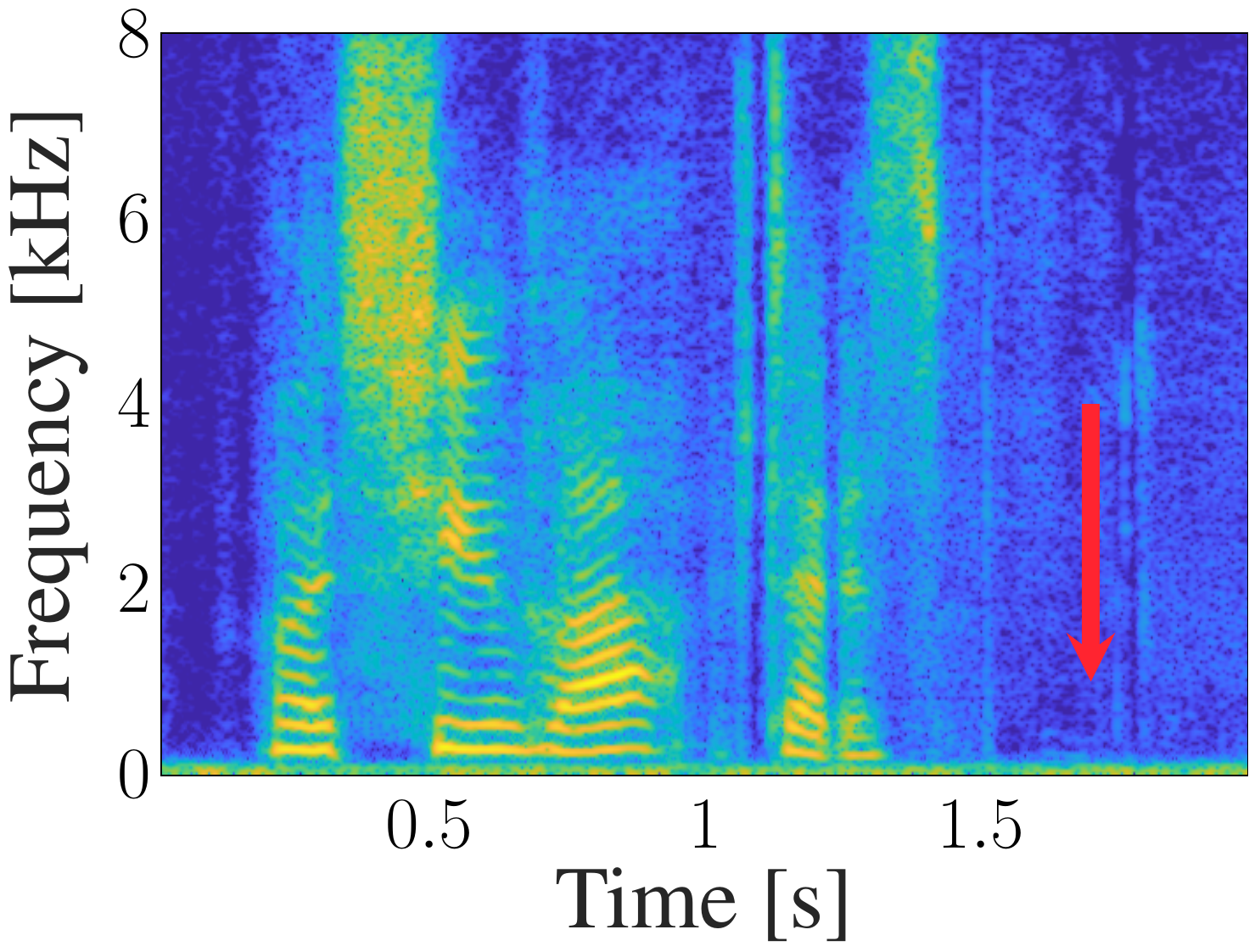}
			\caption{PHASEN}
			\label{fig:wb_phasen_mag}
		\end{subfigure}
	}
	\vspace{1mm}
	\centering
	\centerline{
		\begin{subfigure}[b]{.23\textwidth}
			\centering
			\includegraphics[width=\columnwidth]{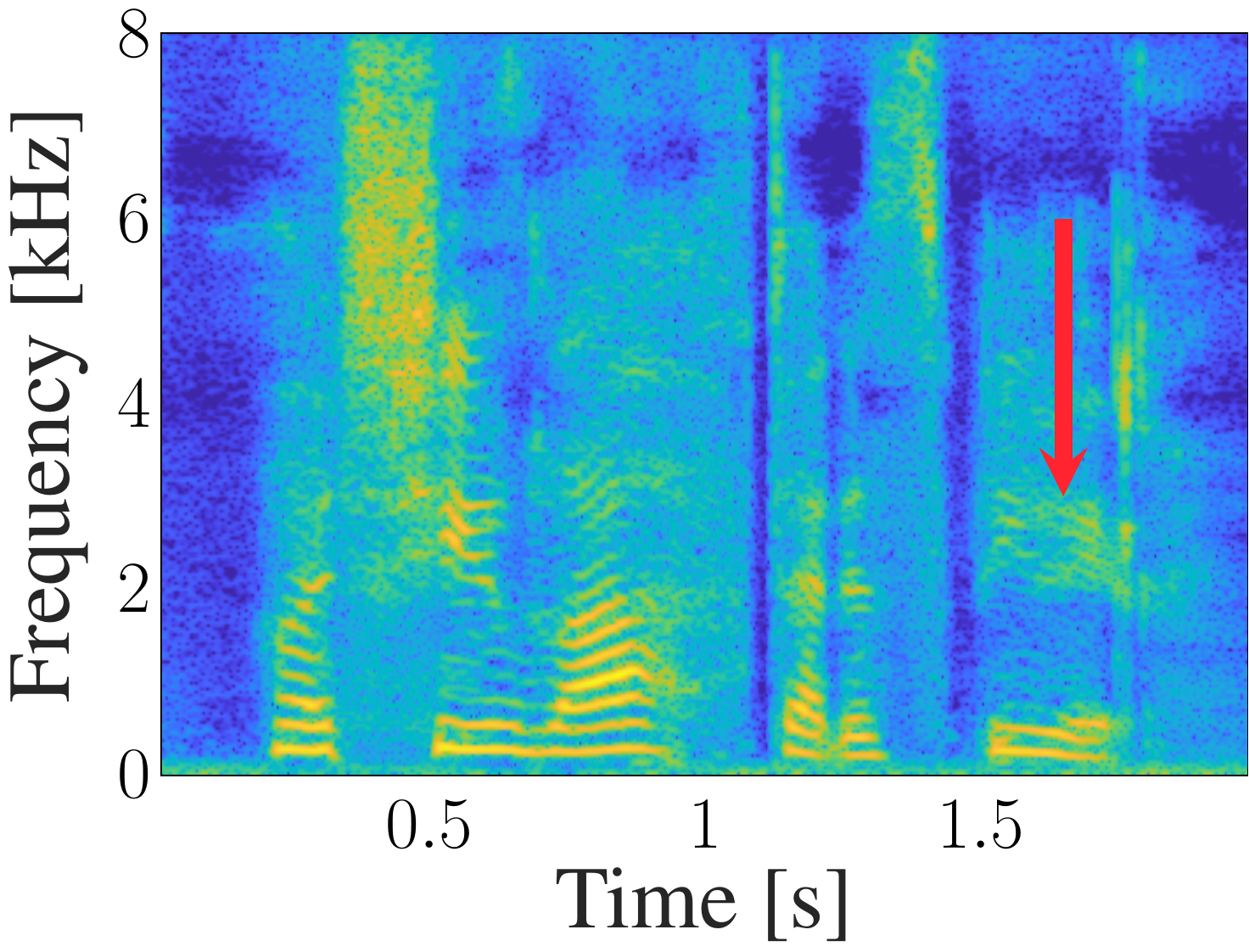}
			\caption{PFPL}
			\label{fig:wb_pfpl_mag}
		\end{subfigure}
		\hspace{12mm}
		\begin{subfigure}[b]{.23\textwidth}
			\centering
			\includegraphics[width=\columnwidth]{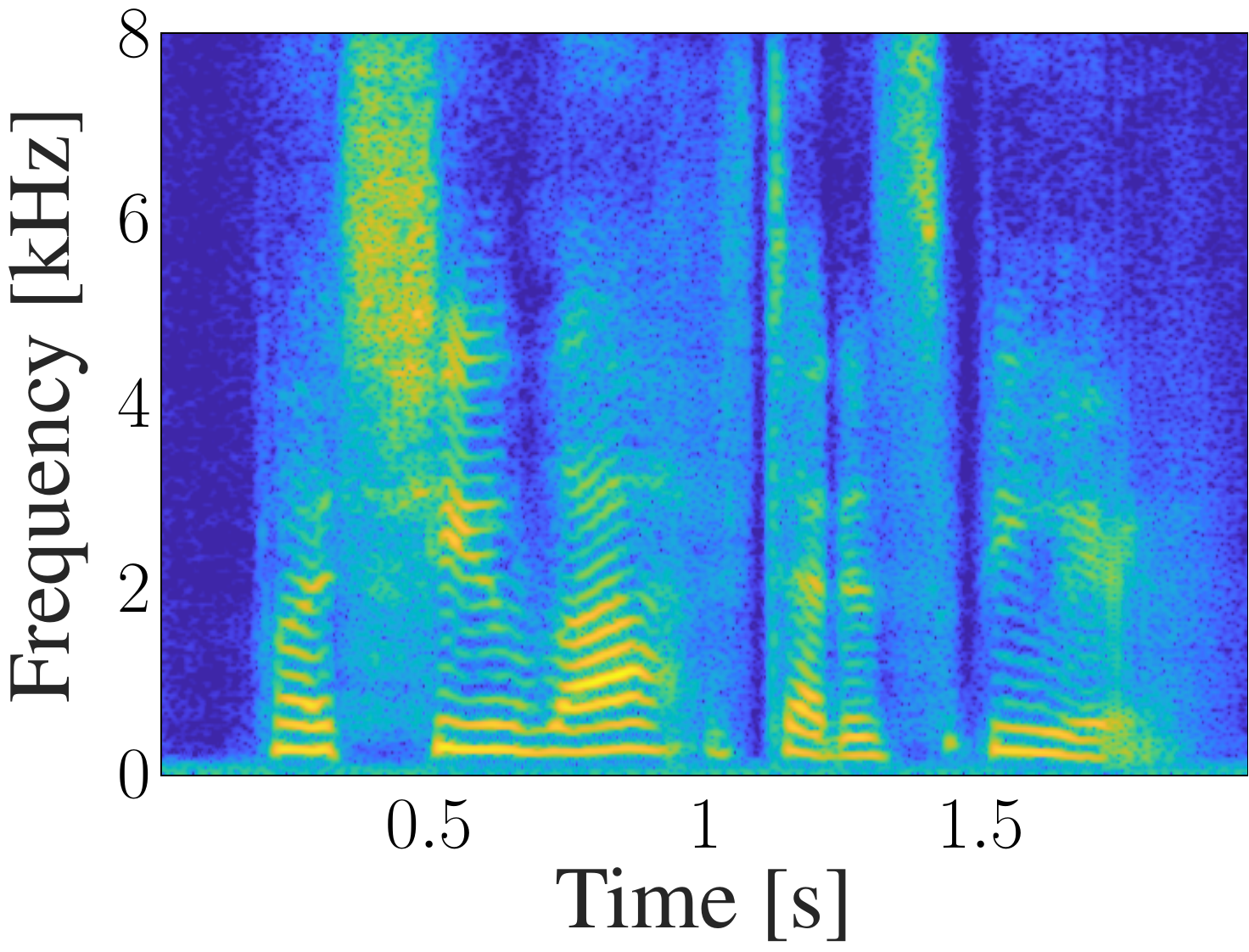}
			\caption{CMGAN}
			\label{fig:wb_cmgan_mag}
		\end{subfigure}
		\hspace{10mm}
		\begin{subfigure}[b]{.23\textwidth}
			\centering
			\includegraphics[width=\columnwidth]{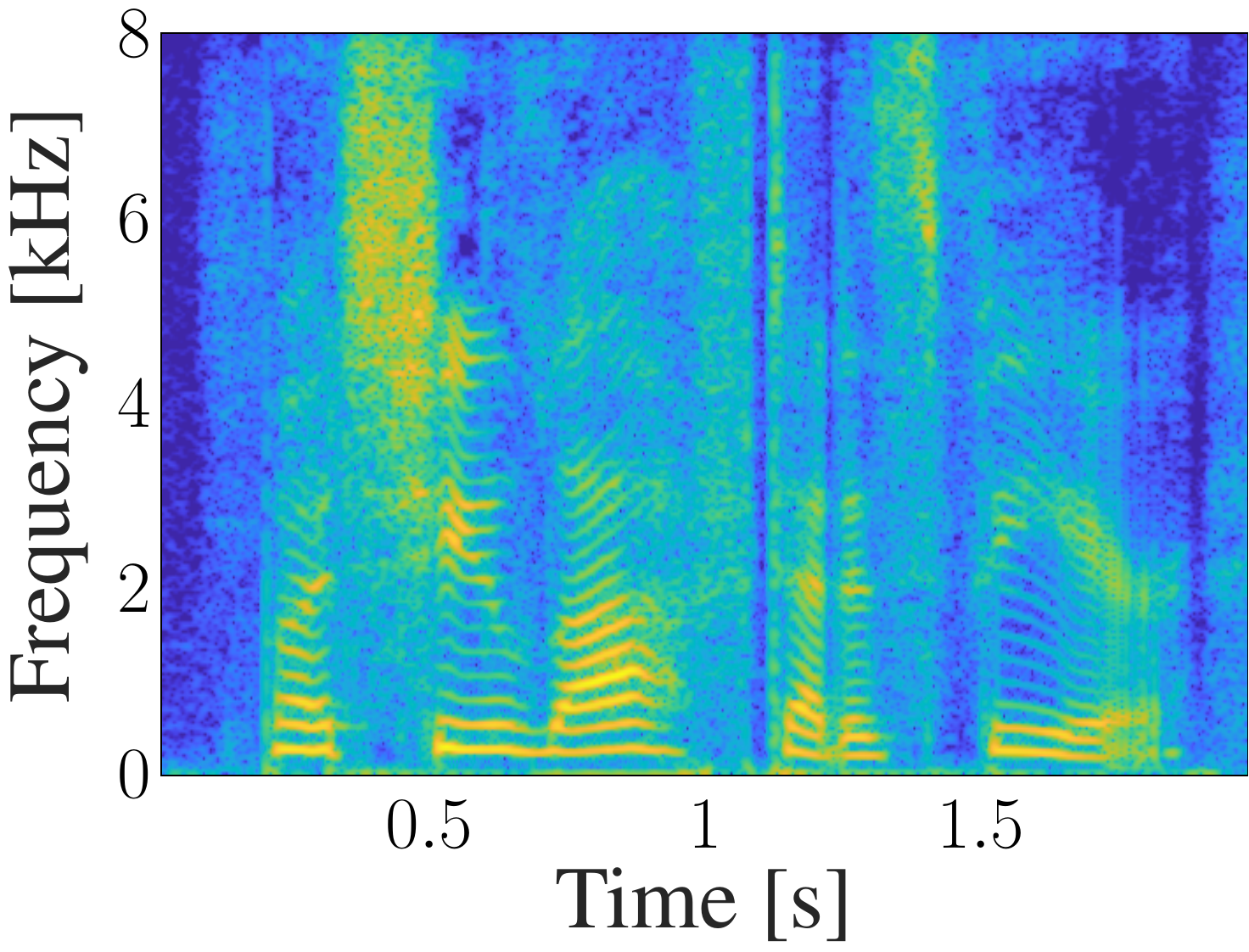}
			\caption{Clean}
			\label{fig:wb_clean_mag}
		\end{subfigure}
	}
	\vspace{1mm}
	\centering
	\centerline{
		\begin{subfigure}[b]{.23\textwidth}
			\centering			
			\includegraphics[width=\columnwidth]{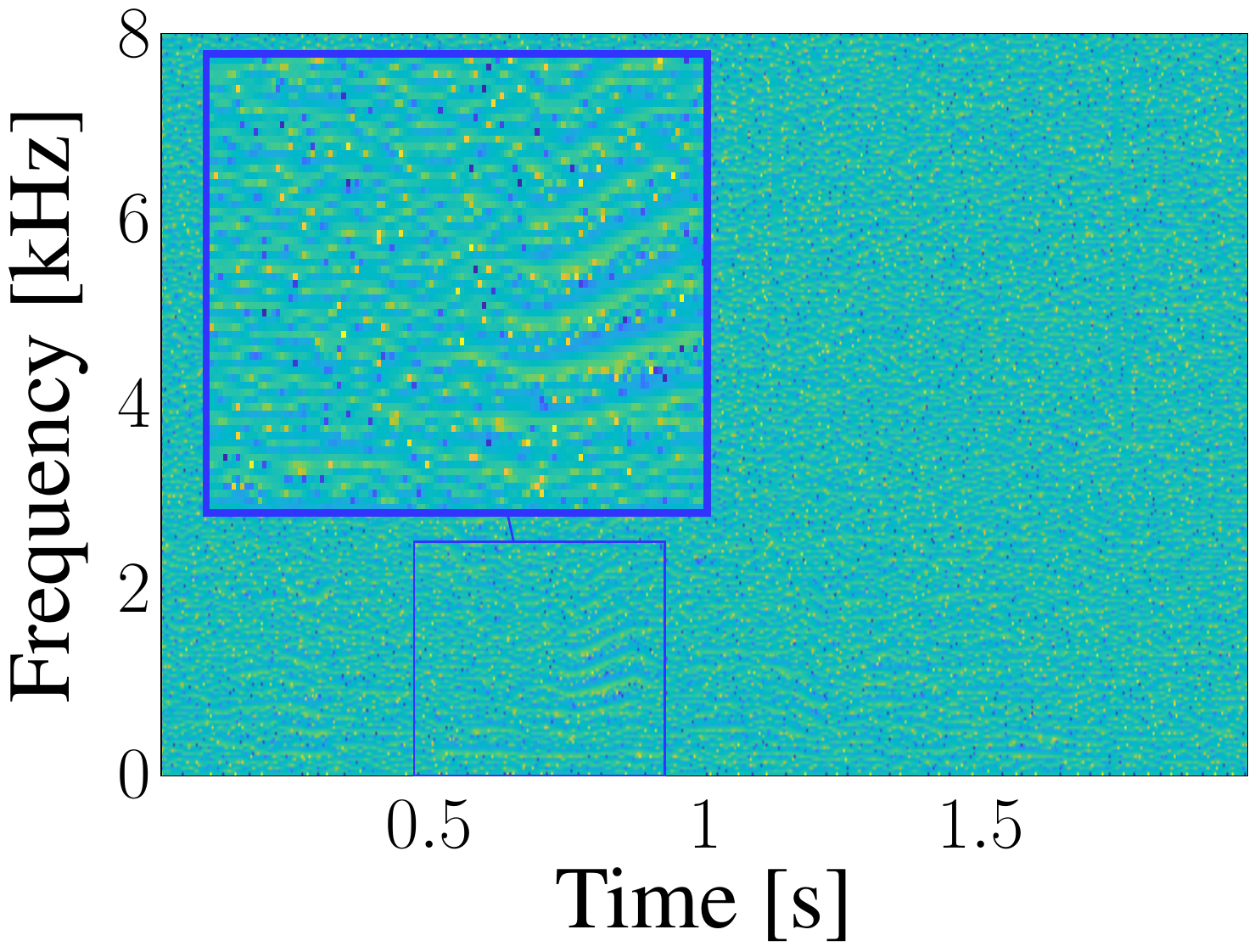}
			\caption{Noisy}
			\label{fig:wb_noisy_phase}
		\end{subfigure}
		\hspace{3.7mm}
		\begin{subfigure}[b]{.23\textwidth}
			\centering
			\includegraphics[width=\columnwidth]{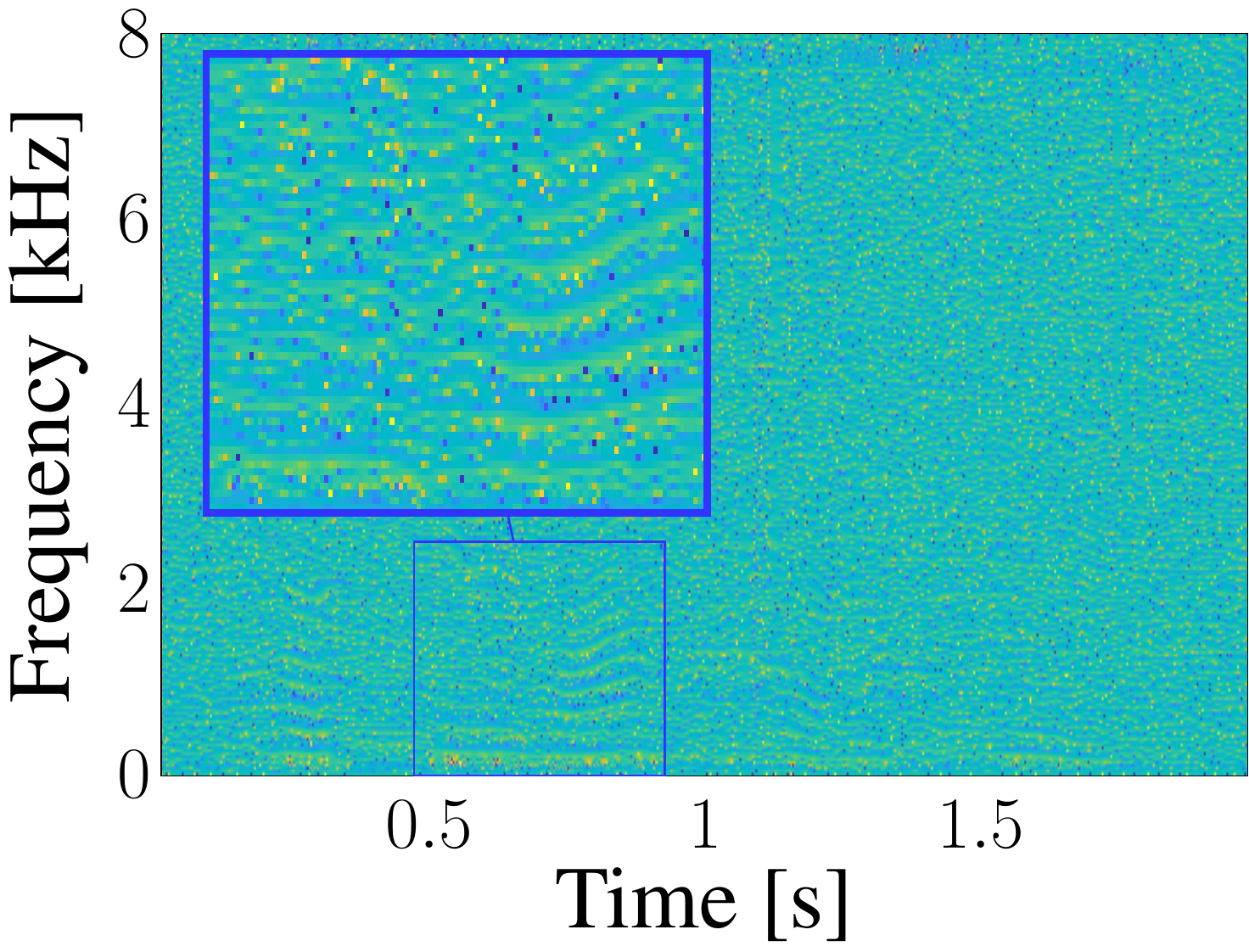}
			\caption{MetricGAN+}
			\label{fig:wb_metricgan_phase}
		\end{subfigure}
		\hspace{2.7mm}
		\begin{subfigure}[b]{.23\textwidth}
			\centering		
			\includegraphics[width=\columnwidth]{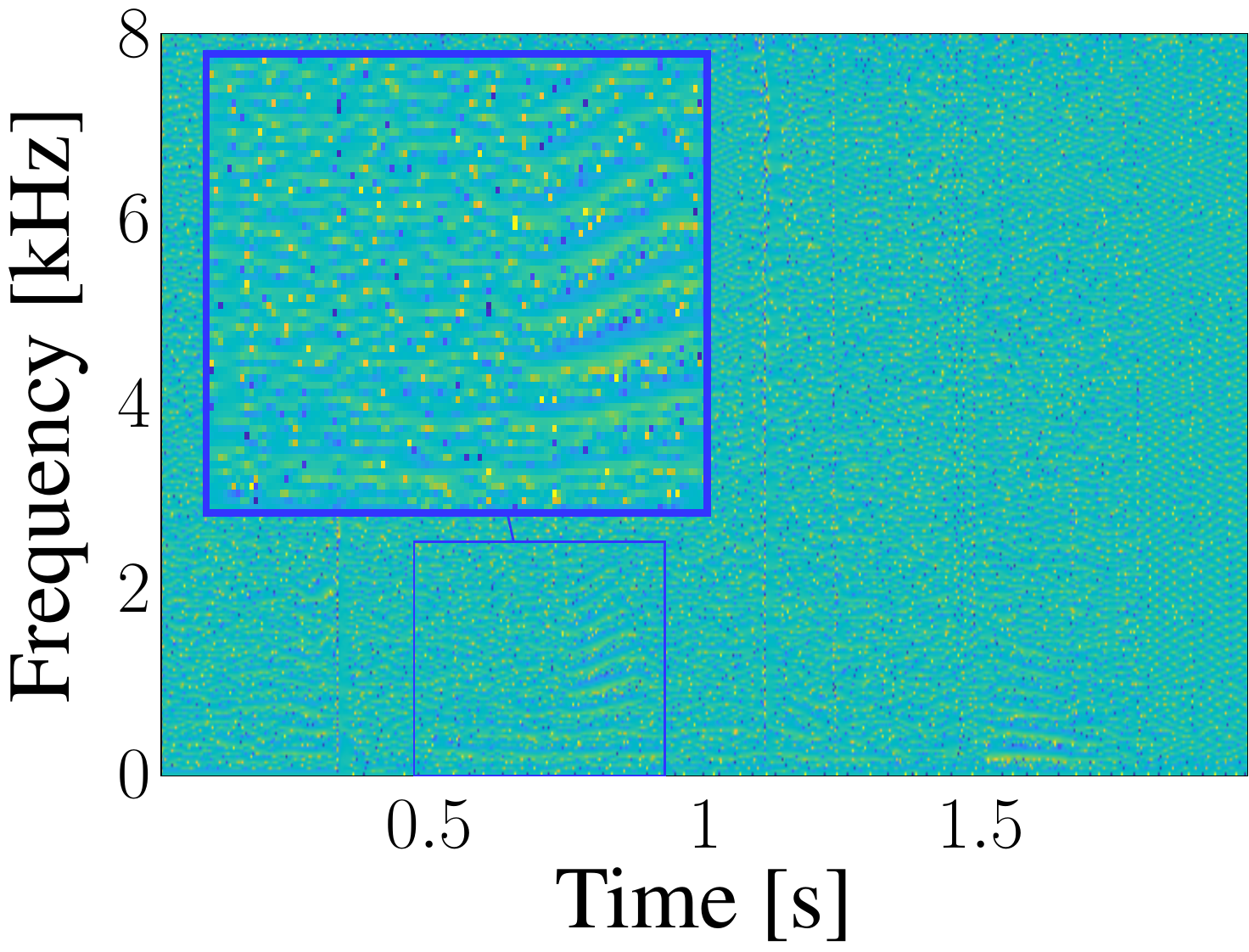}
			\caption{DEMUCS}
			\label{fig:wb_demucs_phase}
		\end{subfigure}
		\hspace{3.7mm}
		\begin{subfigure}[b]{.23\textwidth}
			\centering
			\includegraphics[width=\columnwidth]{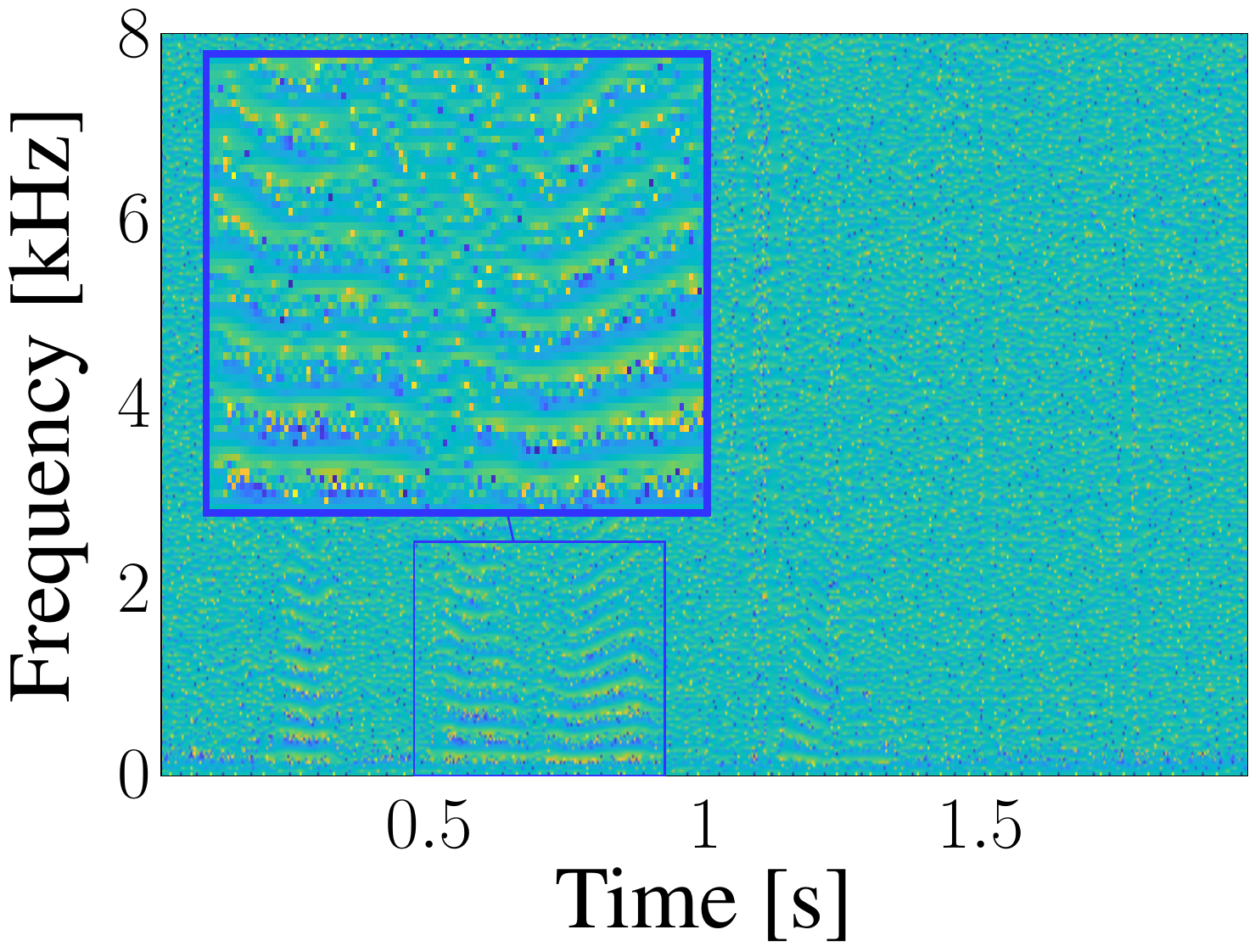}
			\caption{PHASEN}
			\label{fig:wb_phasen_phase}
		\end{subfigure}
	}
	\vspace{1mm}
	\centering
	\centerline{
		\begin{subfigure}[b]{.23\textwidth}
			\centering
			\includegraphics[width=\columnwidth]{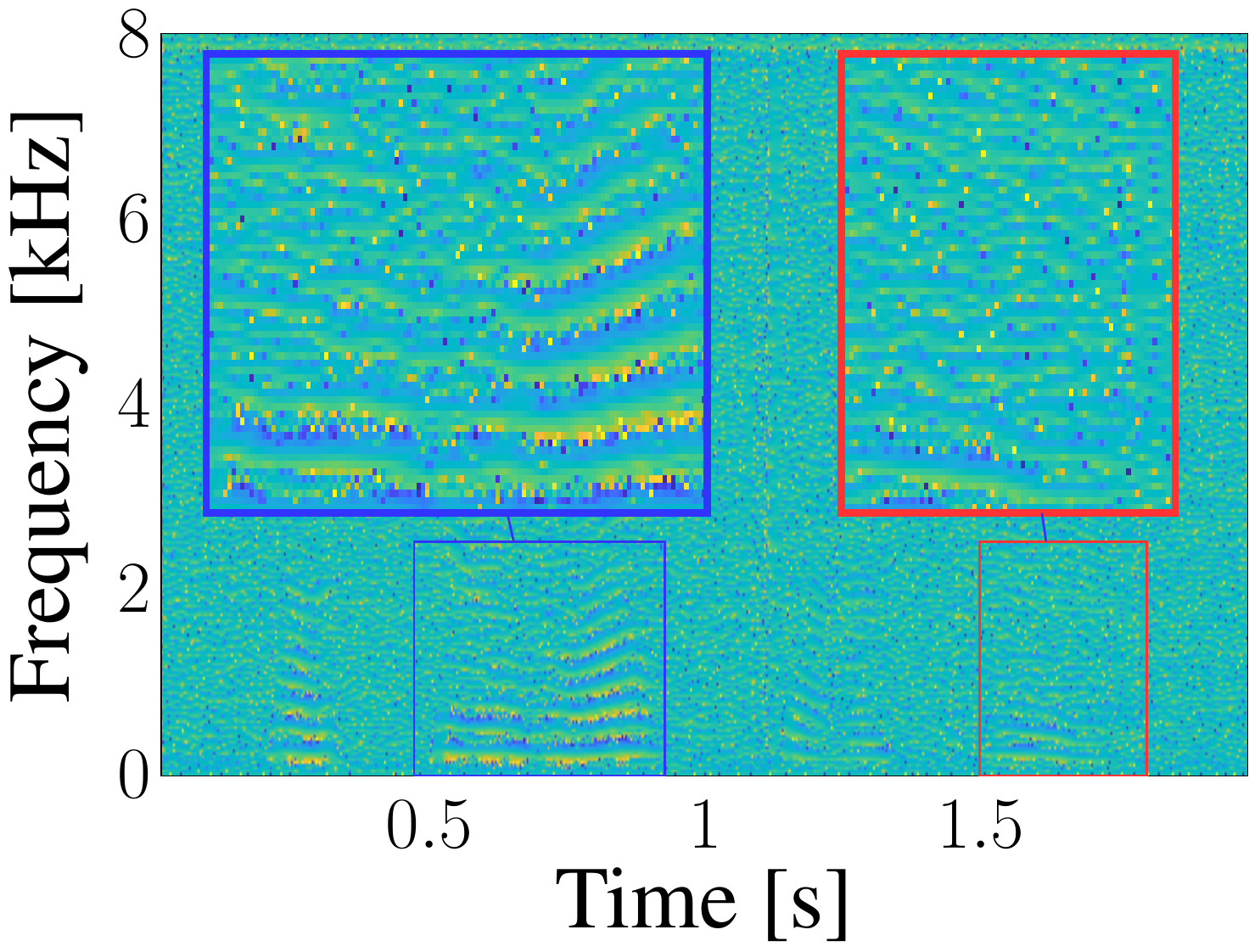}
			\caption{PFPL}
			\label{fig:wb_pfpl_phase}
		\end{subfigure}
		\hspace{12mm}
		\begin{subfigure}[b]{.23\textwidth}
			\centering
			\includegraphics[width=\columnwidth]{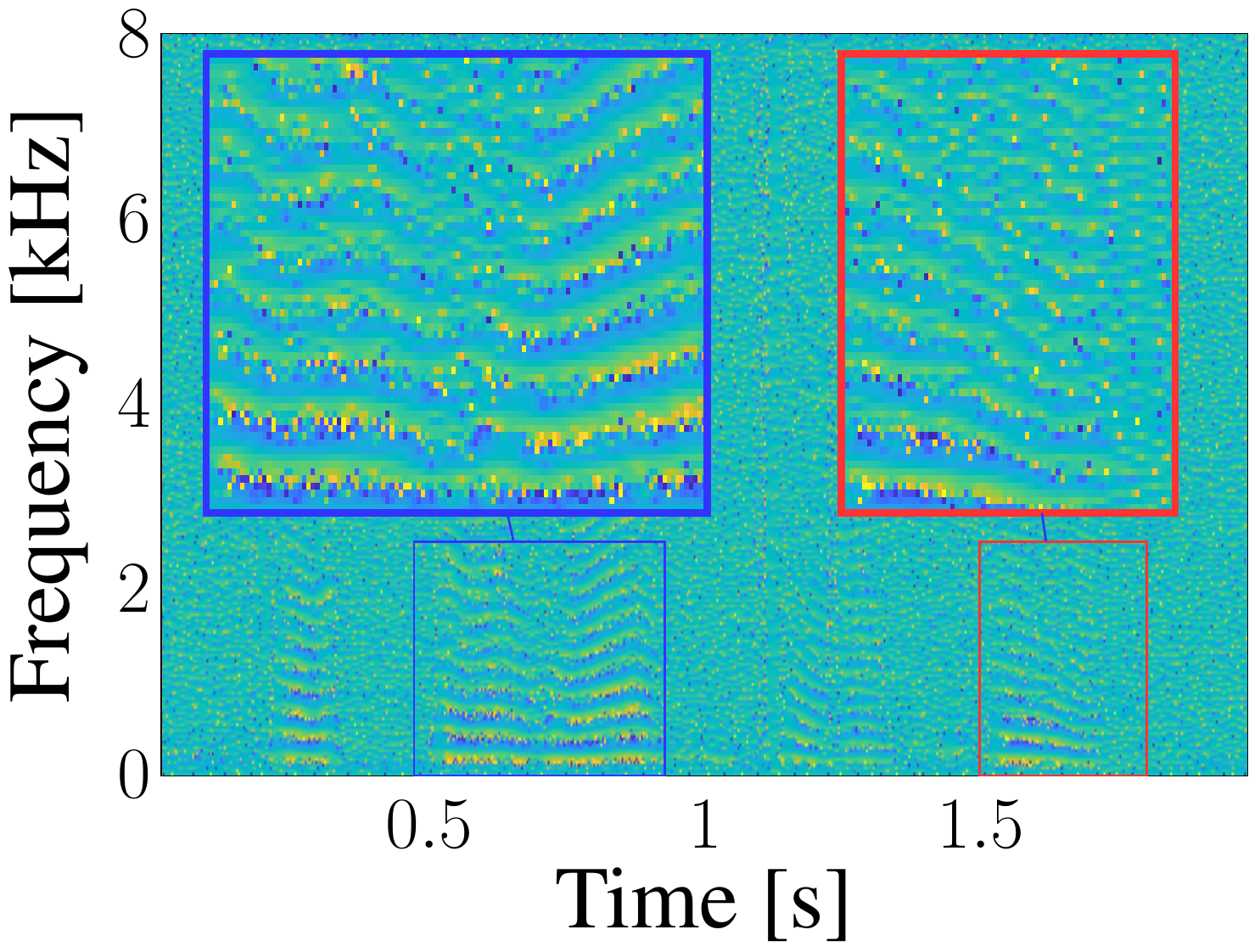}
			\caption{CMGAN}
			\label{fig:wb_cmgan_phase}
		\end{subfigure}
		\hspace{10mm}
		\begin{subfigure}[b]{.23\textwidth}
			\centering
			\includegraphics[width=\columnwidth]{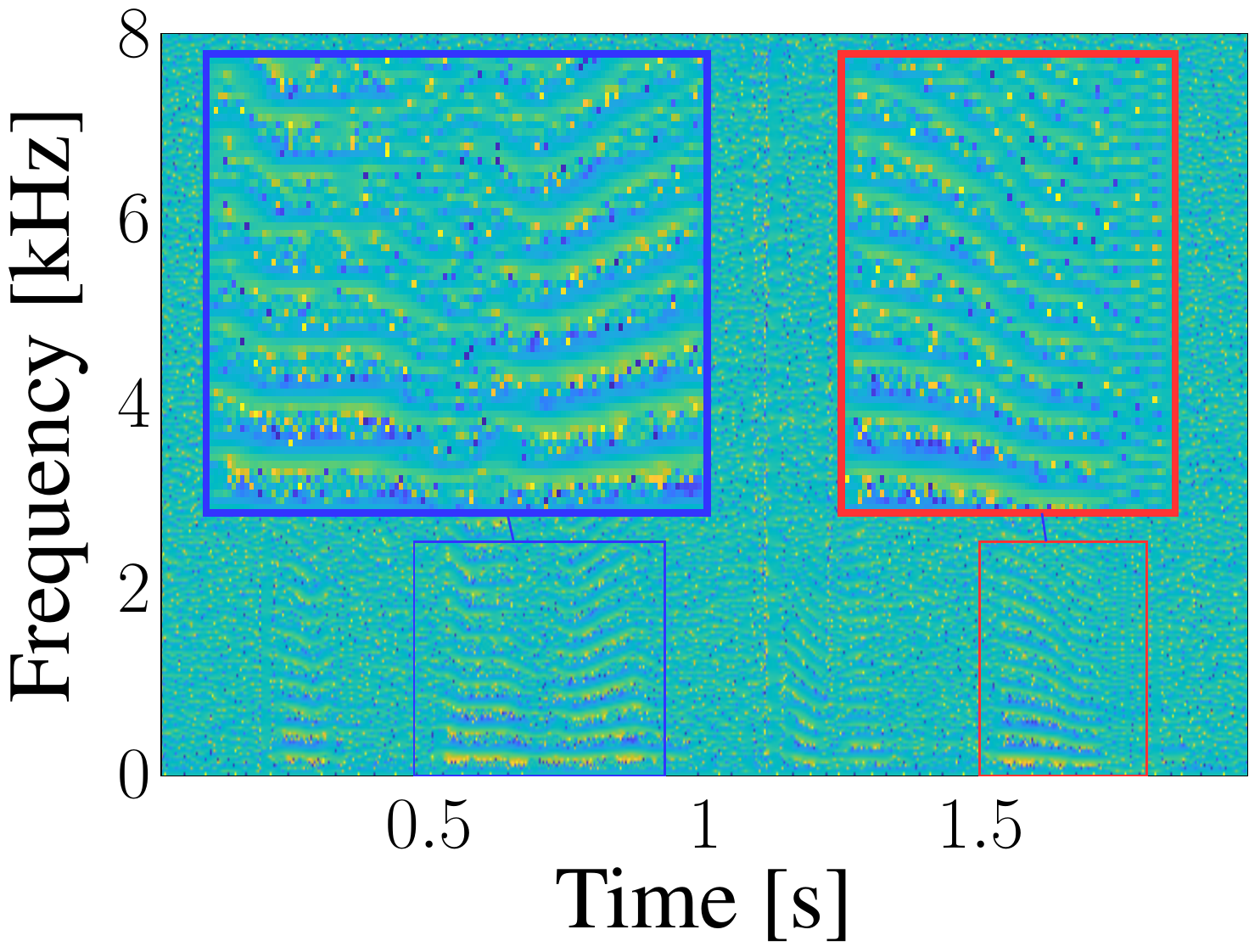}
			\caption{Clean}
			\label{fig:wb_clean_phase}
		\end{subfigure}
	}
	\caption{Visualization of subjective approaches under a wide-band cafe noise (DEMAND dataset) at SNR = 0~dB. (a-g) represent the time-domain signal, while (h-n) are the TF-magnitude representations in dB and (o-u) are the reconstructed BPD of the given TF-phase representations. (\myarrowblue) and (\myarrowred) reflect the distortions in time and TF-magnitude representations, respectively. \label{fig:wb_subjective}}
	\vspace{-2mm}
\end{figure*}

	Additionally, this section presents a visualization of the CMGAN in comparison to subjective evaluation methods. A wide-band non-stationary cafe noise from the DEMAND dataset (SNR = 0~dB) and a narrow-band high-frequency stationary doorbell noise from the Freesound dataset (SNR = 3~dB) are used to evaluate the methods. Both noises are added to sentences from the DNS challenge. Comparisons are made between time-domain, TF-magnitude, and TF-phase representations for comprehensive performance analysis. Since the phase is unstructured, we utilize the baseband phase difference (BPD) approach proposed in \cite{krawczyk2014bpd} to enhance the phase visualization. From Fig.~\ref{fig:wb_subjective}, MetricGAN+, DEMUCS, and PHASEN show the worst performance by confusing speech with noise, particularly in the 1.5 to 2 seconds interval (similar speech and noise powers). The distortions and missing speech segments are annotated in the time and TF-magnitude representations by (\myarrowblue) and (\myarrowred), respectively. Moreover, the denoised phase in methods employing only magnitude (MetricGAN+) and time-domain (DEMUCS) is very similar to the noisy input, in contrast to clear enhancement in complex TF-domain methods (PHASEN, PFPL, and CMGAN). PFPL and CMGAN exhibit the best performance, with better phase reconstruction in CMGAN (1.5 to 2 seconds interval).
	
\begin{figure*}[t!]
	\captionsetup[subfigure]{justification=centering}
	\centering
	\centerline{
		\begin{subfigure}[b]{.23\textwidth}
			\centering			
			\includegraphics[width=\columnwidth]{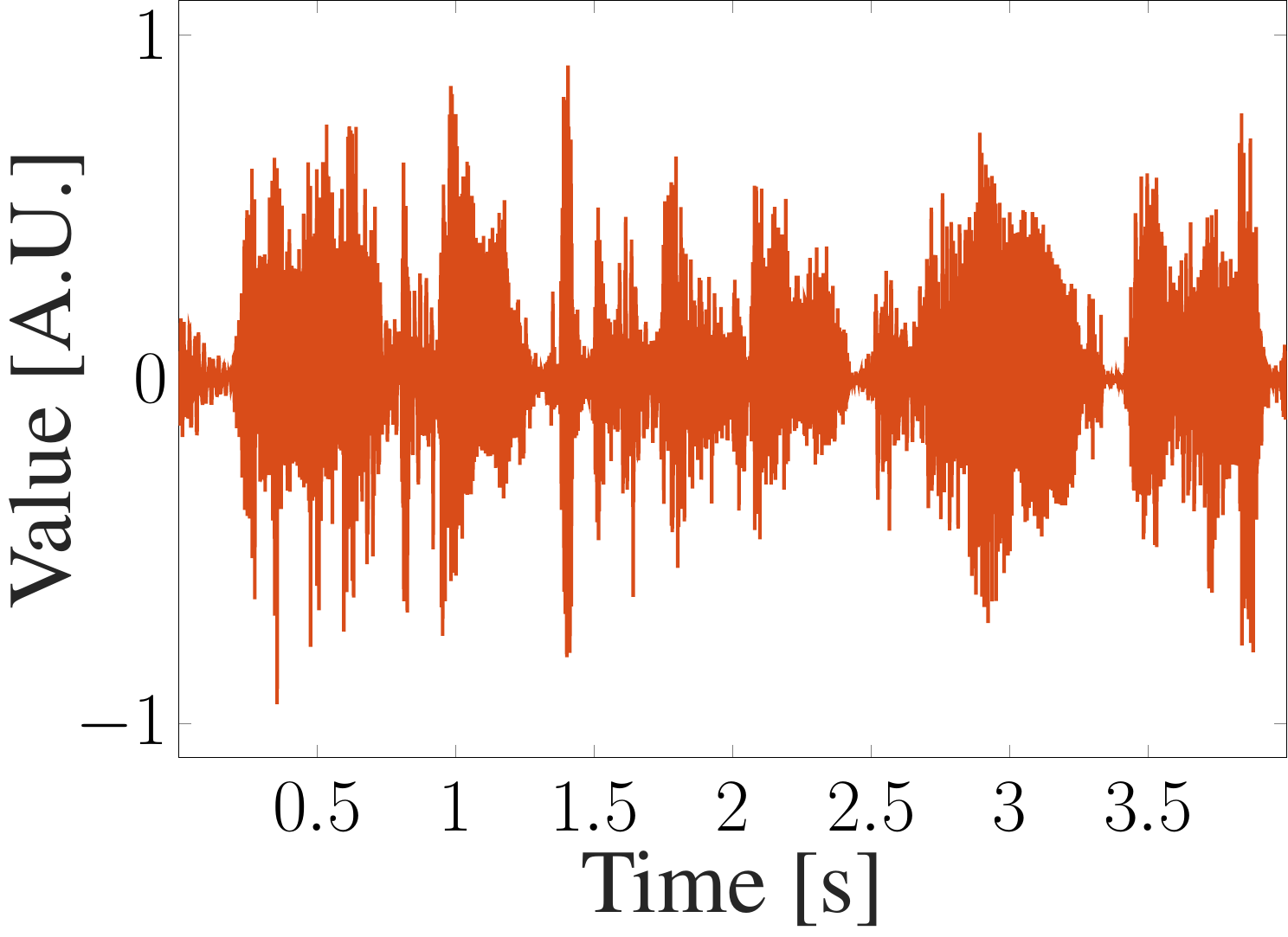}
			\caption{Noisy}
			\label{fig:nb_noisy_time}
		\end{subfigure}
		\hspace{3.7mm}
		\begin{subfigure}[b]{.23\textwidth}
			\centering
			\includegraphics[width=\columnwidth]{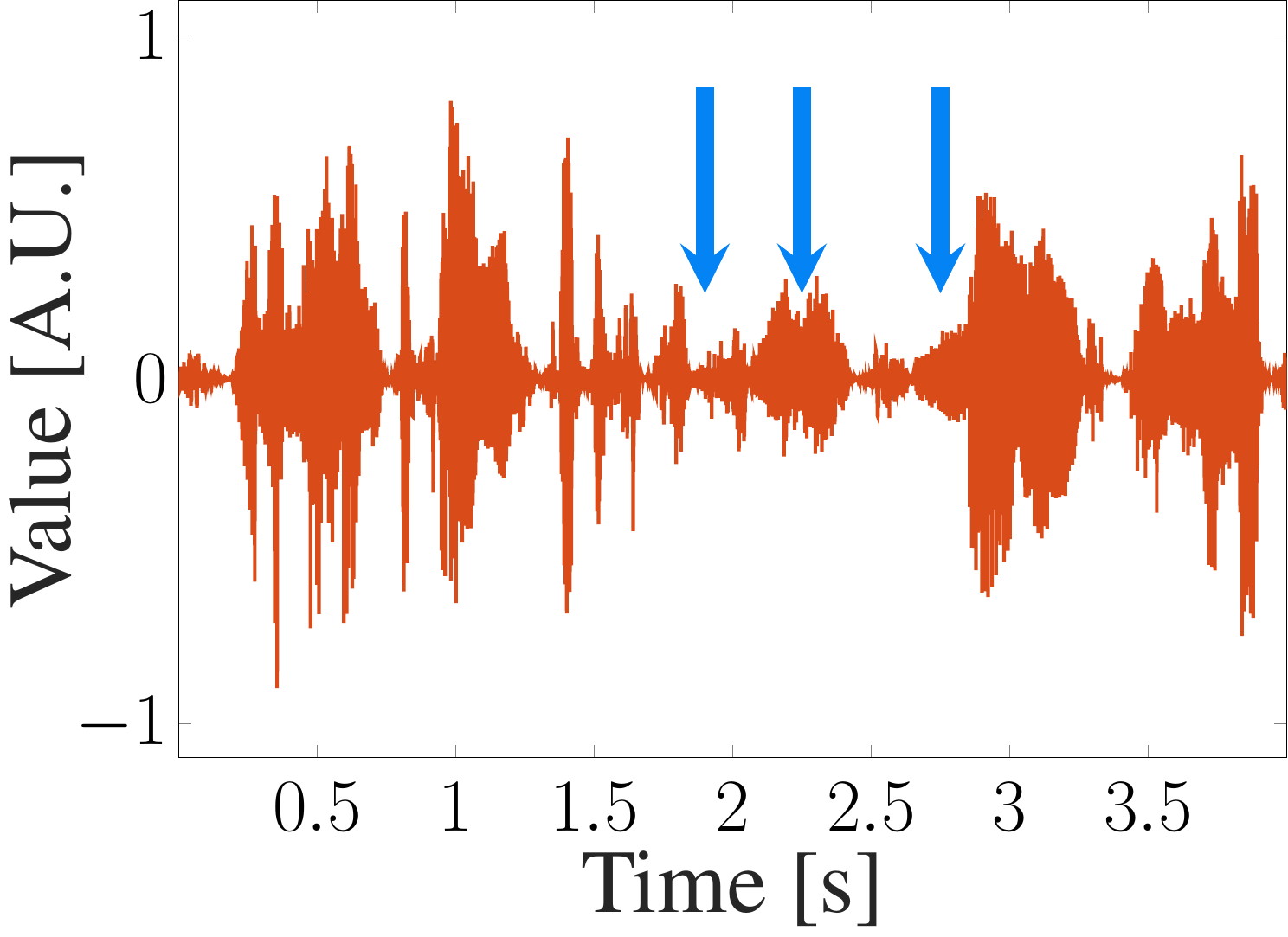}
			\caption{MetricGAN+}
			\label{fig:nb_metricgan_time}
		\end{subfigure}
		\hspace{2.7mm}
		\begin{subfigure}[b]{.23\textwidth}
			\centering		
			\includegraphics[width=\columnwidth]{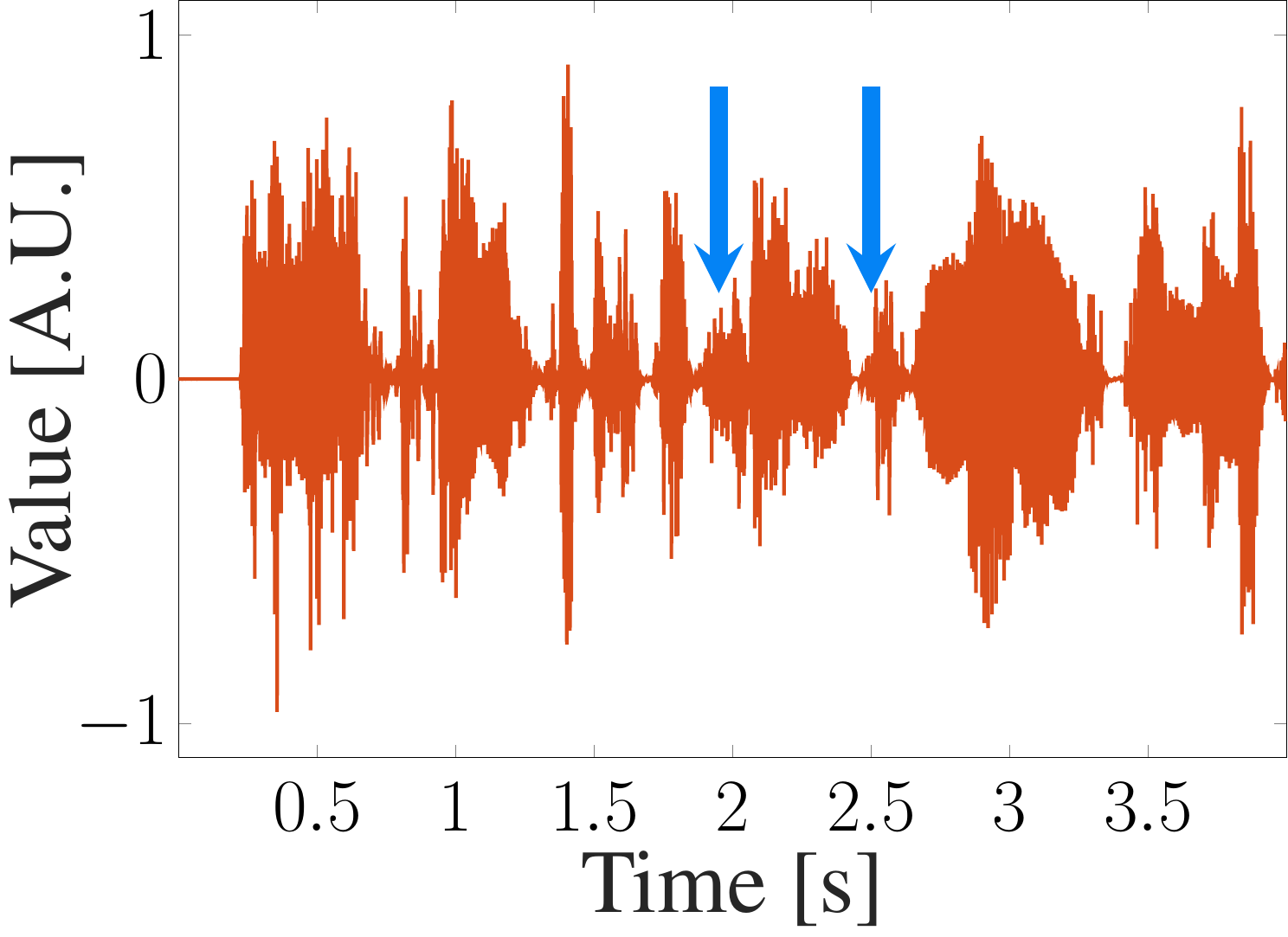}
			\caption{DEMUCS}
			\label{fig:nb_demucs_time}
		\end{subfigure}
		\hspace{2.7mm}
		\begin{subfigure}[b]{.23\textwidth}
			\centering
			\includegraphics[width=\columnwidth]{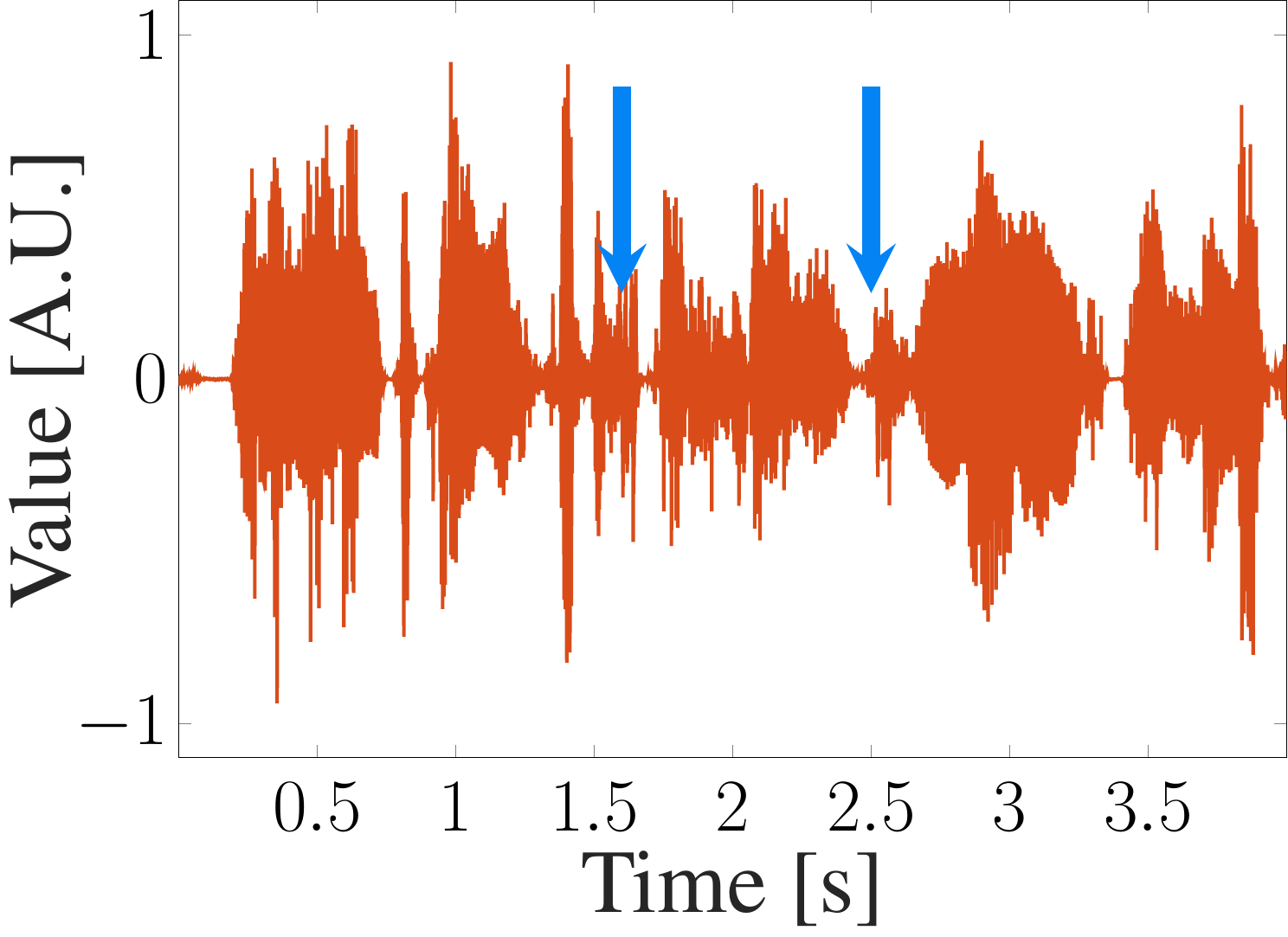}
			\caption{PHASEN}
			\label{fig:nb_phasen_time}
		\end{subfigure}
	}
	\vspace{1mm}
	\centering
	\centerline{
		\begin{subfigure}[b]{.23\textwidth}
			\centering
			\includegraphics[width=\columnwidth]{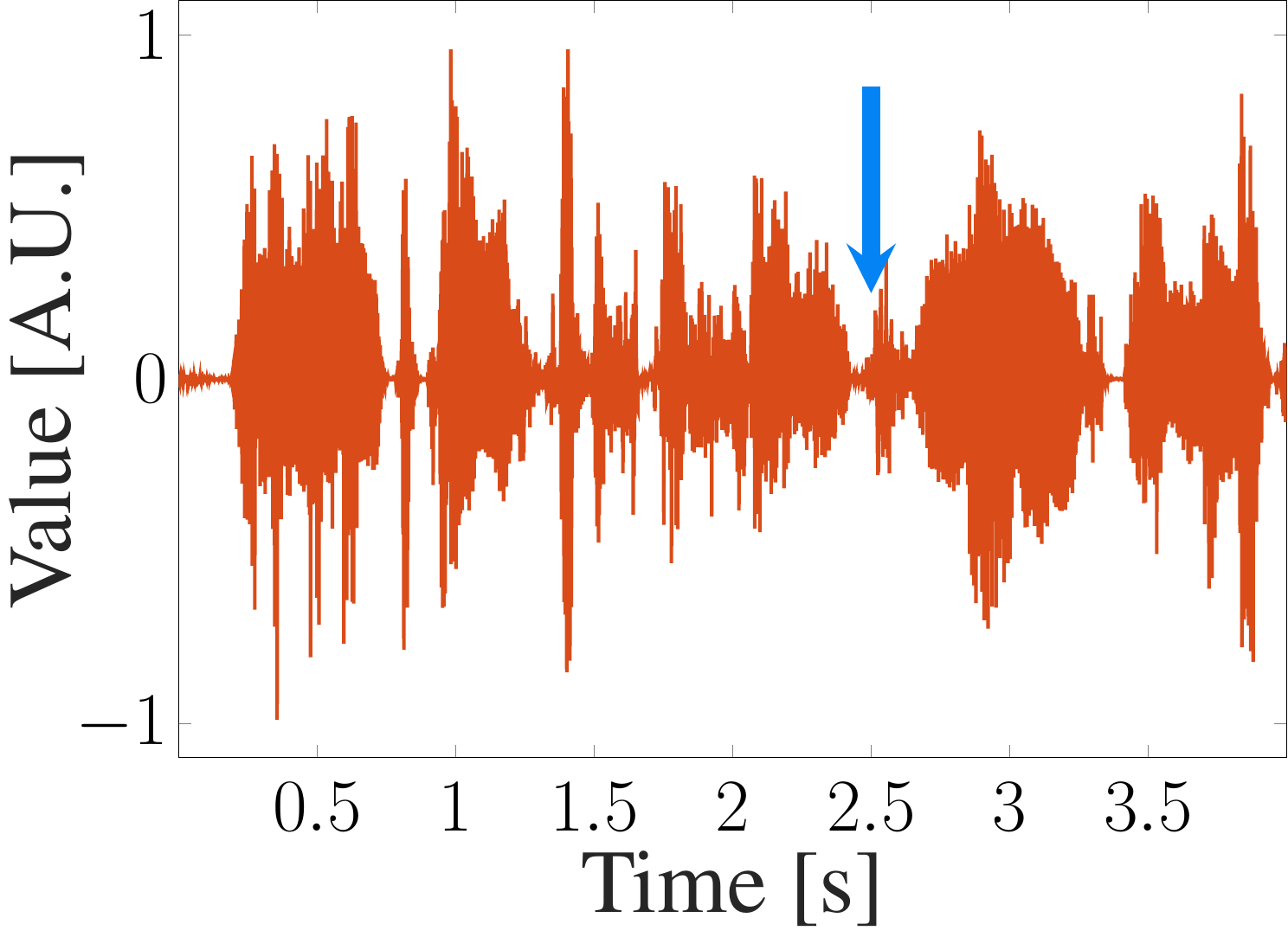}
			\caption{PFPL}
			\label{fig:nb_pfpl_time}
		\end{subfigure}
		\hspace{12mm}
		\begin{subfigure}[b]{.23\textwidth}
			\centering
			\includegraphics[width=\columnwidth]{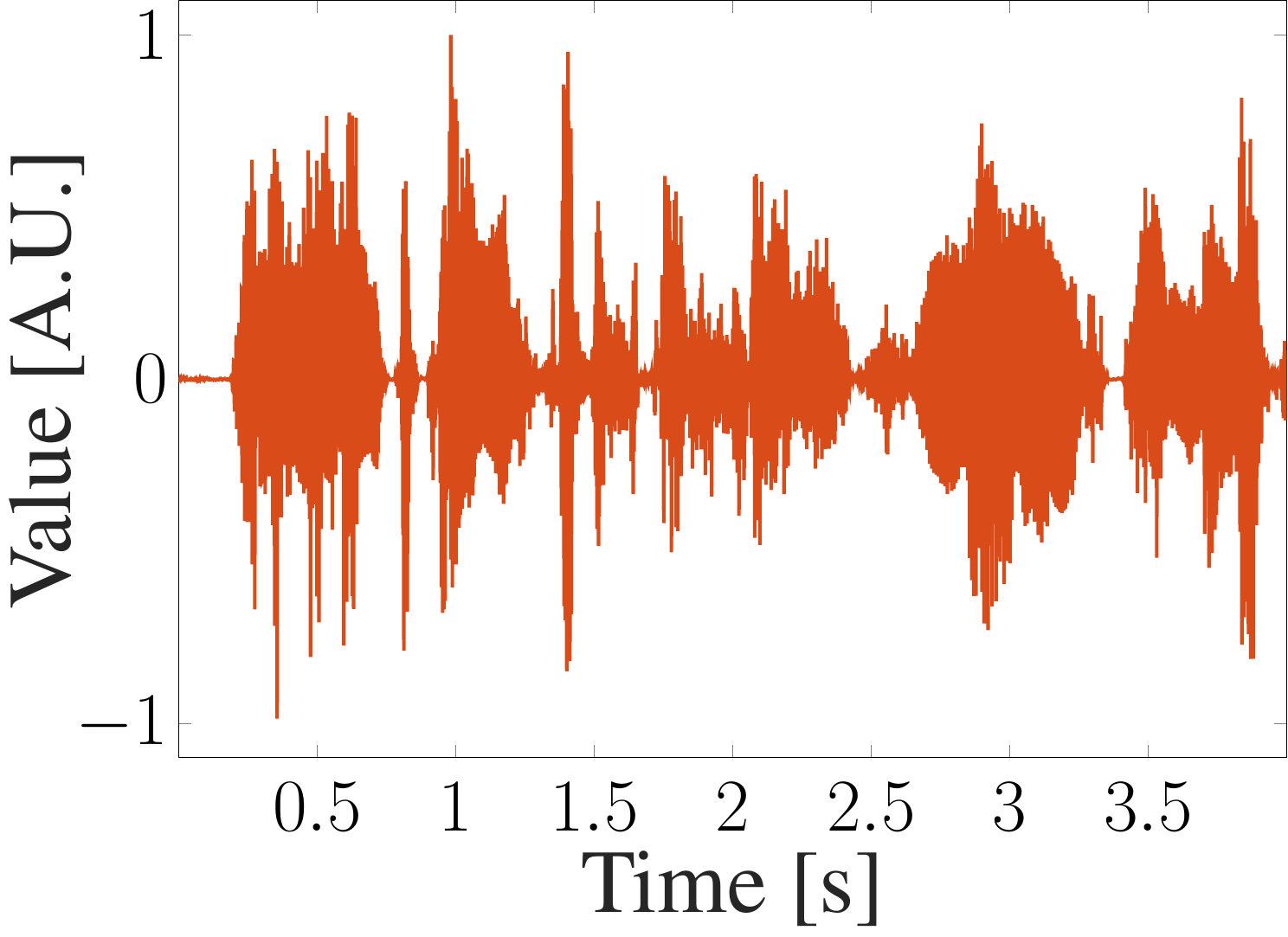}
			\caption{CMGAN}
			\label{fig:nb_cmgan_time}
		\end{subfigure}
		\hspace{10mm}
		\begin{subfigure}[b]{.23\textwidth}
			\centering
			\includegraphics[width=\columnwidth]{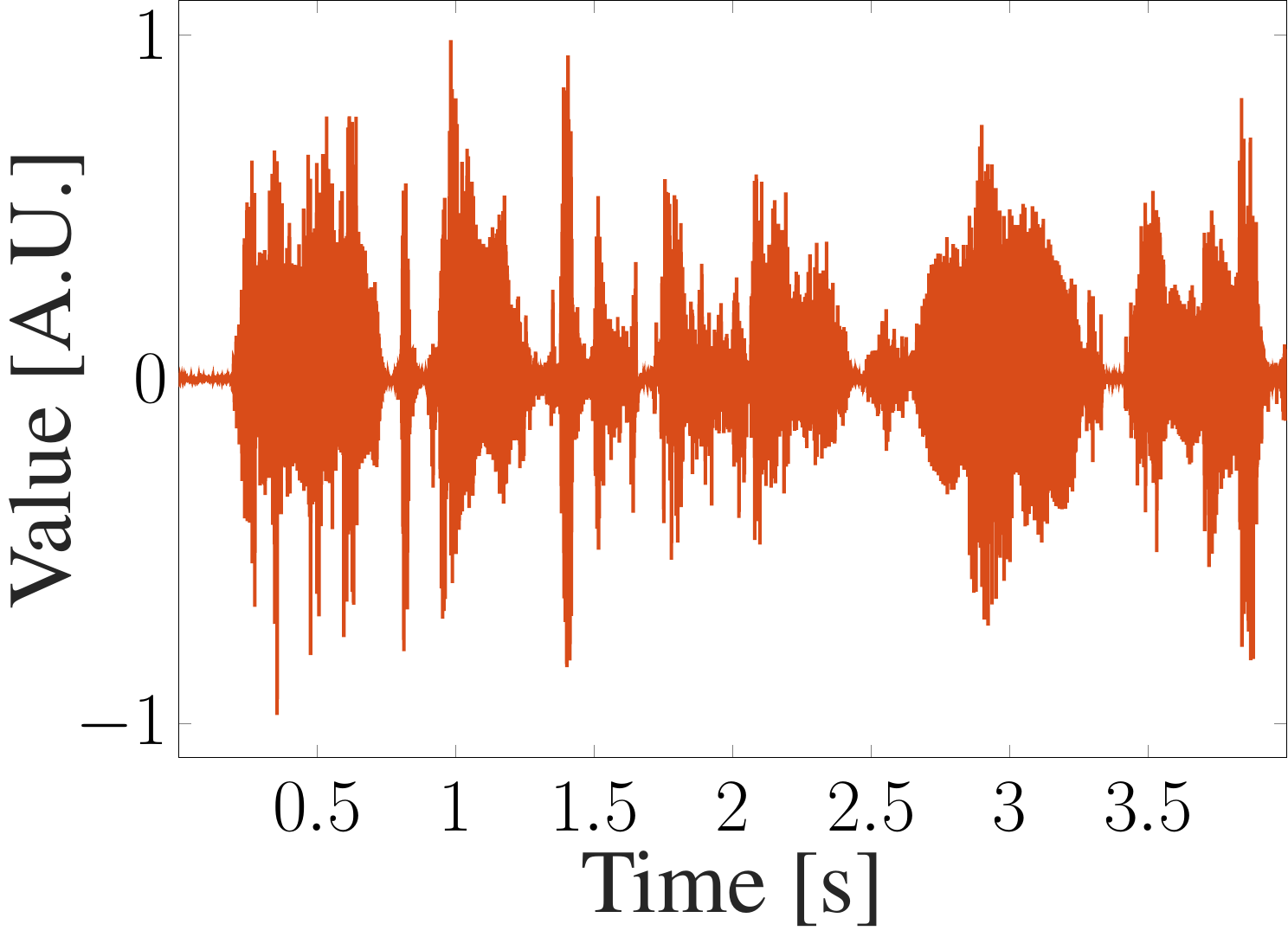}
			\caption{Clean}
			\label{fig:nb_clean_time}
		\end{subfigure}
	}
	\vspace{1mm}
	\centering
	\centerline{
		\begin{subfigure}[b]{.23\textwidth}
			\centering			
			\includegraphics[width=\columnwidth]{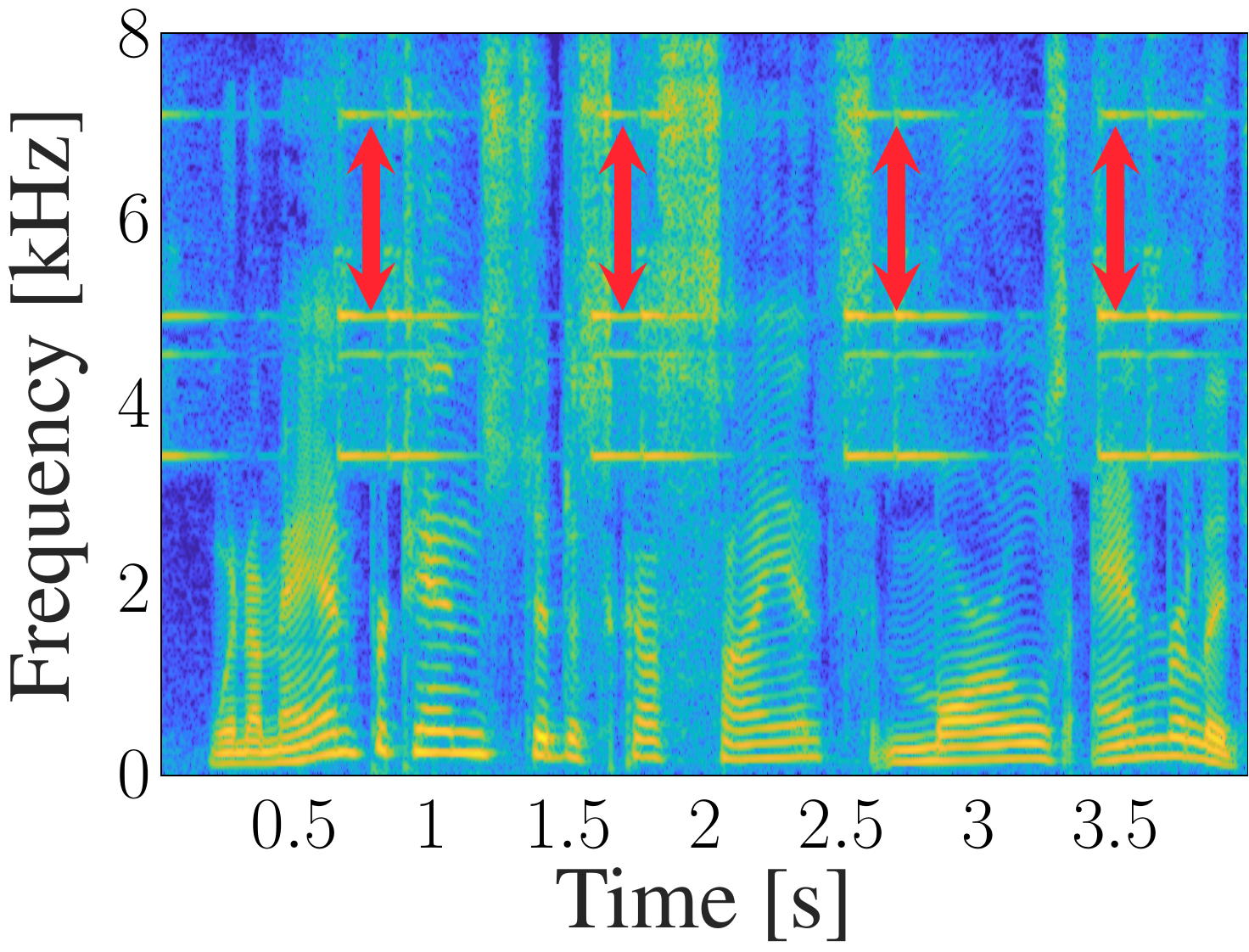}
			\caption{Noisy}
			\label{fig:nb_noisy_mag}
		\end{subfigure}
		\hspace{3.7mm}
		\begin{subfigure}[b]{.23\textwidth}
			\centering
			\includegraphics[width=\columnwidth]{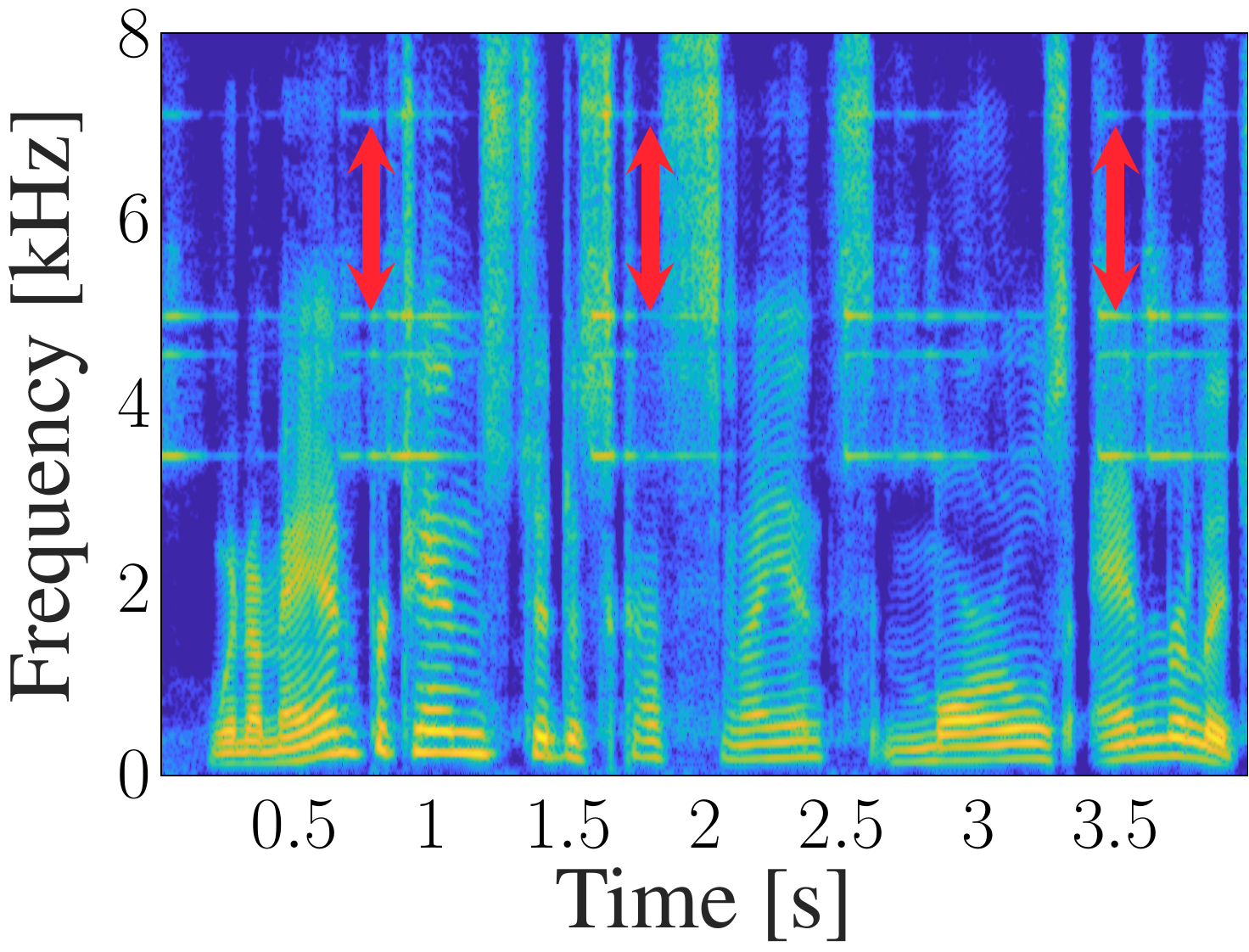}
			\caption{MetricGAN+}
			\label{fig:nb_metricgan_mag}
		\end{subfigure}
		\hspace{2.7mm}
		\begin{subfigure}[b]{.23\textwidth}
			\centering		
			\includegraphics[width=\columnwidth]{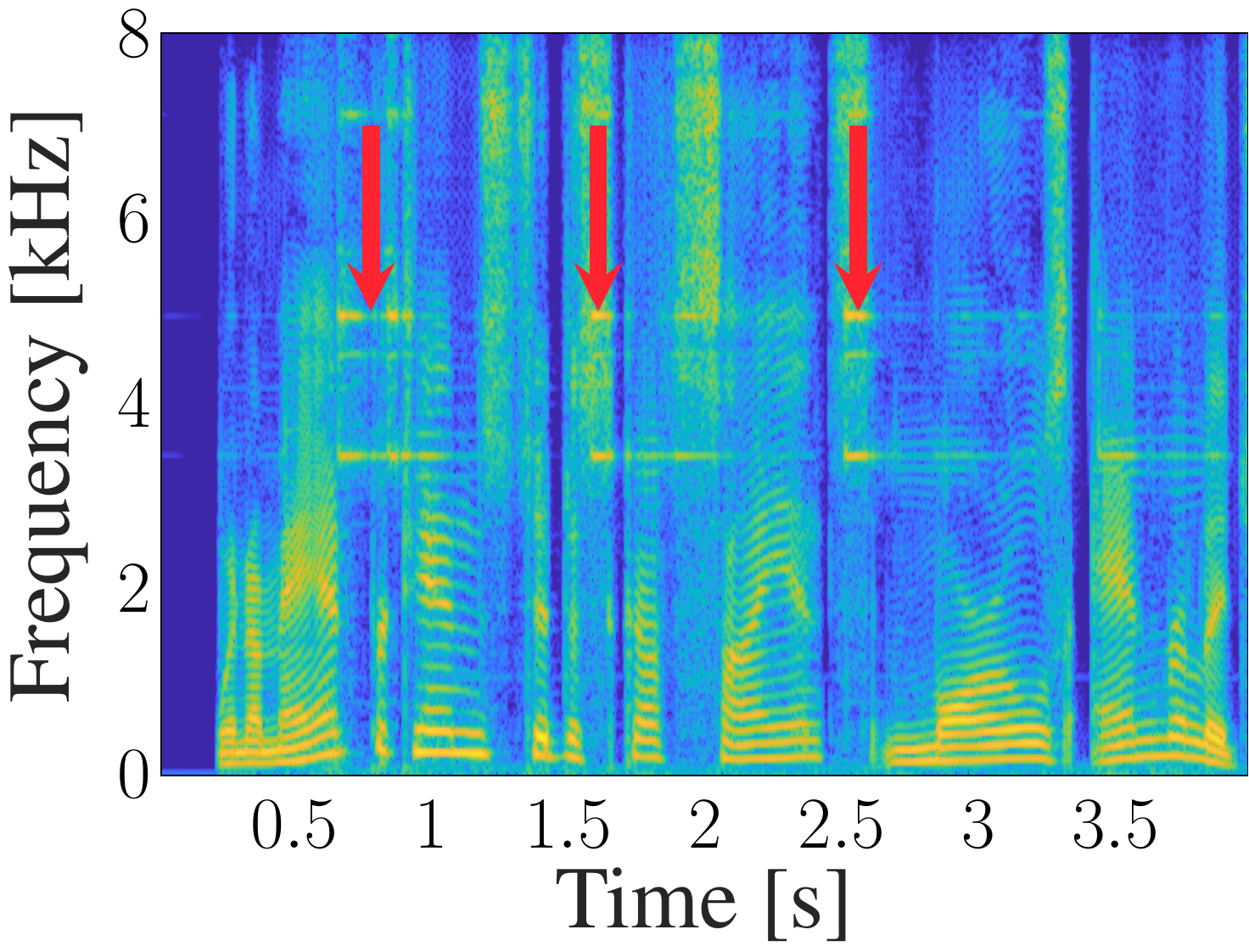}
			\caption{DEMUCS}
			\label{fig:nb_demucs_mag}
		\end{subfigure}
		\hspace{2.7mm}
		\begin{subfigure}[b]{.23\textwidth}
			\centering
			\includegraphics[width=\columnwidth]{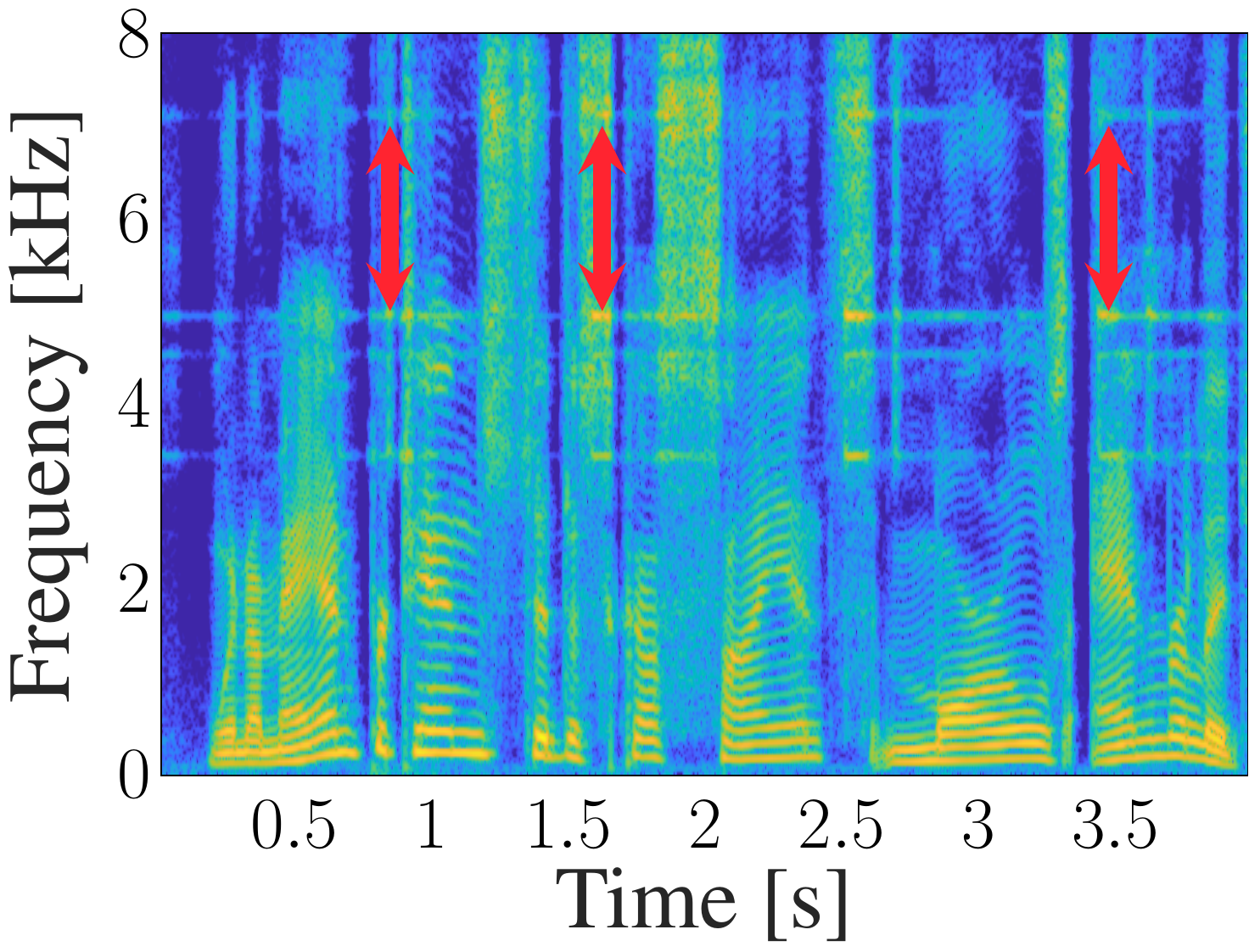}
			\caption{PHASEN}
			\label{fig:nb_phasen_mag}
		\end{subfigure}
	}
	\vspace{1mm}
	\centering
	\centerline{
		\begin{subfigure}[b]{.23\textwidth}
			\centering
			\includegraphics[width=\columnwidth]{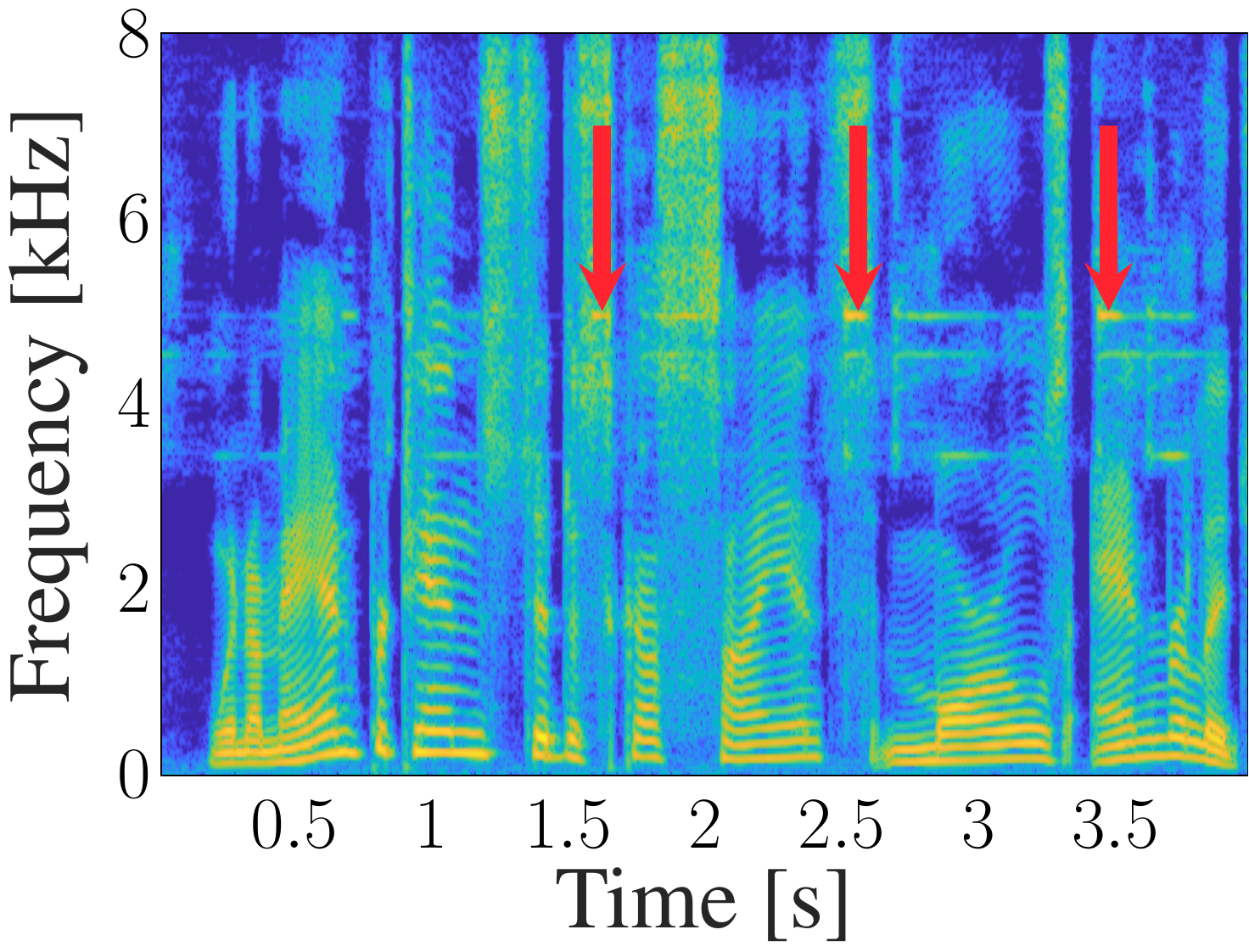}
			\caption{PFPL}
			\label{fig:nb_pfpl_mag}
		\end{subfigure}
		\hspace{12mm}
		\begin{subfigure}[b]{.23\textwidth}
			\centering
			\includegraphics[width=\columnwidth]{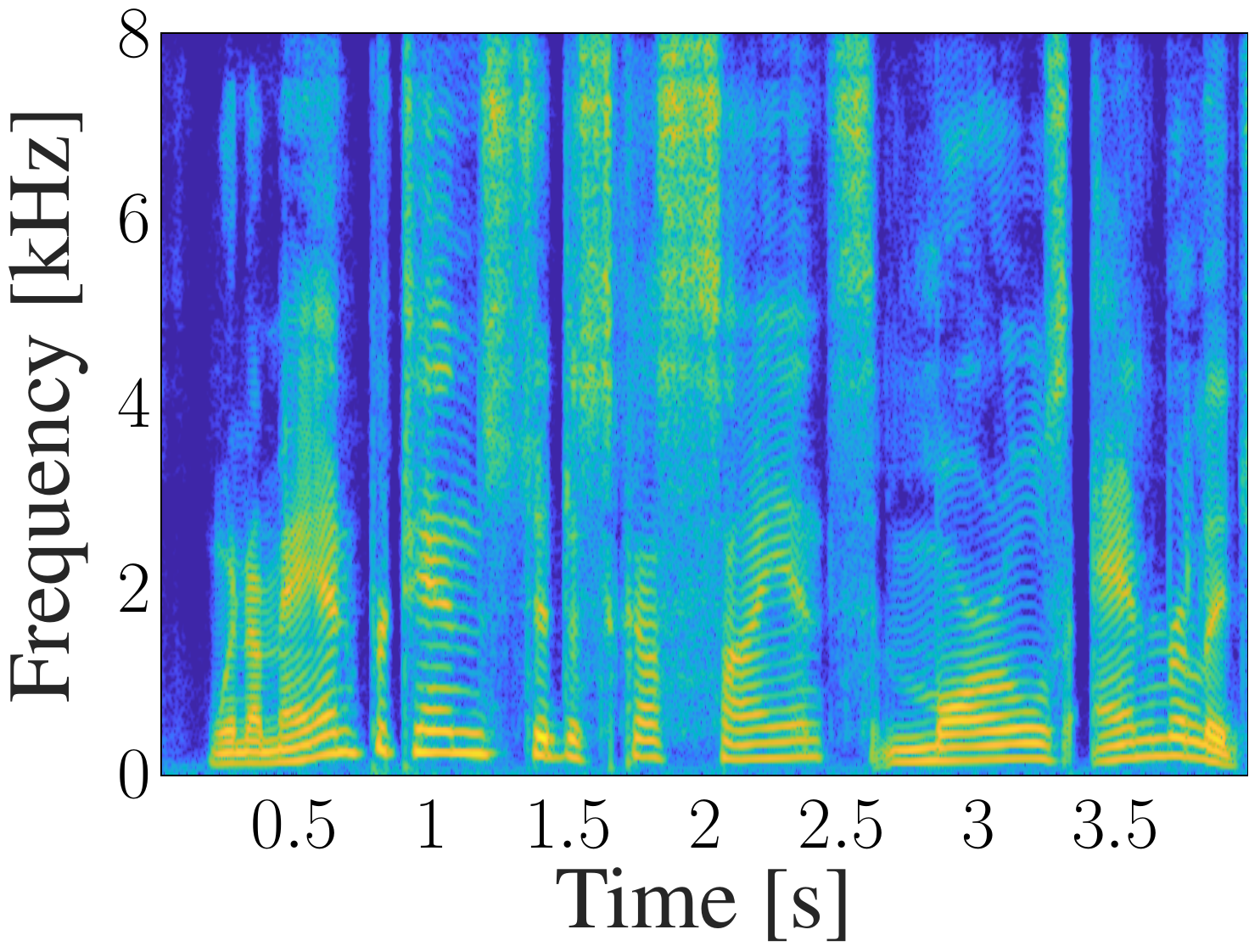}
			\caption{CMGAN}
			\label{fig:nb_cmgan_mag}
		\end{subfigure}
		\hspace{10mm}
		\begin{subfigure}[b]{.23\textwidth}
			\centering
			\includegraphics[width=\columnwidth]{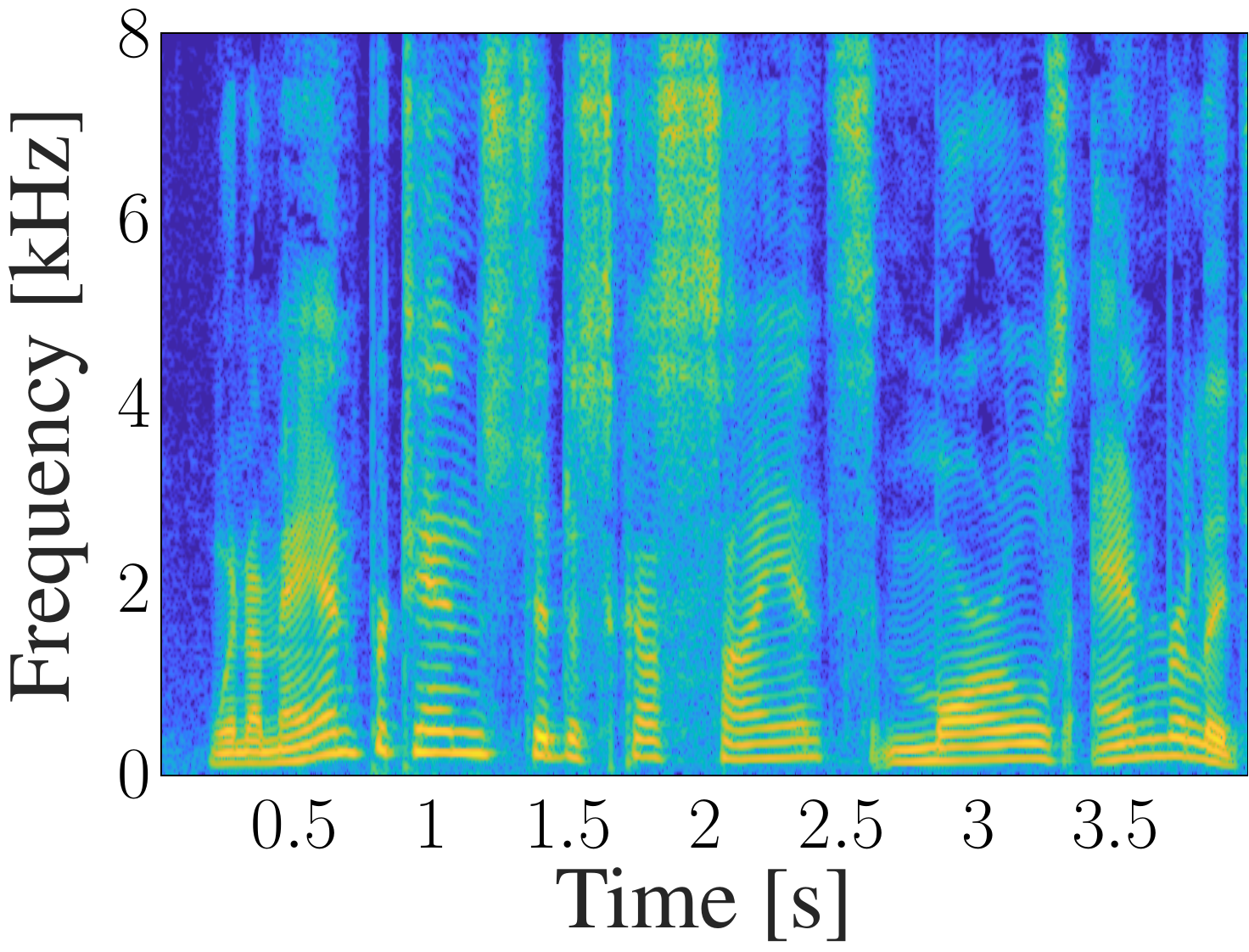}
			\caption{Clean}
			\label{fig:nb_clean_mag}
		\end{subfigure}
	}
	\vspace{1mm}
	\centering
	\centerline{
		\begin{subfigure}[b]{.23\textwidth}
			\centering			
			\includegraphics[width=\columnwidth]{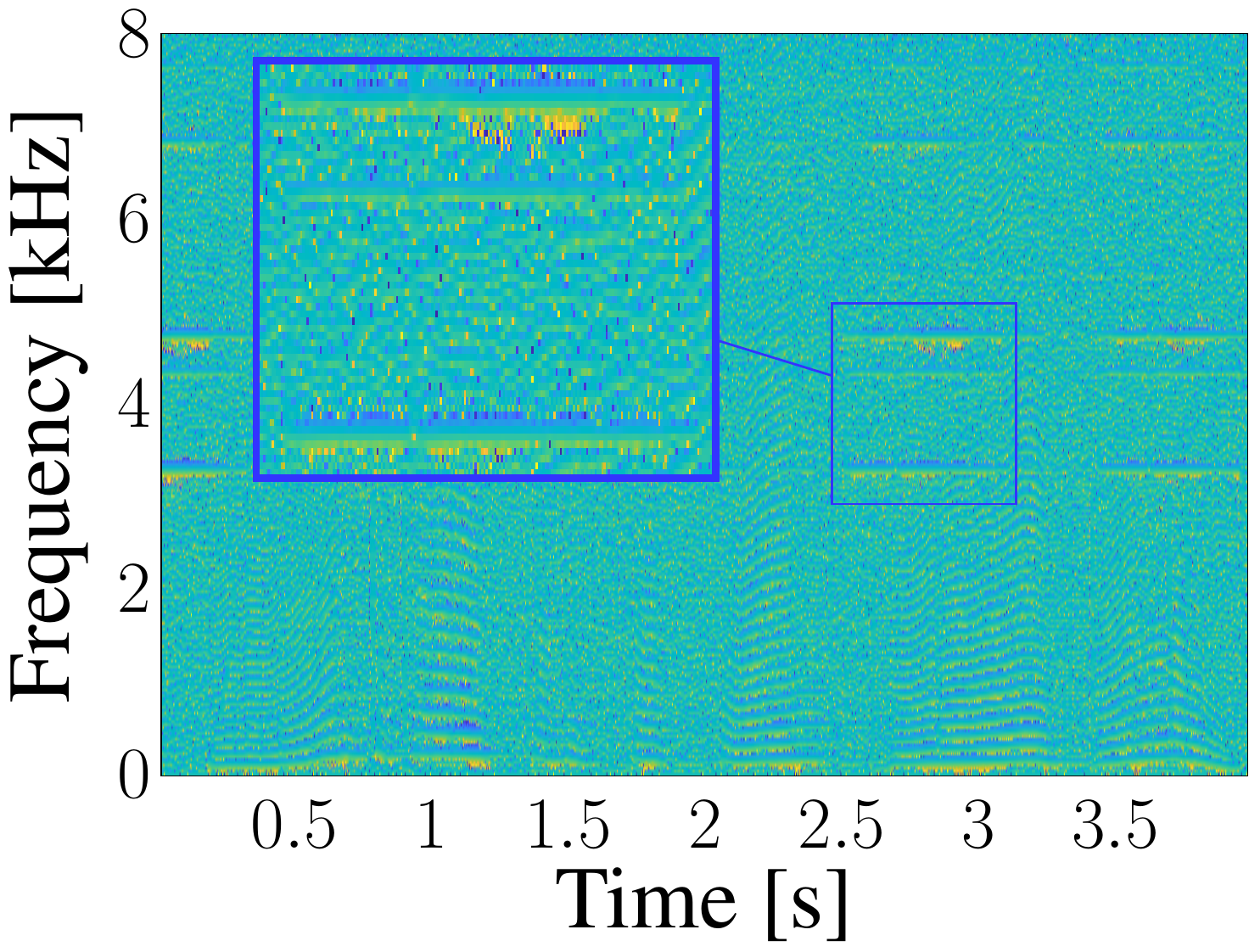}
			\caption{Noisy}
			\label{fig:nb_noisy_phase}
		\end{subfigure}
		\hspace{3.7mm}
		\begin{subfigure}[b]{.23\textwidth}
			\centering
			\includegraphics[width=\columnwidth]{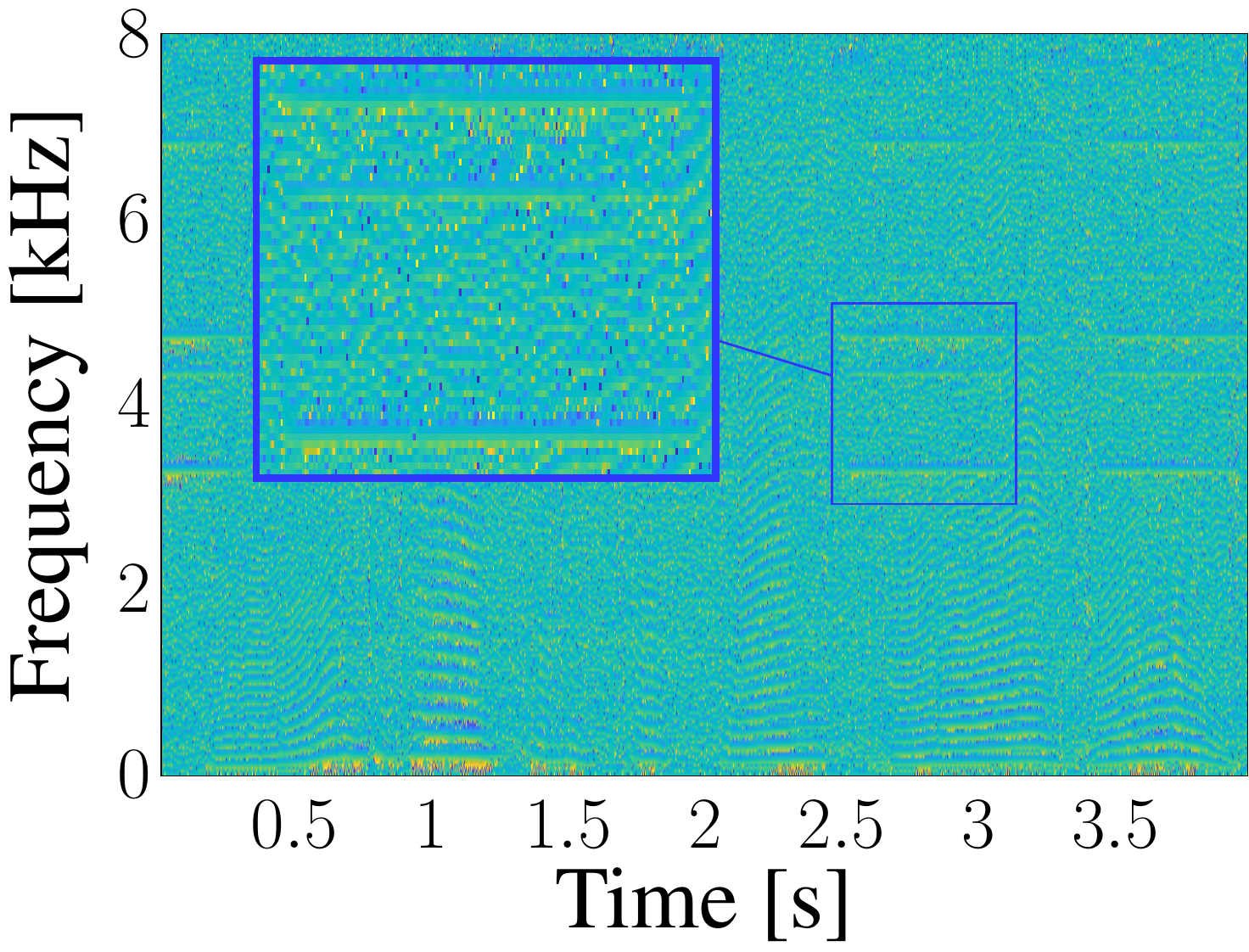}
			\caption{MetricGAN+}
			\label{fig:nb_metricgan_phase}
		\end{subfigure}
		\hspace{2.7mm}
		\begin{subfigure}[b]{.23\textwidth}
			\centering		
			\includegraphics[width=\columnwidth]{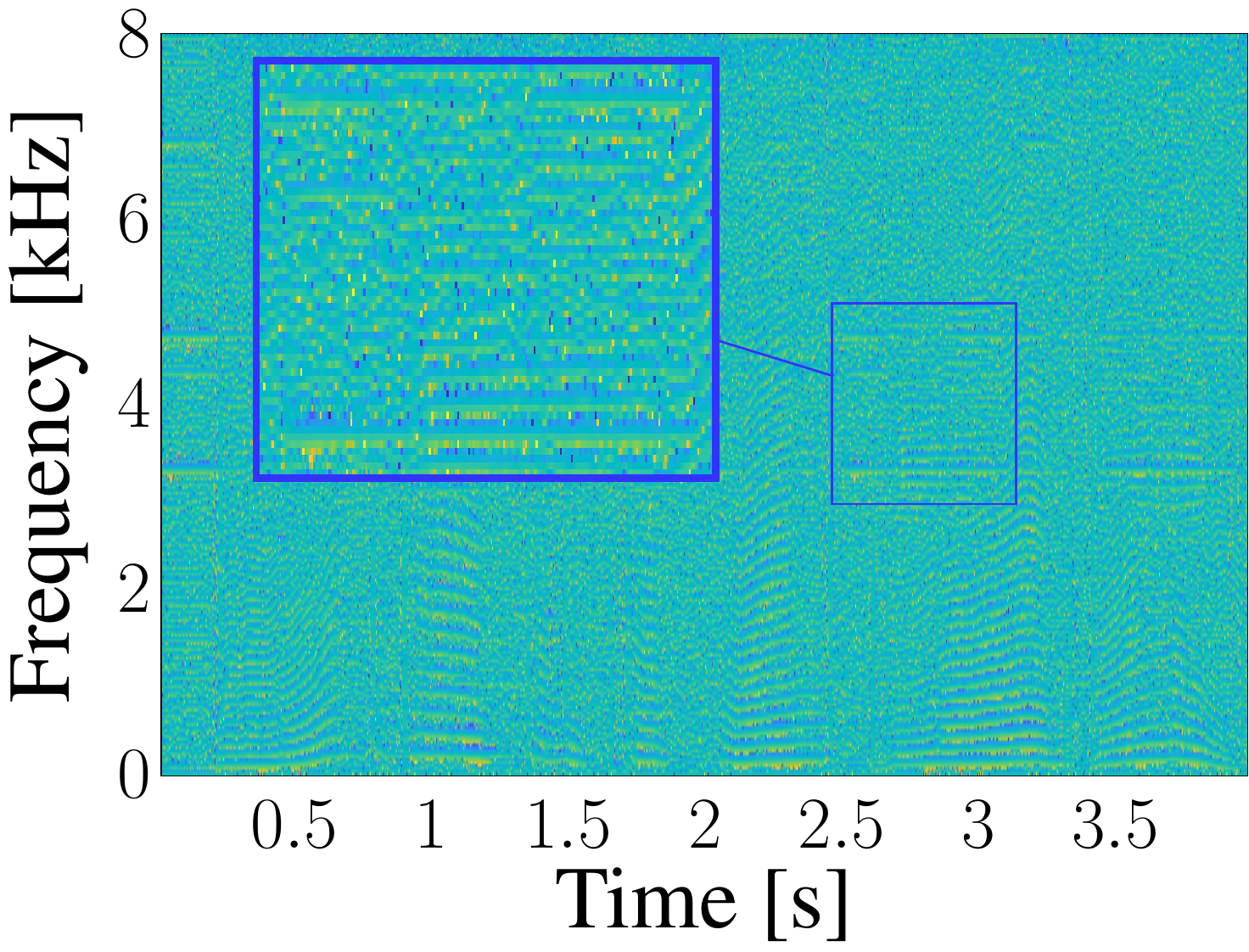}
			\caption{DEMUCS}
			\label{fig:nb_demucs_phase}
		\end{subfigure}
		\hspace{3.7mm}
		\begin{subfigure}[b]{.23\textwidth}
			\centering
			\includegraphics[width=\columnwidth]{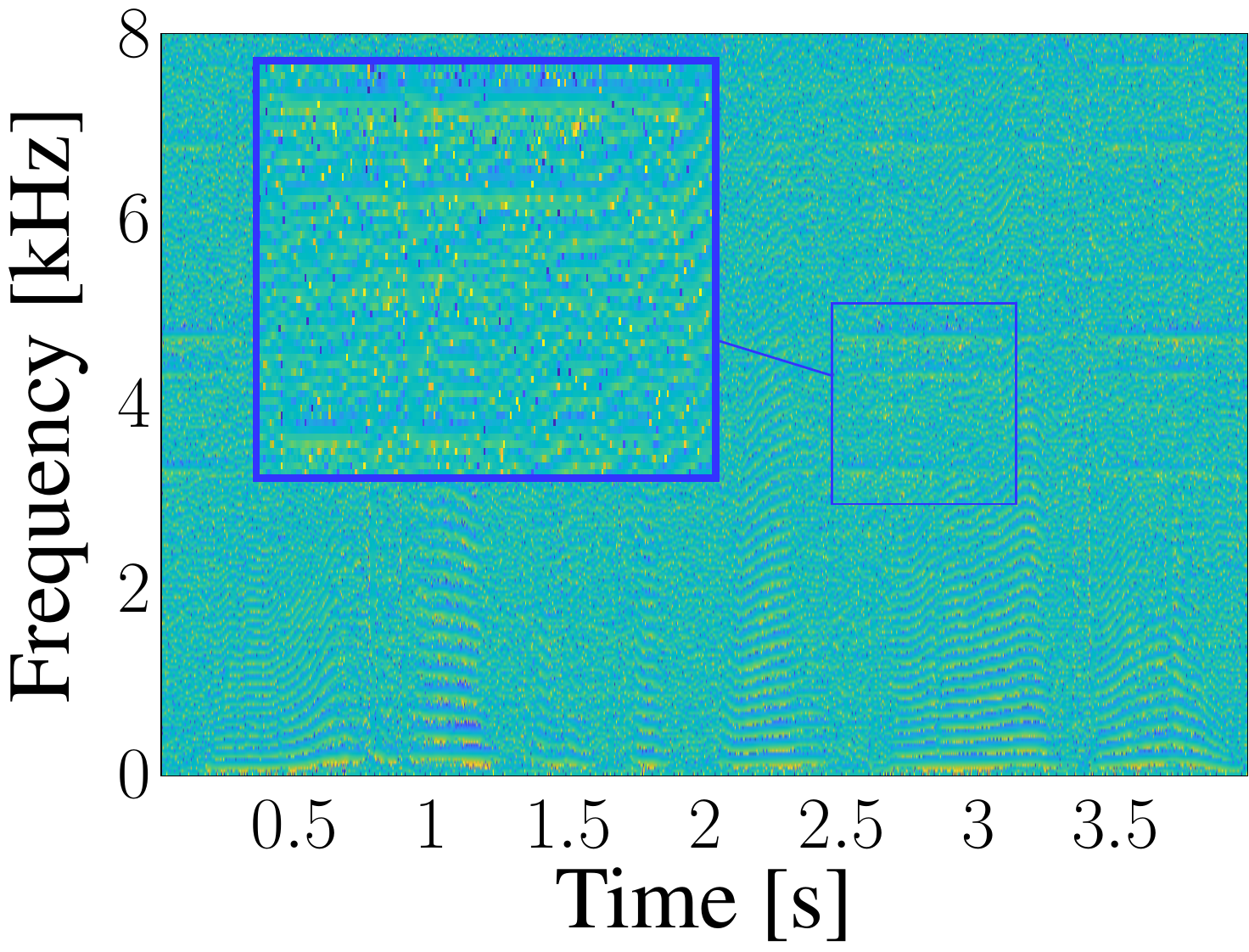}
			\caption{PHASEN}
			\label{fig:nb_phasen_phase}
		\end{subfigure}
	}
	\vspace{1mm}
	\centering
	\centerline{
		\begin{subfigure}[b]{.23\textwidth}
			\centering
			\includegraphics[width=\columnwidth]{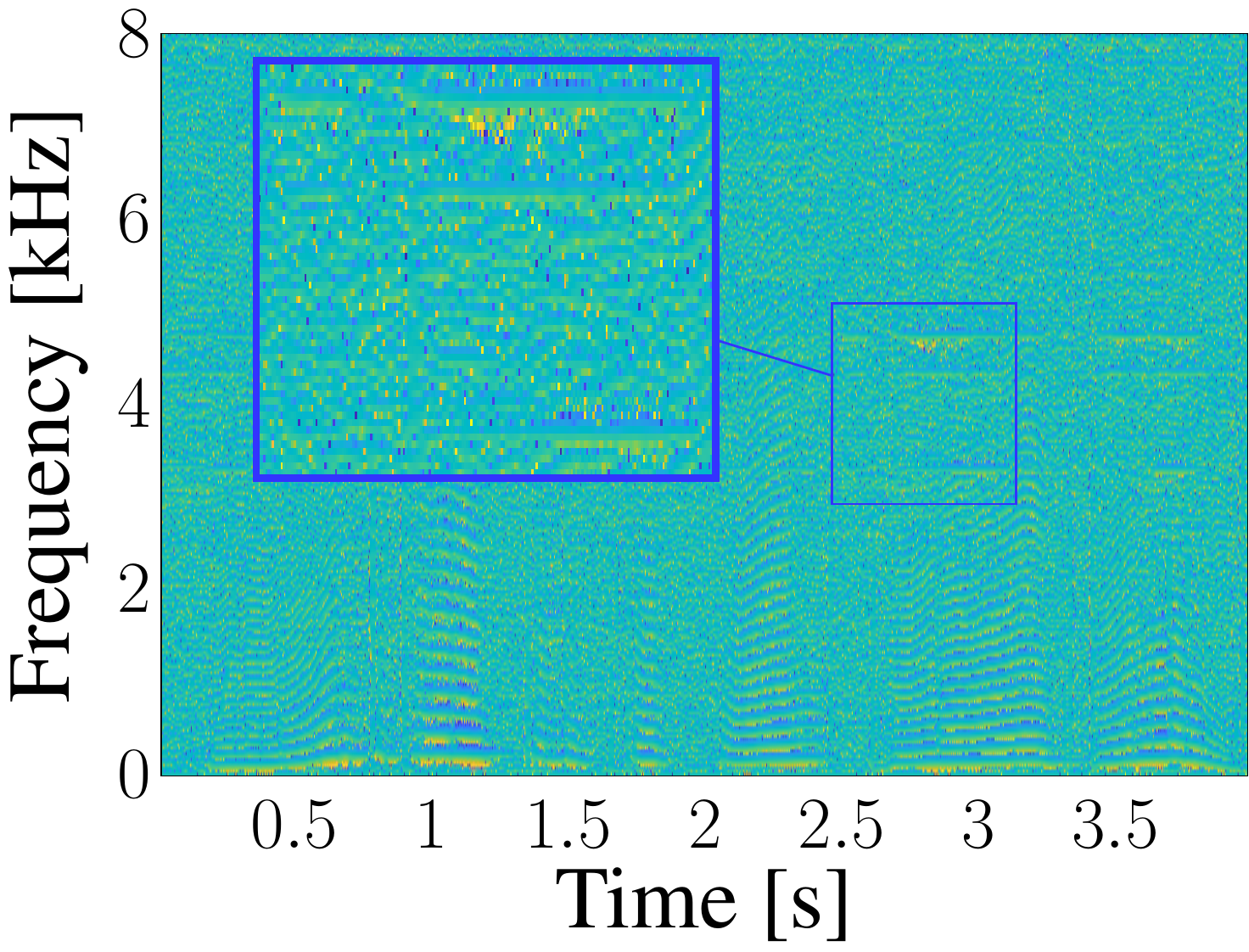}
			\caption{PFPL}
			\label{fig:nb_pfpl_phase}
		\end{subfigure}
		\hspace{12mm}
		\begin{subfigure}[b]{.23\textwidth}
			\centering
			\includegraphics[width=\columnwidth]{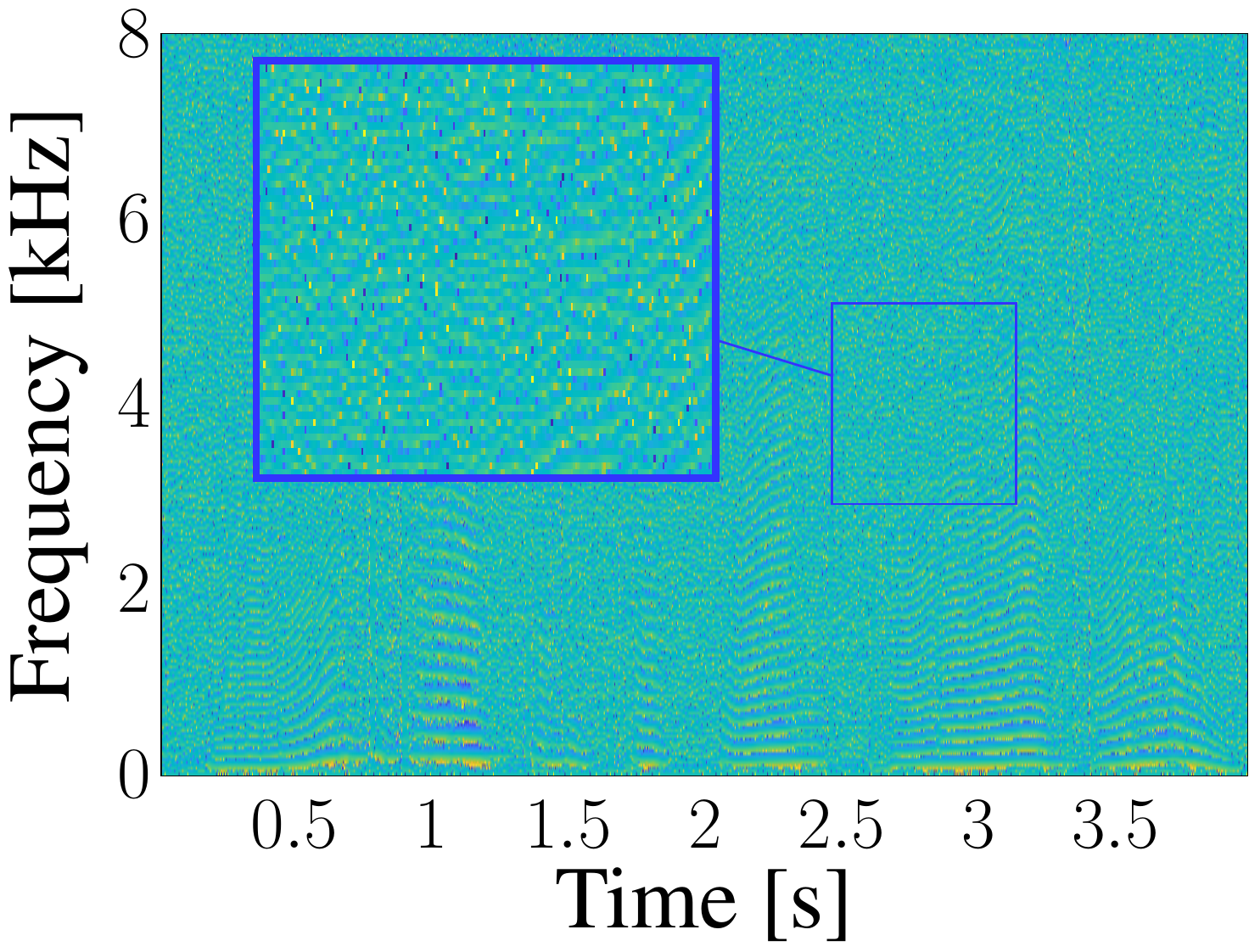}
			\caption{CMGAN}
			\label{fig:nb_cmgan_phase}
		\end{subfigure}
		\hspace{10mm}
		\begin{subfigure}[b]{.23\textwidth}
			\centering
			\includegraphics[width=\columnwidth]{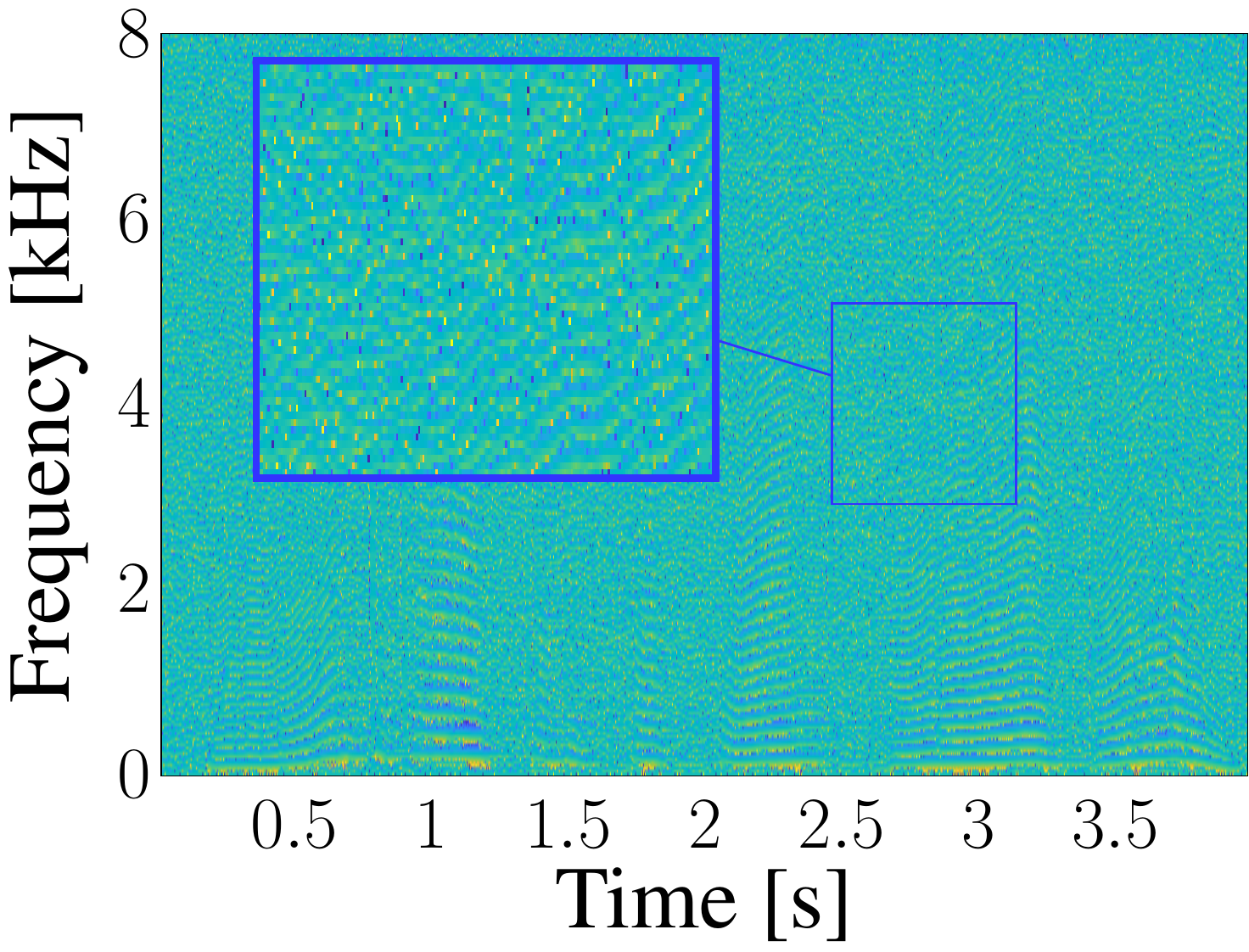}
			\caption{Clean}
			\label{fig:nb_clean_phase}
		\end{subfigure}
	}
	\caption{Visualization of subjective approaches under a narrow-band doorbell noise (Freesound dataset) at SNR = 3~dB. (a-g) represent the time-domain signal, while (h-n) are the TF-magnitude representations in dB and (o-u) are the reconstructed BPD of the given TF-phase representations. (\myarrowblue) and (\myarrowred) reflect the distortions in time and TF-magnitude representations, respectively. \label{fig:nb_subjective}}
\end{figure*}

	In general, stationary noises are less challenging than non-stationary counterparts. However, stationary noises are underrepresented in the training data. As depicted in Fig.~\ref{fig:nb_subjective}, methods such as MetricGAN+ and PHASEN are showing a poor generalization performance, with doorbell distortions clearly visible at frequencies (3.5, 5, and 7 kHz). On the other hand, the performance is slightly better in DEMUCS and PFPL, whereas CMGAN perfectly attenuates all distortions. Note that high-frequency distortions are harder to spot in the time-domain than in TF-magnitude and TF-phase representations.

\vfill
	
\end{document}